%% file: main_publication_ediss.tex
\documentclass[goettingen, gauss, print]{thesis_ediss}

\title{Observations of large-scale solar flows}
\author{Bastian Proxauf}
\town{Freistadt, {\"O}sterreich}

\thesisadvisorycommitteea{Prof. Dr. Laurent Gizon}
\instthesisadvisorycommitteea{Max-Planck-Institut f{\"u}r Sonnensystemforschung und \\
Institut f{\"u}r Astrophysik, Georg-August-Universit{\"a}t G{\"o}ttingen, G{\"o}ttingen, Deutschland}
\thesisadvisorycommitteeb{Prof. Dr. Andreas Tilgner}
\instthesisadvisorycommitteeb{Institut f{\"u}r Geophysik, Georg-August-Universit{\"a}t G{\"o}ttingen, G{\"o}ttingen, Deutschland}
\thesisadvisorycommitteec{Dr. Bj{\"o}rn L{\"o}ptien}
\instthesisadvisorycommitteec{Max-Planck-Institut f{\"u}r Sonnensystemforschung, G{\"o}ttingen, Deutschland}
\refereea{Prof. Dr. Laurent Gizon}
\instrefereea{Max-Planck-Institut f{\"u}r Sonnensystemforschung und \\
Institut f{\"u}r Astrophysik, Georg-August-Universit{\"a}t G{\"o}ttingen, G{\"o}ttingen, Deutschland}
\refereeb{Prof. Dr. Andreas Tilgner}
\instrefereeb{Institut f{\"u}r Geophysik, Georg-August-Universit{\"a}t G{\"o}ttingen, G{\"o}ttingen, Deutschland}
\commissiona{Prof. Dr. Ulrich Christensen}
\instcommissiona{Max-Planck-Institut f{\"u}r Sonnensystemforschung, G{\"o}ttingen, Deutschland}
\commissionb{Prof. Dr. Stefan Dreizler}
\instcommissionb{Institut f{\"u}r Astrophysik, Georg-August-Universit{\"a}t G{\"o}ttingen, G{\"o}ttingen, Deutschland}
\commissionc{Prof. Dr. Wolfram Kollatschny}
\instcommissionc{Institut f{\"u}r Astrophysik, Georg-August-Universit{\"a}t G{\"o}ttingen, G{\"o}ttingen, Deutschland}
\commissiond{PD Dr. Olga Shishkina}
\instcommissiond{Max-Planck-Institut f{\"u}r Dynamik und Selbstorganisation, G{\"o}ttingen, Deutschland}

\submittedyear{2020}
\publicationyear{2021}
\examinationdate{10.06.2020}
\isbn{}

\newcommand{\comment}[1]{}
\usepackage[ngerman,english]{babel}
\usepackage{printlen}
\usepackage{color}
\usepackage{rotating}
\usepackage{pdfpages}

\usepackage{graphicx}
\usepackage{txfonts}
\usepackage{amsmath, amssymb}
\usepackage{mathtools}
\usepackage{commath}
\usepackage{natbib}
\usepackage{siunitx}
\usepackage{array}
\usepackage[dvipsnames]{xcolor}

\DeclareSIUnit[]\solarradius{R_\odot}
\DeclareSIUnit[]\year{yr}

\usepackage[]{hyperref}

\defcitealias{Loeptien2018}{LGBS18}
\defcitealias{Liang2019}{LGBD19}

\usepackage{xspace}

\input{texfiles/main_convection_power/definitions.tex}

\usepackage{caption} 
\makeatletter
\def\blfootnote{\xdef\@thefnmark{}\@footnotetext}
\makeatother

\begin{document}

\includepdf[pages={1}, scale=1.01]{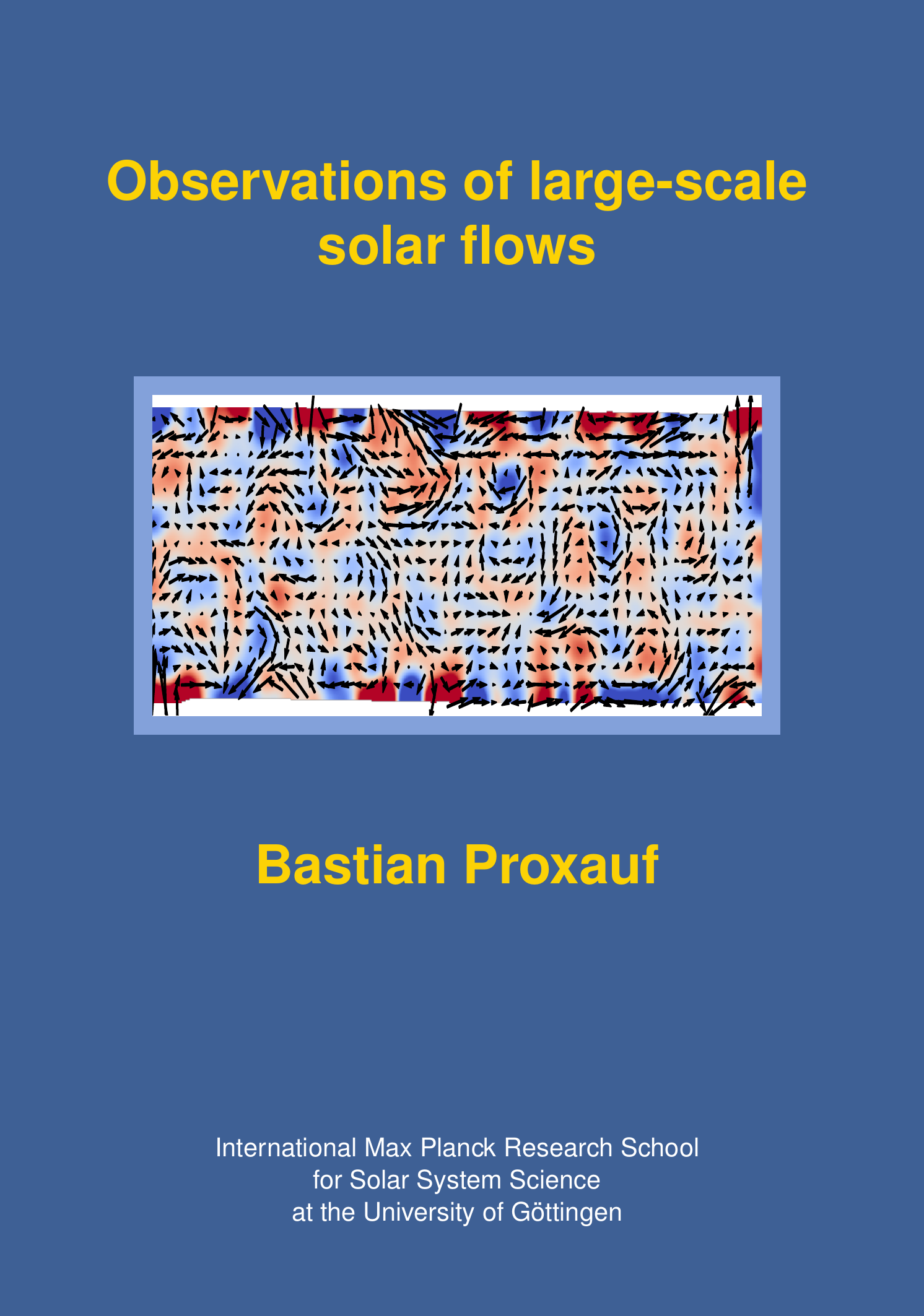}

\selectlanguage{english}
\maketitle

\pagestyle{empty}
\include{secondpage}
\pagestyle{plain}

\tableofcontents
\listoffigures
\listoftables

\chapter*{Summary\markboth{Summary}{Summary}}
\addcontentsline{toc}{chapter}{Summary}

The aim of this thesis is to observationally characterize various large-scale solar flows, including the recently detected solar Rossby waves (waves of radial vorticity), large-scale convection, and flows around active regions. These large-scale flows likely interact with the solar differential rotation and, through a dynamo process, with the solar magnetic field.

To study these flows I use several years of observations from the Helioseismic and Magnetic Imager (HMI) aboard the Solar Dynamics Observatory (SDO). These data are processed using two complementary techniques to obtain horizontal flows on the solar surface and in the solar interior: local correlation tracking, which is limited to the solar surface, and ring-diagram analysis, which is able to probe the near-surface layers in the solar interior (the observational depth limit is roughly $\SI{16}{\mega\metre}$) at a lower temporal and spatial resolution.

First, I study the latitudinal and radial dependence of solar equatorial Rossby waves. For this, the radial vorticity is computed from the horizontal flows and a spectral analysis is applied via a spherical harmonics transform in latitude and longitude and a Fourier transform in time. In the top $\SI{9}{\mega\metre}$ below the surface, the radial dependence of the vorticity eigenfunctions is consistent with a variation of the form $r^{m-1}$, expected from models, where $r$ is the radial coordinate and $m$ is the longitudinal wavenumber. However, systematic errors in the ring-diagram analysis prevent me from constraining the radial eigenfunctions deeper in the solar interior. The latitudinal dependence of the mode eigenfunctions is determined via a correlation analysis between the equator and other latitudes, and via a singular value decomposition. The real part of the eigenfunctions decreases away from the equator and switches sign at absolute latitudes between $20$ and $\SI{30}{\degree}$, in agreement with previous results. The imaginary part of the eigenfunctions has a small, but nonzero, amplitude at all latitudes, which may be indicative of attenuation.

Second, using the horizontal flow maps, I study the energy spectrum of large-scale convection in the context of existing results inferred by time-distance helioseismology and simulations. These results had revealed a huge discrepancy for the velocity of large-scale convection in the solar interior (root-mean-square values of roughly $1$ and $\SI{100}{\metre\per\second}$, respectively). This disagreement, the convective conundrum, is crucial with regard to current models of solar convection. Several issues are found in the existing analysis, such as different conventions for spherical harmonics transforms, missing multiplicative factors, and inconsistent comparisons. The correction of these issues reduces the discrepancy between energy spectra of convection from time-distance helioseismology and simulations, but does not eliminate it entirely. Additionally, new, consistent results from local correlation tracking and ring-diagram analysis are presented, which are closer to the results derived from time-distance helioseismology than those from simulations.

\chapter*{Zusammenfassung\markboth{Zusammenfassung}{Zusammenfassung}}
\addcontentsline{toc}{chapter}{Zusammenfassung}

Das Ziel dieser Thesis ist es, mithilfe von Beobachtungen verschiedene gro{\ss}skalige Str{\"o}mungen zu charakterisieren, insbesondere die vor kurzem entdeckten solaren Rossby-Wellen (Wellen der radialen Vortizit{\"a}t), gro{\ss}skalige Konvektion, und Str{\"o}mungen um aktive Regionen. Diese gro{\ss}skaligen Str{\"o}mungen wechselwirken wahrscheinlich mit der differenziellen Sonnenrotation und, {\"u}ber einen Dynamo-Prozess, mit dem Sonnenmagnetfeld.

Um diese Str{\"o}mungen zu erforschen, verwende ich mehrj{\"a}hrige Beobachtungen des Helioseismic and Magnetic Imager (HMI) an Bord des Solar Dynamics Observatory (SDO). Diese Daten werden mit zwei sich erg{\"a}nzenden Methoden zur Messung von Str{\"o}mungen auf der Sonnenoberfl{\"a}che und im Sonneninneren verarbeitet: Lokalem Korrelationstracking, welches auf die Sonnenoberfl{\"a}che beschr{\"a}nkt ist, und Ring-Diagramm-Analyse, mit welcher die oberfl{\"a}chennahen Schichten im Sonneninneren (das Tiefenlimit liegt bei circa $\SI{16}{\mega\metre}$) mit niedrigerer zeitlicher und r{\"a}umlicher Aufl{\"o}sung erforscht werden k{\"o}nnen.

Zun{\"a}chst erforsche ich die latitudinale und radiale Abh{\"a}ngigkeit von solaren {\"a}quatorialen Rossby-Wellen. Dazu wird die radiale Vortizit{\"a}t aus den horizontalen Str{\"o}mungen berechnet und eine Spektralanalyse {\"u}ber eine sph{\"a}rische harmonische Transformation in der Latitude und Longitude und eine Fourier-Transformation in der Zeit durchgef{\"u}hrt. In den oberen $\SI{9}{\mega\metre}$ unterhalb der Oberfl{\"a}che ist die radiale Abh{\"a}ngigkeit der Vortizit{\"a}tseigenfunktionen konsistent mit einer von Modellen erwarteten {\"A}nderung der Form $r^{m-1}$, wobei $r$ die radiale Koordinate und $m$ die longitudinale Wellenzahl ist. Allerdings k{\"o}nnen die radialen Eigenfunktionen tiefer im Sonneninneren aufgrund von systematischen Fehlern in der Ring-Diagramm-Analyse nicht zuverl{\"a}ssig bestimmt werden. Die Latitudenabh{\"a}ngigkeit der Eigenfunktionen der Moden wird {\"uber} eine Korrelations-Analyse zwischen dem {\"A}quator und anderen Latituden, und {\"u}ber eine Singul{\"a}rwertzerlegung bestimmt. Der Realteil der Eigenfunktionen nimmt vom {\"A}quator weg ab und {\"a}ndert sein Vorzeichen bei absoluten Latituden zwischen $20$ und $\SI{30}{\degree}$. Dies stimmt mit vorherigen Ergebnissen {\"u}berein. Der Imagin{\"a}rteil der Eigenfunktionen besitzt eine kleine Amplitude ungleich Null bei allen Latituden, was eventuell auf einen D{\"a}mpfungsprozess deutet.

Anschlie{\ss}end erforsche ich mithilfe von Karten der horizontalen Str{\"o}mungen das Energiespektrum von gro{\ss}skaliger Konvektion im Kontext vorhandener Ergebnisse, die durch Zeit-Distanz-Helioseismologie und Simulationen erhalten wurden. Diese Ergebnisse hatten eine riesige Diskrepanz f{\"u}r die Geschwindigkeit von gro{\ss}skaliger Konvektion im Sonneninneren offenbart (quadratische Mittelwerte von circa $1$ beziehungsweise $\SI{100}{\metre\per\second}$). Diese Diskrepanz, das konvektive Dilemma, ist von essenzieller Bedeutung in Bezug auf aktuelle Modelle der Sonnenkonvektion. In der vorhandenen Analyse wurden einige Probleme gefunden, beispielsweise unterschiedliche Konventionen f{\"u}r sph{\"a}rische harmonische Transformationen, fehlende multiplikative Faktoren, und inkonsistente Vergleiche. Das Beheben dieser Probleme reduziert die Diskrepanz zwischen den Energiespektren der Konvektion von Zeit-Distanz-Helioseismologie und Simulationen, entfernt sie allerdings nicht vollst{\"a}ndig. Zus{\"a}tzlich werden neue, konsistente Ergebnisse von lokalem Korrelationstracking und Ring-Diagramm-Analyse pr{\"a}sentiert, welche n{\"a}her an den Ergebnissen der Zeit-Distanz-Helioseismologie als jenen der Simulationen liegen.

\include{introduction}

\include{main_rossby_waves}

\include{main_convection_power}

\include{discussion}

\bibliography{literature}
\bibstyle{thesis}

\chapter*{Scientific contributions\markboth{Scientific contributions}{Scientific contributions}}
\addcontentsline{toc}{chapter}{Scientific contributions}

{\bf\large Refereed publications}\\
\begin{itemize}
\item B.~L{\"o}ptien, A.~C.~Birch, T.~L.~Duvall, L.~Gizon, \textbf{B.~Proxauf}, and J.~Schou: \textit{Measuring solar active region inflows with local correlation tracking of granulation}, Astronomy and Astrophysics, 606, A28 (2017)
\item B.~L{\"o}ptien, L.~Gizon, A.~C.~Birch, J.~Schou, \textbf{B.~Proxauf}, T.~L.~Duvall, R.~S.~Bogart, and U.~R.~Christensen: \textit{Global-scale equatorial Rossby waves as an essential component of solar internal dynamics}, Nature Astronomy, 2, 568-573 (2018)
\item \textbf{B.~Proxauf}, L.~Gizon, B.~L{\"o}ptien, J.~Schou, A.~C.~Birch, and R.~S.~Bogart: \textit{Exploring the latitude and depth dependence of solar Rossby waves using ring-diagram analysis}, Astronomy and Astrophysics, 634, A44 (2020)
\end{itemize} ~\\
{\bf\large Paper in preparation}\\
\begin{itemize}
\item A.~C.~Birch, T.~L.~Duvall, L.~Gizon, S.~Hanasoge, B.~W.~Hindman, \textbf{B.~Proxauf}, and K.~R.~Sreenivasan: \textit{Revisiting helioseismic constraints on solar convection}
\end{itemize}~\\
{\bf\large Conference contributions}\\
\begin{itemize}
\item Rocks \& Stars II, G{\"o}ttingen, Germany, 13 -- 16~September 2017 \newline
Oral presentation: \textit{Depth dependence of solar horizontal flows from SDO ring-diagram analysis} 
\item Annual Meeting of the German Astronomical Society, G{\"o}ttingen, Germany, 18 -- 22~September 2017 \newline
Poster: \textit{Solar near-surface flows from ring-diagram helioseismology}
\item XXXth General Assembly of the International Astronomical Union, Vienna, Austria, 20 -- 31~August 2018 \newline
Poster: \textit{On the depth dependence of solar equatorial Rossby waves}
\item 2018 Solar Dynamics Observatory Workshop \textit{Catalyzing Solar Connections}, Ghent, Belgium, 29~October -- 2~November 2018 \newline
Poster and e-poster: \textit{On the depth dependence of solar equatorial Rossby waves}
\item 2nd Max Planck Partner Group Workshop on Solar Physics, Mumbai, India, 17 -- 21~March 2019 \newline
Oral presentation: \textit{Exploring the latitude and depth dependence of solar Rossby waves}
\item 234th Meeting of the American Astronomical Society, St. Louis, USA, 9 -- 13~June 2019 \newline
Oral presentation: \textit{On the latitude dependence of Rossby waves in the Sun}
\end{itemize}

\chapter*{Acknowledgements\markboth{Acknowledgements}{Acknowledgements}}
\addcontentsline{toc}{chapter}{Acknowledgements}

First, I wish to thank Laurent Gizon, who firmly and patiently guided me through my doctorate, always encouraging me to think differently about particular questions and to increase my knowledge on various subjects. Furthermore I wish to express my sincere gratitude to Bj{\"o}rn L{\"o}ptien, whose kind and thorough supervision helped me whenever I had questions. Without their patience and tireless support, this thesis would not have been possible.

I am also deeply indebted to Aaron Birch, for his creativity in thinking about problems, his never-ending helpfulness, and particularly his contributions and support for Chap.~\ref{chap_main_convection}. Jesper Schou is thanked for sharing his knowledge on a variety of subjects and countless interesting discussions. Rick Bogart's help and expertise on all matters related to ring-diagram analysis supported me during the initial phase of my studies and contributed to a significant part of the data analysis in this thesis.

I want to express my gratitude to Andreas Tilgner for being in my thesis advisory committee, evaluating and assuring the progress of my studies, and for being the second referee of this thesis. I also would like to thank Ulrich Christensen, Stefan Dreizler, Wolfram Kollatschny and Olga Shishkina for their membership in my thesis defense committee.

Many thanks go to Zhi-Chao Liang for an independent confirmation of the results in Chap.~\ref{chap_main_rossby}. My deep thankfulness is also directed at Nils Gottschling and Paul-Louis Poulier for numerous, detailed discussions about local correlation tracking and active regions, and at Yuto Bekki for several discussions about Rossby waves and large-scale convection. I deeply owe Vincent B{\"o}ning and Chris Goddard for several new, interesting insights and their overall support on a variety of matters. I also greatly appreciate a number of chats and useful discussions I had with Cilia Damiani, Robert Cameron, Hannah Schunker, Tom Duvall, Christian Baumgartner, Dan Yang, Felix Mackebrandt, and Jan Langfellner. Ray Burston is thanked in his role as the administrator of the data record management system DRMS.

I also thank the International Max Planck Research School for Solar System Science for enabling me to pursue my doctoral studies in this wonderful program. My thankfulness goes in particular to Sonja Schuh as the program coordinator, whose door was always open for everything related to the smooth progression of my PhD studies.

At the end, I want to say thanks to my family and friends, without whom my PhD studies would not have been possible. Their help and interest in my studies was invaluable and boosted my motivation. Finally, my deepest gratitude is dedicated to my personal sun, Erika Avenda{\~n}o-Guzm{\'a}n, who was always there for me, even when the skies where cloudy in G{\"o}ttingen. Her endless love and support, which could be seen without any telescope, brightened any dark days I had and made the sunny ones unforgettable.

\end{document}

%% file: texfiles/main_convection_power/definitions.tex
\newcommand{\shravan}{HDS2012\xspace}
\newcommand{\gizon}{GB2012\xspace}
\newcommand{\greer}{GHFT2015\xspace}
\newcommand{\arfm}{HGS2016\xspace}
\newcommand{\miesch}{M2008\xspace}
\newcommand{\roudier}{R2012\xspace}

\newcommand{\rmd}{ {\ \mathrm d} }

\newcommand{\bx}{\mathbf{x}}
\newcommand{\bk}{\mathbf{k}}

\chardef\us=`\_

%% file: secondpage.tex
\vspace*{\fill}

\noindent\textbf{Cover figure:} Large-scale flows on the solar surface as derived by tracking the motion of small convection cells (granules) on observations from the Helioseismic and Magnetic Imager instrument aboard the Solar Dynamics Observatory spacecraft. The flows have been averaged over roughly $27$~days (one solar rotation). The flow velocities are given by the arrows and are shown as a function of latitude (covering $\pm\SI{65}{\degree}$) and longitude (covering $\SI{360}{\degree}$). Additionally, the flow vorticity (a measure for local twists in the velocity field) is shown by the color image, with blue and red indicating a clockwise/counter-clockwise flow curvature.

%% file: introduction.tex
\chapter{Introduction}
\label{chap_introduction}

\urlstyle{same}

\blfootnote{Disclaimer: Several figures in this introduction originate from existing publications and have been reproduced with permission. Figures~\ref{fig_hathaway2015_butterfly_ssn}, \ref{fig_howe2009_rotation}, \ref{fig_hathaway2015_butterfly_mag} and \ref{fig_convection_scales} (top right panel) have been reproduced under the \textit{Creative Commons CC BY license 4.0} (see \textit{\url{https://creativecommons.org/licenses/by/4.0/legalcode}}). Figures~\ref{fig_komm2018_zonal_meridional}, \ref{fig_rossby_waves} (right panel), \ref{fig_hanasoge2016_lct}, \ref{fig_gizon2010_ell_nu} and \ref{fig_rda_concept} have been reproduced under licenses provided by the respective journals via \textit{RightsLink}. Figures~\ref{fig_convection_zone_waves}, \ref{fig_convection_scales} (top left panel) and \ref{fig_hmi_data} have been reproduced under reproduction rights granted for educational/academic purposes. Figures~\ref{fig_rossby_waves} (left panel) and \ref{fig_convection_scales} (bottom panel) have been reproduced under reproduction rights granted by the \textnormal{American Astronomical Society} and \textnormal{IOP Publishing}, with the consent of the authors of the respective publications.}

\section{The dynamic Sun}

While the Sun may appear as a static star to the human eye, it is in fact highly dynamic and variable. For example, very soon after the advent of the first telescopes, around 1610, Galileo Galilei observed dark spots on the solar surface moving across the visible disk. It soon became clear that this motion is due to a rotation of the Sun and Galilei was able to calculate the rotation rate of these \textnormal{sunspots}. Only a few years later, in 1630, Christoph Scheiner noticed that the sunspots rotate slower at higher latitudes and faster close to the equator and thus introduced the concept of \textnormal{differential rotation} to the solar community, i.e. the rotation rate decreases with latitude. Based on his own measurements of the mean \textnormal{synodic} sunspot rotation period of $27.2753$~days, in 1863, Richard Carrington invented an ordering system of \textnormal{Carrington rotations} (CRs), which is still in use nowadays.

\begin{figure}
\centering
\begin{minipage}[t]{\textwidth}
\includegraphics[width=\textwidth]{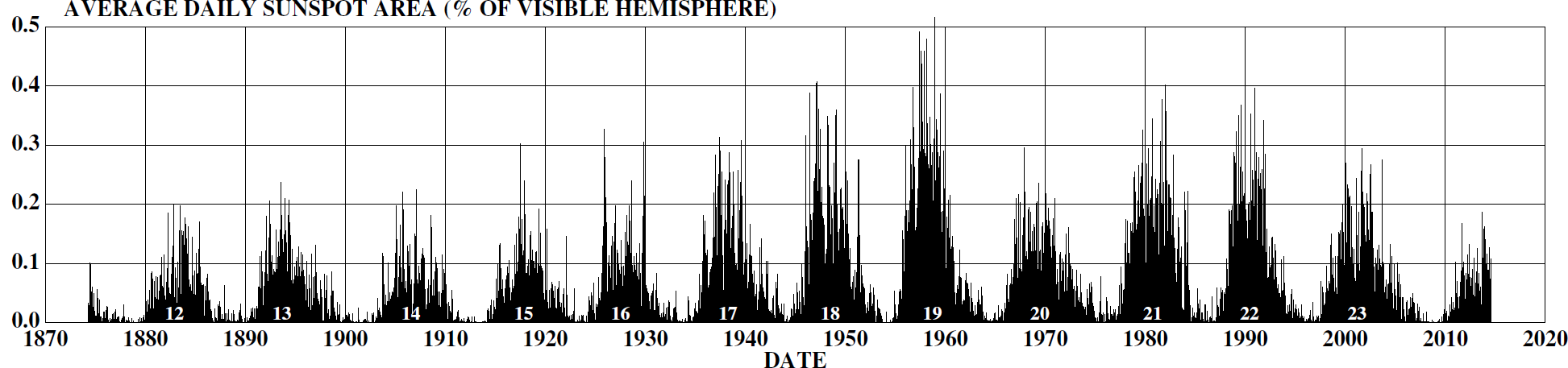}
\end{minipage}

\vspace{0.5cm}
\begin{minipage}[t]{\textwidth}
\includegraphics[width=\textwidth]{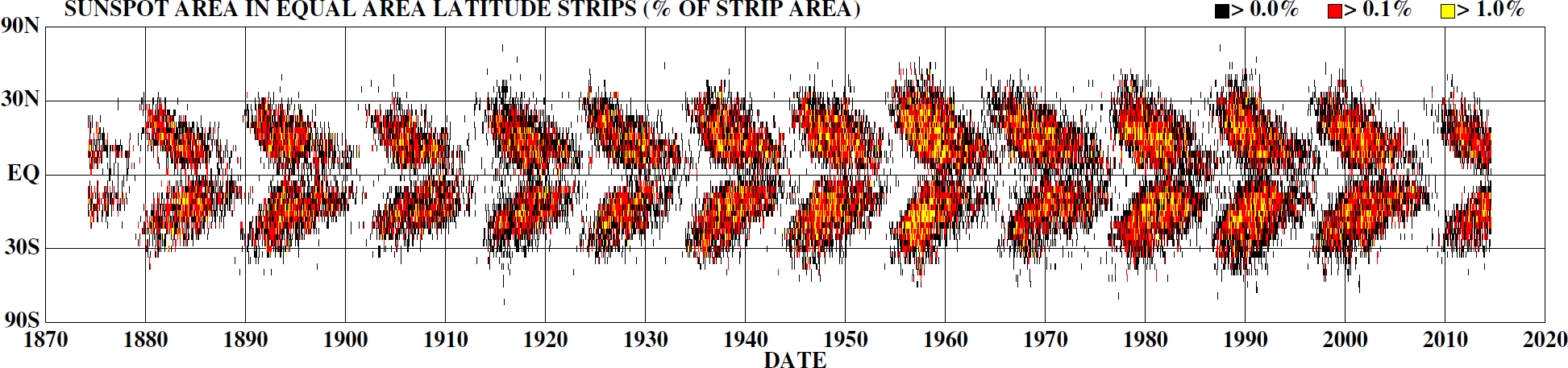}
\end{minipage}
\caption[Sunspot cycle and sunspot butterfly diagram]{
The sunspot cycle. Top: Fractional sunspot area of the visible solar disk versus time. Bottom: The solar butterfly diagram. Fractional sunspot area in equal area latitude strips versus time and latitude. From \citet{Hathaway2015a}, with permission.
}
\label{fig_hathaway2015_butterfly_ssn}
\end{figure}

Almost at the same time, in 1843, Samuel Heinrich Schwabe observed that the number of sunspots visible on the Sun varies with a period of roughly $10$~years \citep{Schwabe1844}. These sunspot or \textnormal{solar cycles} actually have an average period of rather $11$~years and they are the most easily visible manifestation of solar variability (Fig.~\ref{fig_hathaway2015_butterfly_ssn}, top panel). $65$~years later, George Ellery Hale discovered from the splitting of spectral lines due to the \textnormal{Zeeman effect} that the sunspots are intimately linked to the solar magnetic field \citep{Hale1908}. Hale also noticed that sunspots at any given latitude are typically bipolar, with the two polarities of the sunspots being opposite between opposite hemispheres and between successive cycles (\textnormal{Hale's law}), while Alfred Harrison Joy found that the leading polarity is typically closer to the equator than the trailing one, with an angle increasing with latitude (\textnormal{Joy's law}, \citealt{Hale1919}). Despite these huge successes, at that time observations of the Sun were unfortunately always limited to the solar surface.

This changed in 1962, with further evidence for solar variability, when Robert Leighton observed that the Sun oscillates with periods predominantly around $5$~min, or equivalently frequencies around $3$~mHz \citep{Leighton1962}. This discovery formed the basis of \textnormal{helioseismology}, the study of the Sun using waves. Similar to seismology on Earth, the waves carry information about the matter they traverse and their frequency is shifted along the wave path. The characterization of the waves and the analysis of oscillation \textnormal{power spectra} enabled us to look into the solar interior and thus increased our knowledge about the Sun dramatically.

\begin{figure}
\centering
\includegraphics[width=\textwidth]{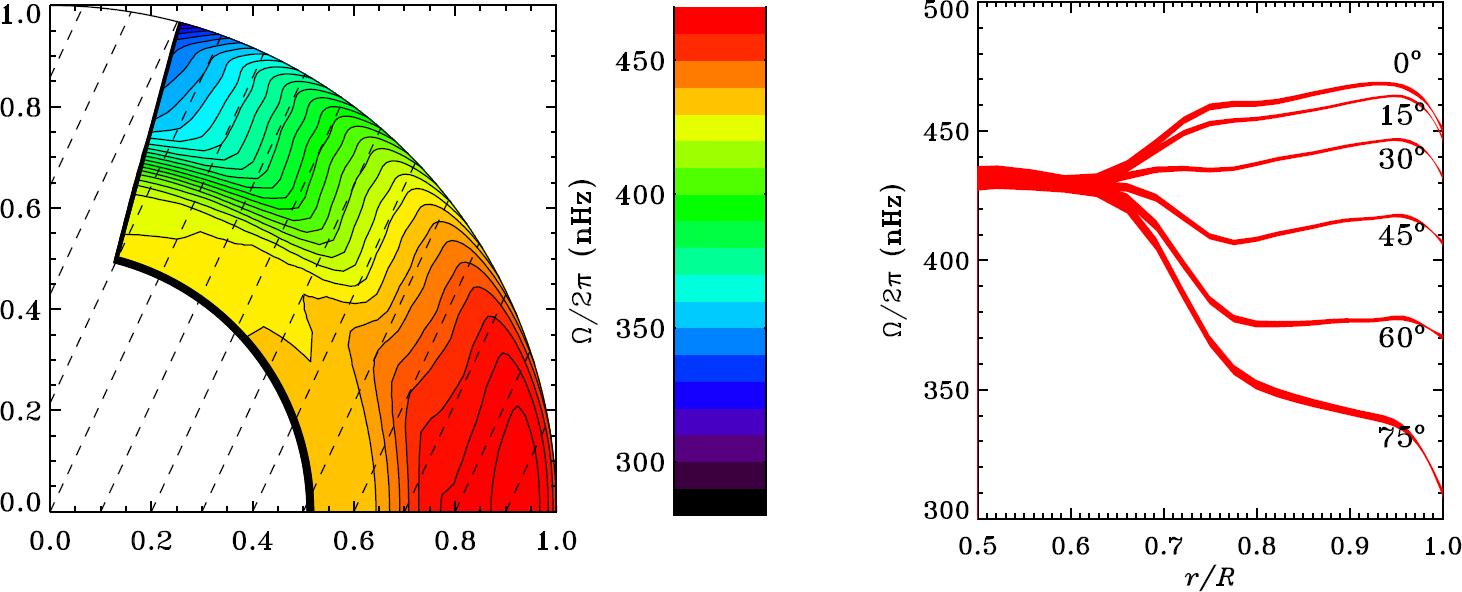}
\caption[Solar differential rotation]{
The solar differential rotation. Left: Contours of the rotation rate in a meridional plane (the solar rotation axis is pointing upwards). The dashed lines indicate a $\SI{25}{\degree}$ angle from the rotation axis. Right: Rotation rate versus radius, for different latitudes. From \citet{Howe2009}, with permission. 
}
\label{fig_howe2009_rotation}
\end{figure}

\begin{figure}
\centering
\includegraphics[width=\textwidth]{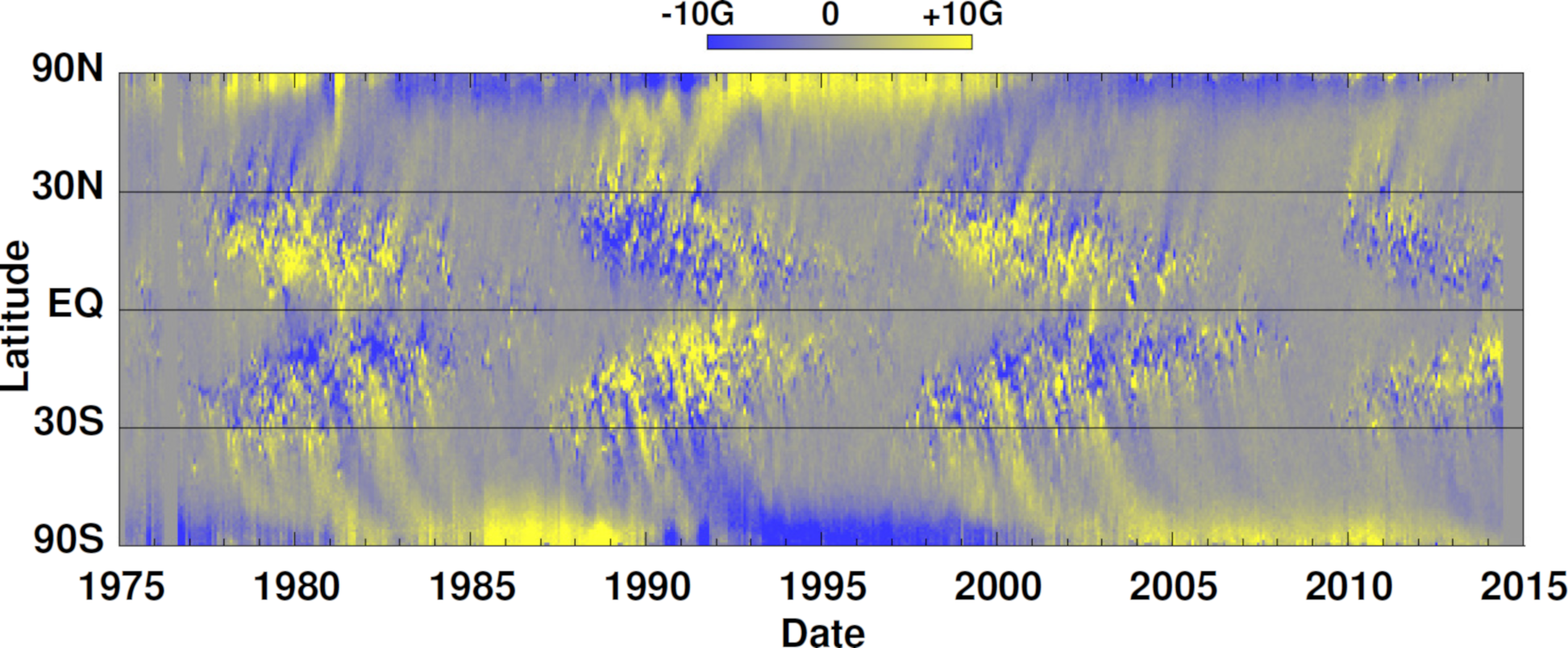}
\caption[Magnetic butterfly diagram]{
The solar magnetic butterfly diagram. Radial magnetic field, averaged over longitude, versus time and latitude. The image also visualizes Hale's law and Joy's law (see text). From \citet{Hathaway2015a}, with permission.
}
\label{fig_hathaway2015_butterfly_mag}
\end{figure}

\begin{figure}
\centering
\begin{minipage}[t]{\textwidth}
\includegraphics[width=\textwidth]{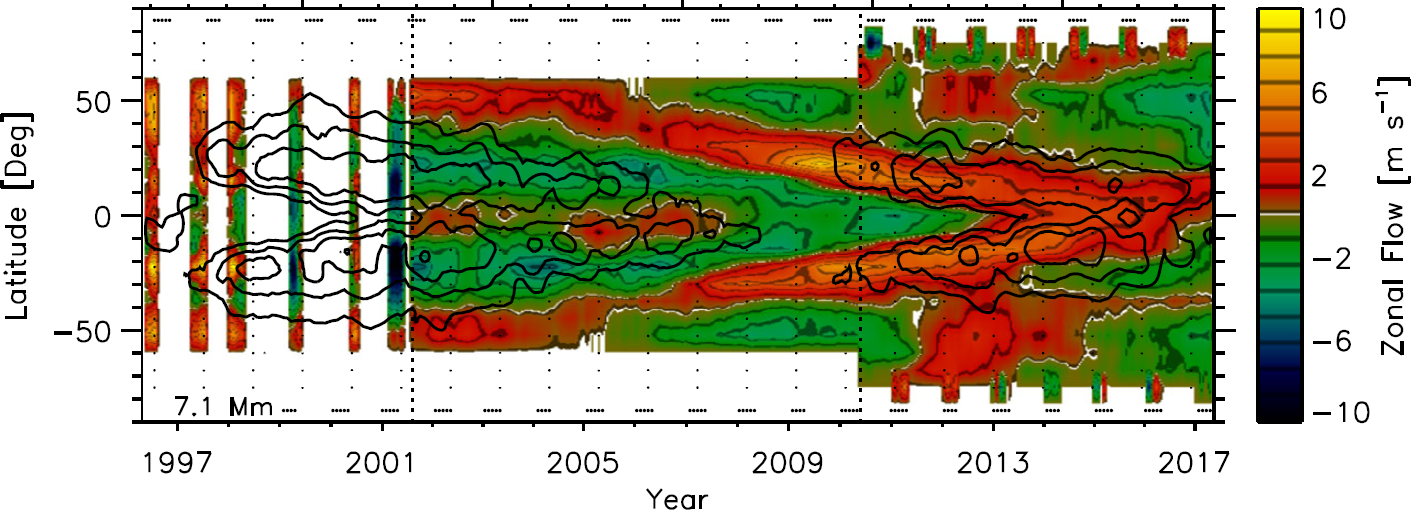}
\end{minipage}

\vspace{0.5cm}
\begin{minipage}[t]{\textwidth}
\includegraphics[width=\textwidth]{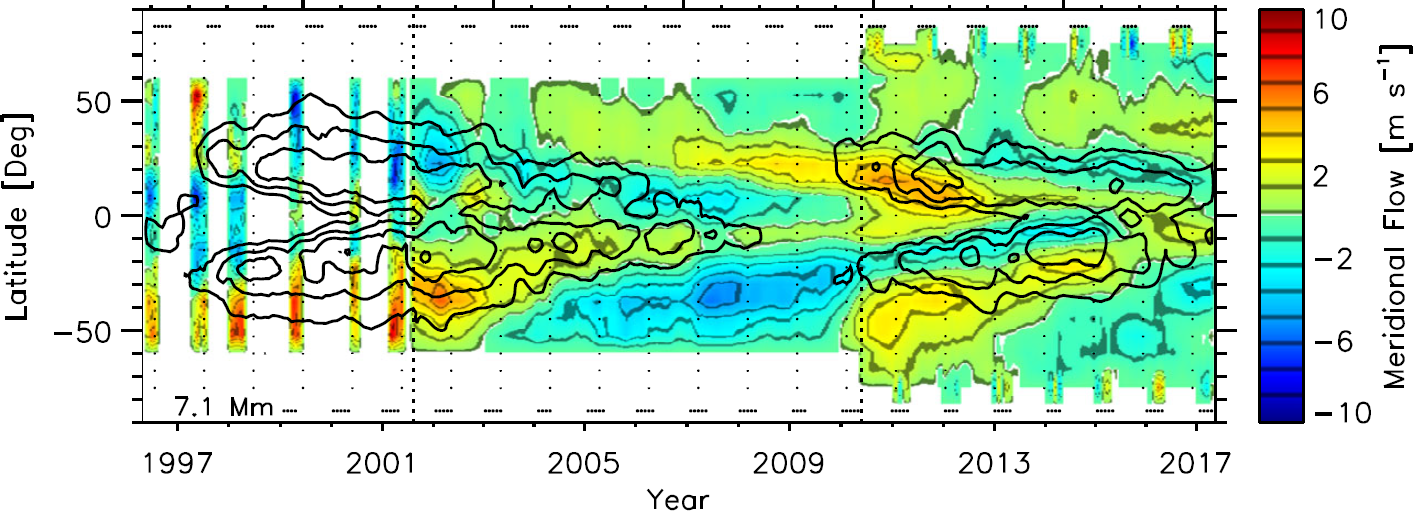}
\end{minipage}
\caption[Torsional oscillations and residual meridional flow]{
Top: Torsional oscillations, i.e. zonal flows as a function of time and latitude after subtraction of the time-independent component and averaged over longitude. Bottom: Residual meridional flow, i.e. the analogue for the meridional flows. Positive velocities indicate \textnormal{prograde} and northward motions, respectively. The images combine data at a depth of $\SI{7.1}{\mega\metre}$ from three different instruments, with different observation periods (vertical dashed lines). The solid black lines show magnetic field contours ($5$, $10$, $20$ and $40$~G), smoothed over five solar rotations. From \citet{Komm2018}, with permission.
}
\label{fig_komm2018_zonal_meridional}
\end{figure}

We now know the interior rotation profile for a significant part of Sun (Fig.~\ref{fig_howe2009_rotation}), in particular that the differential rotation rate increases with depth close to the surface (in the \textnormal{near-surface shear layer}) and that the rotation becomes uniform around $\SI{0.7}{\solarradius}$ (at the so-called \textnormal{tachocline}), see e.g. \citet{Howe2000} and the reviews by \citet{Thompson2003} and \citet{Howe2009}. Additionally we know that there is a $\sim\SI{10}{\metre\per\second}$ poleward belt flow, the \textnormal{meridional flow} \citep{Hathaway1996}. The sunspot area and the magnetic field as a function of time and latitude (Fig.~\ref{fig_hathaway2015_butterfly_ssn}, bottom panel, and Fig.~\ref{fig_hathaway2015_butterfly_mag}), the so-called sunspot and magnetic \textnormal{butterfly diagrams}, are routinely recorded nowadays. Both the rotation and the meridional flow vary along with the solar cycle in the form of bands of faster- and slower-than-average velocities (Fig.~\ref{fig_komm2018_zonal_meridional}), called \textnormal{torsional oscillations} \citep{Howard1980} and \textnormal{residual meridional flow} \citep{Snodgrass1996, Beck2002}, respectively. This indicates that there is a link between flows and magnetic activity. Helioseismology also allows us to define standard solar reference models such as \textnormal{Model S} \citep{Christensen-Dalsgaard1996} as well as to determine such fundamental parameters as the age of the Sun.

Finally and maybe most importantly, our knowledge about the energy transport in the Sun has improved significantly thanks to helioseismology, through interior density and sound speed profiles. These results enabled us to locate the \textnormal{base of the solar convection zone} at roughly $\SI{0.7}{\solarradius}$, close to the tachocline \citep{Christensen-Dalsgaard1991}. Below this region, energy generated by nuclear hydrogen fusion in the solar core is carried by photons, while above \textnormal{convection} (plasma motions carrying heat) dominates the energy transport. At the same time we think that the majority of magnetic flux originates at the base of the convection zone and moves toward the surface in the form of \textnormal{flux tubes}. There it appears in the form of patches of high magnetic field (\textnormal{active regions}) and their intensity counterparts, sunspots (which appear dark as they are cooler than their surroundings due to the magnetic field inhibiting the convection), see \citet{Parker1955} and \citet{Cheung2010}.

The deep connection between the solar activity and flows then naturally raises the question as to how the magnetic field and the differential rotation are maintained via a \textnormal{solar dynamo} process and also how large-scale flows come into play there. For a review on large-scale dynamics in the convection zone, we refer the reader to \citet{Miesch2005}. Apart from the rotation and the meridional circulation, such large-scale flows include for example convective motions, flows around active regions and a new, recently observed type of waves known as \textnormal{Rossby waves}. As this thesis is indeed about observations of large-scale flows in the solar interior, in the following sections we want to give further details on each of them.

\section{Flows in and on the Sun}

\subsection{Rossby waves}
\label{sect_intro_rossby_waves}

Rossby waves were first described in detail by \citet{Rossby1939} and \citet{Rossby1940}. They can exist on rotating fluid bodies and are a type of \textnormal{inertial waves}. As such their restoring force is the Coriolis force. In particular, most relevant for the existence of Rossby waves is that the strength of the Coriolis force, quantified by the \textnormal{Coriolis parameter} $f = 2 \Omega \sin\lambda$ (with $\Omega$ the angular rotation rate related to a rotation vector $\boldsymbol{\Omega} = \Omega \boldsymbol{\hat{z}}$), depends on the latitude $\lambda$. Let us briefly see why this causes an oscillatory motion. Here, we describe the Rossby waves in the framework of the \textnormal{shallow-water-approximation}, i.e. we consider a fluid whose horizontal length scale vastly exceeds its vertical length scale. The vertical flow velocity is considered to be small compared to the horizontal flow velocity. The flow is assumed to be incompressible, i.e. the fluid should be divergence-free, and only one depth layer (the surface) is taken into account.

Assume that we have a small fluid parcel that rotates with the body. We assume that the parcel initially does not have any \textnormal{relative vorticity}, i.e. $\boldsymbol{\zeta}_r = \boldsymbol{\nabla} \times \boldsymbol{u} = \boldsymbol{0}$ (for any velocity $\boldsymbol{u}$). However, the rotation itself causes a \textnormal{planetary vorticity} $\boldsymbol{\zeta}_p = 2 \boldsymbol{\Omega}$. Under the assumption that all motions occur only horizontally on the surface of the body, the relevant contribution to $\boldsymbol{\zeta}_p$ is essentially the locally vertical (radial) component $f \boldsymbol{\hat{r}}$. Therefore when the parcel is perturbed and displaced in latitude (say locally northward), this results in a change of the planetary vorticity $\boldsymbol{\zeta}_p$. However, because the \textnormal{potential vorticity}, closely related to the \textnormal{absolute vorticity} $\boldsymbol{\zeta}_a = \boldsymbol{\zeta}_r + \boldsymbol{\zeta}_p$, must be conserved, this then induces a relative vorticity that is in the opposite direction. In this way the change of the Coriolis force with latitude provides a restoring force, causing the wave motions of the Rossby waves.

\begin{figure}
\centering
\begin{minipage}[t]{0.45\textwidth}
\includegraphics[width=\textwidth]{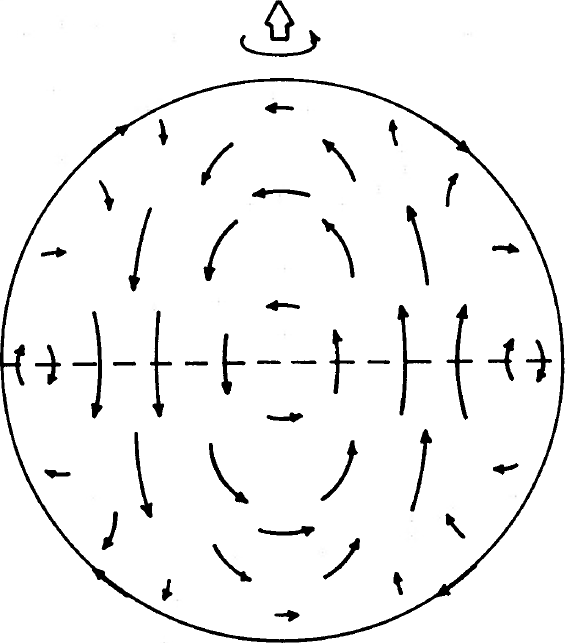}
\end{minipage}
\hspace*{0.5cm}
\begin{minipage}[t]{0.50\textwidth}
\includegraphics[width=\textwidth]{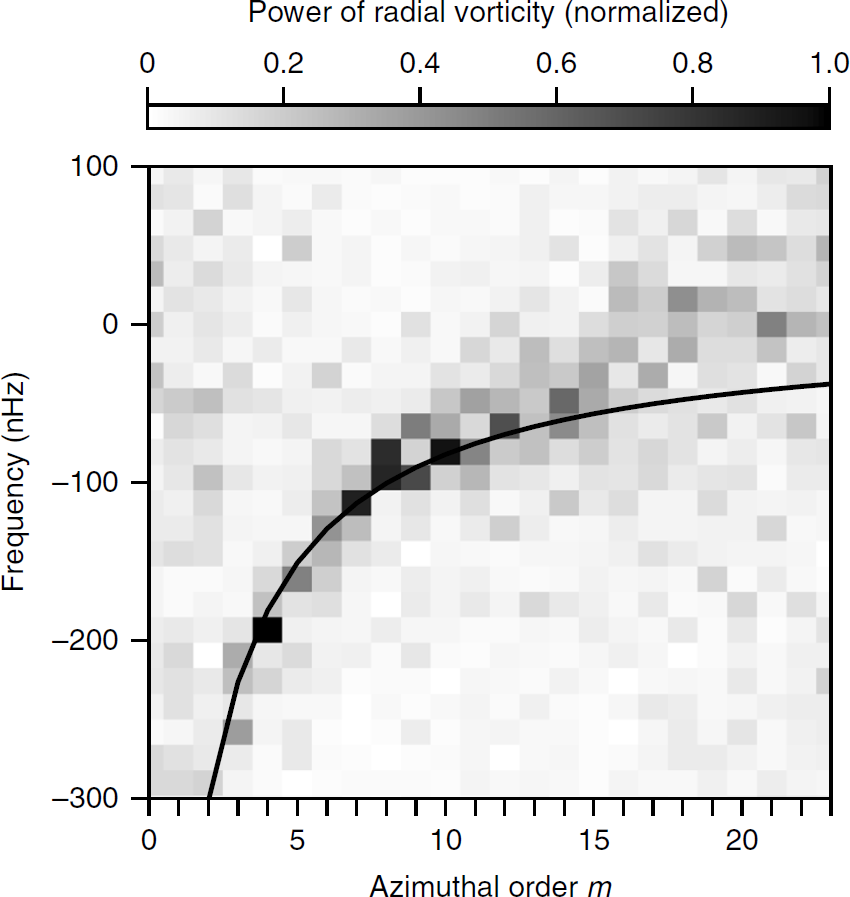}
\end{minipage}
\caption[Rossby waves]{
Left: Schematic flow field for the Rossby mode with $\ell = m = 3$. Rossby waves are retrograde-propagating vortex patterns. From \citet{Saio1982}, \textcopyright AAS. Reproduced with permission. Right: Power spectrum of solar Rossby waves. The power of the radial vorticity is shown as a function of frequency and azimuthal order in the co-rotating reference frame. The solid black line indicates the simple theoretical dispersion relation for the sectoral ($\ell = m$) case (see text). Rossby wave modes are detected for $m \geq 3$. From \citet{Loeptien2018}, with permission.
}
\label{fig_rossby_waves}
\end{figure}

From theory, we know that Rossby waves obey a simple relation between frequency $\omega$ and wavenumber (\textnormal{azimuthal order} $m$, \textnormal{angular degree} $\ell$). Their \textnormal{dispersion relation} is
\begin{equation}
\label{eq_rossby_dispersion}
\omega = -\frac{2\Omega m}{\ell(\ell + 1)}.
\end{equation}
The minus sign shows that the phase speed of the Rossby waves is negative and that these waves thus propagate in the \textnormal{retrograde} direction. The above dispersion relation can be derived from the equation of motion (momentum equation), including the Coriolis term, but it requires three assumptions. First, the fluid body is assumed to rotate uniformly, i.e. $\Omega$ is constant. The second assumption is that the flows are restricted to the surface of the sphere and purely horizontal, i.e. there are no radial motions. Finally, it is assumed that the \textnormal{horizontal divergence} of the flows is zero, i.e. there are no sources or sinks of the flows. This implies that the horizontal velocities are purely vortical and can be written as the curl of a \textnormal{stream function} $\psi(\lambda,\varphi)$ that depends on latitude $\lambda$ and longitude $\varphi$ and which points radially away from the surface. Theory suggests that the flow field associated with single Rossby wave modes (Fig.~\ref{fig_rossby_waves}, left panel) is given by spherical harmonics \citep{Saio1982}. If $\psi(\lambda,\varphi)$ is proportional to \textnormal{sectoral} ($\ell = m$) spherical harmonics (we will see in Sect.~\ref{sect_latfunc_results} that the $\ell = m$ component is the dominant contribution in horizontal Rossby wave eigenfunctions of the radial vorticity), the prograde flow $u_x = \frac{\partial \psi}{\partial \lambda}$ is anti-symmetric in latitude and the northward flow $u_y = \frac{1}{\cos\lambda} \frac{\partial \psi}{\partial \varphi}$ is symmetric. These symmetries can also be seen in the left panel of Fig.~\ref{fig_rossby_waves}.

Rossby waves were first discovered on Earth, where they appear in the atmosphere, but also in the ocean \citep{Chelton1996}. The atmospheric Rossby waves are connected to large-scale meanders observed in the jet stream and to the transport of cold air from the poles toward the equator and of hot air from the tropics toward the poles \citep[e.g.][]{Holton2004}. The oceanic Rossby waves are important for the propagation of ocean-climate signals, such as the El Ni\~{n}o phenomenon \citep{Lachlan-Cope2006}. On Earth, Rossby waves thus play a key role in shaping the weather and climate.

However, while the theoretical existence of Rossby waves on the Sun was already postulated roughly $40$~years ago \citep{Papaloizou1978}, the observational history of solar Rossby waves was for a long time marked by ambiguous detection claims \citep{Kuhn2000, Williams2007, Sturrock2015, McIntosh2017}. Only very recently, \citet{Loeptien2018} provided convincing observational evidence for solar Rossby waves (including an identification via the dispersion relation). \citet{Loeptien2018} used flow measurements obtained from \textnormal{local correlation tracking} (Sect.~\ref{sect_intro_lct}) to study the \textnormal{radial vorticity} field on the Sun and they detected a large-scale (azimuthal order $m \leq 15$) oscillatory pattern near the equator, with lifetimes of several months. The observed dispersion relation of these waves is consistent with the textbook equation (Eq.~\ref{eq_rossby_dispersion}) for the case of sectoral waves, i.e. $\omega = -2\Omega/(m + 1)$, where $\Omega/2\pi = \SI{453.1}{\nano\hertz}$ is the equatorial rotation rate of the Sun (Fig.~\ref{fig_rossby_waves}, right panel). \citet{Loeptien2018} also showed that the eigenfunctions of solar Rossby waves are not the purely sectoral spherical harmonics expected from early theories (Fig.~\ref{fig_rossby_waves}, left panel).

\citet{Liang2019} later confirmed the Rossby wave detection of \citet{Loeptien2018} via \textnormal{time-distance helioseismology} \citep[TD,][]{Duvall1993}. Time-distance helioseismology is a widely used method of \textnormal{local helioseismology} (Sect.~\ref{sect_intro_local_helioseismology}). The basic idea is that, in the presence of a flow, waves travelling between two points on the solar surface propagate faster in the direction of the flow than against it. This directional asymmetry can be measured in the form of \textnormal{travel-time differences} which can be converted into flow velocities by solving an inverse problem. Via different measurement geometries, flows in the prograde or the northward direction and even the horizontal divergence and the radial vorticity can thus be retrieved. Further information about time-distance helioseismology can be found for example in \citet{Gizon2005}. The Rossby wave confirmation by \citet{Liang2019} is crucial since it relies on an independent method and thus shows that the results obtained by \citet{Loeptien2018} are robust. \citet{Hanasoge2019} and \citet{Mandal2019} also detected and characterized Rossby modes with odd $m$ via yet another method called \textnormal{normal-mode coupling}. Another Rossby wave confirmation was provided by \citet{Hanson2020} via ring-diagram analysis.

It has been suggested that Rossby waves could help in maintaining the solar differential rotation \citep{Ward1965} or zonal jets on Jupiter \citep{Liu2011}. However, purely sectoral Rossby waves do not transport angular momentum. \citet{Gilman1969} and \citet{Wolff1987} proposed that the magnetic field could be modulated by Rossby waves. It might also be interesting to study the possible interactions between convection and the Rossby waves, \citep[e.g.][]{Vallis1993}. While much of this is currently not much more than speculation, for sure the discovery of solar Rossby waves opens a new way to probe the solar interior. Similar to other, well-known types of waves commonly used in helioseismology, mode frequencies and eigenfunctions can be measured for Rossby waves. This might allow us to test the validity of existing Rossby wave theories and to study the effects of differential rotation and potentially the magnetic field on this type of waves.

\subsection{Convective flows}

\begin{figure}
\centering
\includegraphics[width=\textwidth]{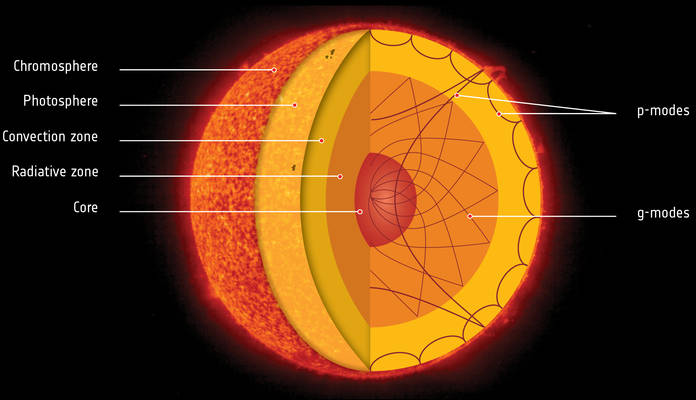}
\caption[Structure of the solar interior]{
Interior solar structure. The energy is transported outwards from the solar core through radiation in the inner and through convection in the outer layer, until it reaches the visible surface, the \textnormal{photosphere}. By using solar oscillations that pass through the optically thick matter below the photosphere, probing different regions, we can study the solar interior (see text). From \protect\textit{\url{https://www.esa.int/Science_Exploration/Space_Science/Gravity_waves_detected_in_Sun_s_interior_reveal_rapidly_rotating_core}}, courtesy of ESA/NASA, with permission.
}
\label{fig_convection_zone_waves}
\end{figure}

As briefly mentioned before, the transfer of energy generated via hydrogen fusion inside the Sun relies on two different physical processes. In the inner $\SI{70}{\percent}$ of the solar radius \textnormal{radiative transfer} (i.e. via photons) is the dominant transport mechanism while in the outer $\SI{30}{\percent}$ convection (i.e. bulk plasma motions) carries the energy outwards (Fig.~\ref{fig_convection_zone_waves}). Most interestingly, the convection and the related flows occur in a cell-like form on distinct spatial scales. This has led to a categorization into granules, supergranules and giant cells.

\begin{figure}
\centering
\begin{minipage}[t]{0.49\textwidth}
\includegraphics[width=\textwidth]{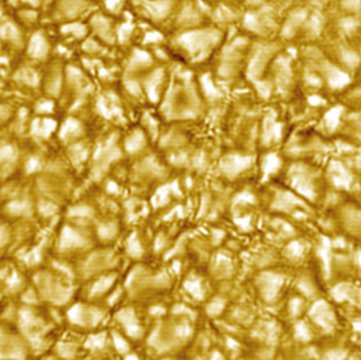}
\end{minipage}
\begin{minipage}[t]{0.49\textwidth}
\includegraphics[width=\textwidth]{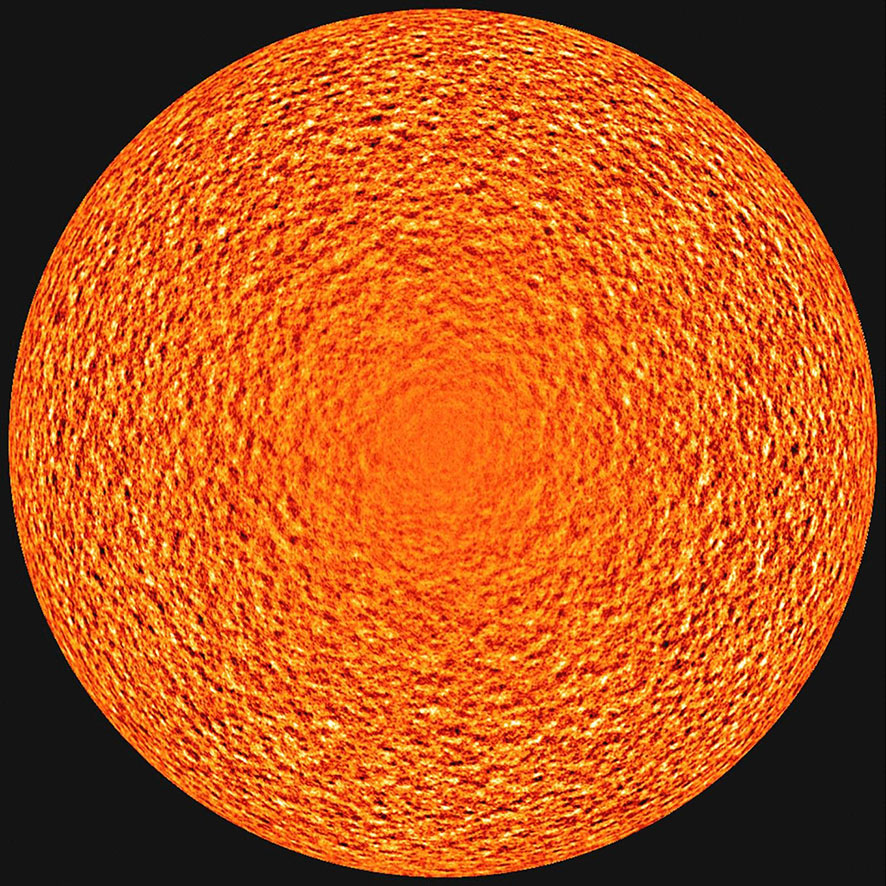}
\end{minipage}

\vspace{0.5cm}
\begin{minipage}[t]{\textwidth}
\includegraphics[width=\textwidth]{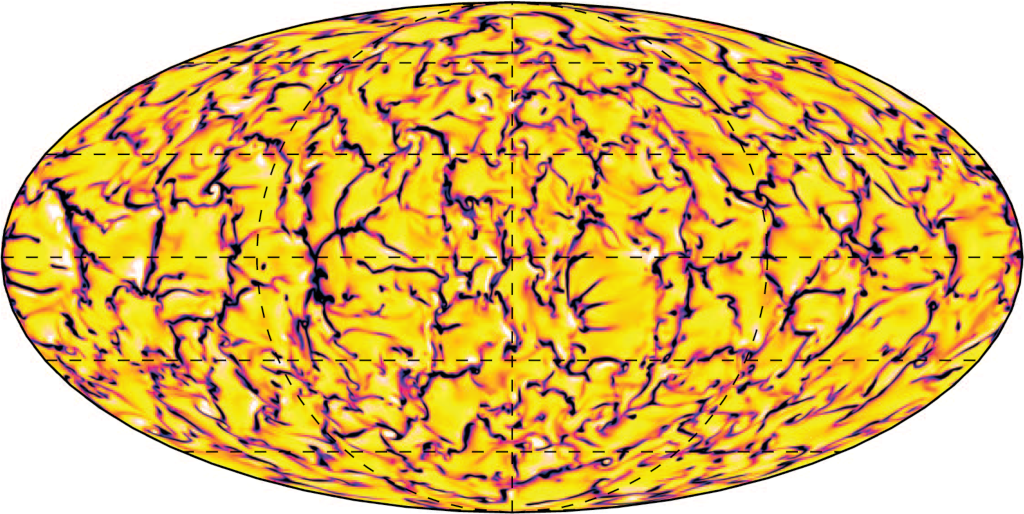}
\end{minipage}
\caption[Convective scales of the Sun]{
Convective scales of the Sun. Top left: Granulation appears as a small-scale cell-like structure in solar intensity images. From \protect\textit{\url{https://apod.nasa.gov/apod/ap051106.html}}, courtesy of NOAO/AURA/NSF, with permission. Top right: Supergranulation as seen in Dopplergrams of the full solar disk. The pattern is more prominent close to the limb as supergranular flows are mostly horizontal. From \citet{Rincon2018}, with permission. Bottom: Giant cells as seen in a radial velocity image from simulations of the Sun. The observational evidence for giant cells is still sparse. From \citet{Miesch2008}, \textcopyright AAS. Reproduced with permission.
}
\label{fig_convection_scales}
\end{figure}

\textnormal{Granulation} (Fig.~\ref{fig_convection_scales}, top left panel) is the smallest scale of convection and was first observed by \citet{Herschel1801}. The term refers to the grainy patterns seen on intensity images of the Sun. Granules are visible as small bright cells with a diameter of roughly $1$-$\SI{2}{\mega\metre}$ \citep{Rieutord2010}, or equivalently an angular degree of $\ell \sim 2000$-$4000$. These cells are relatively shallow and separated by dark narrow lanes, the \textnormal{intergranular network}. Granules have a vertical extent of roughly $\SI{300}{\kilo\metre}$ or less \citep{Nordlund2009}. They have short lifetimes of $\sim{}10$~min. The flows on this scale have velocities typically around $1$-$\SI{3}{\kilo\metre\per\second}$ as seen in simulations \citep{Stein1998, Nordlund2009} and observations \citep{Oba2017}, although in rare cases granules can also reach very high velocities up to $\sim\SI{10}{\kilo\metre\per\second}$. Plasma moves upwards until it reaches the surface, where it diverges horizontally and is radiatively cooled. In this process, the ionized hydrogen captures free electrons and releases ionization energy in the form of photons. The still partially ionized plasma then concentrates in cooler downflow lanes, sinks into the solar interior and is heated and ionized anew. Granulation is well reproduced by simulations, see e.g. the review by \citet{Nordlund2009}.

\textnormal{Supergranulation} (Fig.~\ref{fig_convection_scales}, top right panel) occurs on a larger spatial scale around angular degree $\ell \sim 120$ \citep{Hathaway2000}. This means that supergranules have typical length scales on the order of $\SI{30}{\mega\metre}$. Their discovery is attributed to \citet{Hart1954}. Unlike granules, this convective scale is best observed in the \textnormal{line-of-sight} (LOS) velocity, i.e. in \textnormal{Dopplergrams}, where the supergranulation can be seen as a pattern covering the whole visible solar disk. Supergranules also evolve on much longer timescales than granules, with typical lifetimes of $1$-$2$~days. Their flows have amplitudes of approximately $\SI{300}{\metre\per\second}$ in the horizontal direction and are much weaker in the vertical direction \citep{Rincon2018}. The flows can be easily observed in maps of the horizontal divergence. There is a wide variety of open questions concerning the Sun's supergranulation, as described in the review by \citet{Rincon2018}. Contrary to granulation, which is relatively well understood and successfully reproduced in simulations, the origin of the supergranulation is not clear yet. Although thermal convection is the most likely explanation of its existence, we do not yet understand why supergranulation stands out as a distinct scale of convection. Also, it is currently unknown how deep exactly the supergranules extend into the convection zone. Supergranules are known to rotate faster than their surroundings and \citet{Gizon2003} suggested that this apparent super-rotation is linked to a wave-like character of this convective scale. Further evidence for this was given by \citet{Schou2003} and \citet{Langfellner2018}.

\textnormal{Giant cells} (Fig.~\ref{fig_convection_scales}, bottom panel) are the largest scale of convection, with horizontal extents of $\SI{200}{\mega\metre}$ ($\ell \sim 20$) or more \citep[e.g.][]{Miesch2008}. Typical velocity scales for the largest cells should be $\SI{100}{\metre\per\second}$ or less. Although scientists hypothesized on the theoretical existence of giant cells not long after the supergranulation pattern had been detected \citep{Simon1968}, this scale of convection continues to remain elusive even nowadays: Although they clearly appear in simulations \citep{Miesch2008}, unfortunately convincing observations for giant cells are sparse at this moment. While \citet{Hathaway2013} claim to have detected evidence for giant convection cells at high latitudes (around $\pm\SI{60}{\degree}$) in flow maps, it is currently unclear whether the observed large-scale features are indeed of convective origin. The authors report lifetimes of at least a few months, in line with theoretical expectations. Giant cells are likely strongly affected by the solar differential rotation, possibly being sheared by it. Likewise they could potentially play an important role in angular momentum transport from the higher latitudes to the equator and could thus help in maintaining the latitudinal rotation gradient \citep{Hathaway2013}.

While the convective energy spectrum at large angular degrees (small spatial scales) and close to the surface is comparatively well understood, the dynamics are much less clear deeper in the convection zone and at large spatial scales. Below, we want to briefly introduce several existing results. These results will be re-evaluated in Chap.~\ref{chap_main_convection}.

\citet{Hanasoge2010} and subsequently \citet{Hanasoge2012} have employed time-distance helioseismology to obtain horizontal flows. They applied a spectral analysis on their obtained horizontal flows to estimate the strength of the convection at $\SI{0.96}{\solarradius}$, up to $\ell \sim 60$ (Fig.~\ref{fig.E_of_k_full}, Original HDS2012). The measured root-mean-square (rms) velocities (on the order of $\SI{1}{\metre\per\second}$) and the energy were roughly two and four orders of magnitude smaller, respectively, than those reported from previous simulations by \citet{Miesch2008} with the \textnormal{Anelastic Spherical Harmonics} code \citep[ASH,][]{Clune1999,Brun2004} at $\SI{0.98}{\solarradius}$ (Fig.~\ref{fig.E_of_k_last}, ASH). The ASH code simulates the entire convection zone in a spherical geometry at low resolution/low $\ell$.

If the \citet{Hanasoge2012} measurements were true, this would have serious consequences for the solar angular momentum transport. It would also imply that current models of convection such as the \textnormal{mixing length theory} (convective parcels travel over a certain \textnormal{mixing length}, keeping their identity, and then release their energy and dissolve into their surroundings, see \citealt{Prandtl1925} and \citealt{Boehm-Vitense1958}) and modern simulations, e.g. with the ASH code, fail to accurately describe the physics occurring inside the Sun. Evidently the consequence would be no less than the need to completely rethink our picture of convection \citep{Gizon2012}. This is also referred to as the \textnormal{convective conundrum}.

\citet{Gizon2012} showed another, independent, simulation result at $\SI{0.98}{\solarradius}$, inferred using the \textnormal{stagger} code (Fig.~\ref{fig.E_of_k_last}, stagger) from \citet{Stein2006}. Stagger simulates layers close to the surface at high resolution/high $\ell$. Additionally, \citet{Gizon2012} presented an energy spectrum (Fig.~\ref{fig.E_of_k_full}, Original R2012) from \citet{Roudier2012}, who used granulation tracking (Sect.~\ref{sect_intro_lct}) to derive horizontal velocities on the solar surface at intermediate $\ell$. The stagger and the \citet{Roudier2012} results were inconsistent with the ASH simulation. A lower theoretical bound from \citet{Miesch2012}, also presented by \citet{Gizon2012}, was above the \citet{Hanasoge2012} estimates.

\citet{Greer2015} investigated the energy spectrum of large-scale convection at $\SI{0.96}{\solarradius}$ (Fig.~\ref{fig.E_of_k_full}, Original GHFT2015) using a particular type of ring-diagram analysis (Sect.~\ref{sect_intro_local_helioseismology}). The resulting energy spectrum was again mostly consistent with the ASH results.

Finally, \citet{Hanasoge2016} summarized the existing results and showed another estimate of the large-scale convective energy from \citet{Hathaway2013}, where the authors used supergranulation tracking (similar to granulation tracking, Sect.~\ref{sect_intro_lct}) to obtain the horizontal velocities. This estimate was larger than that from \citet{Hanasoge2012} by roughly one order of magnitude.

\subsection{Flows around active regions}
\label{sect_intro_flows_active_regions}

We have already established that there is a close connection between large-scale flows and the solar magnetic field. It thus comes as no surprise that there are also large-scale flows surrounding active regions, where the magnetic flux is particularly large. These flows around active regions have been first observed on the solar surface by \citet{Gizon2001}. The authors found that the flows are spatially extended and flow amplitudes were measured to be around $\SI{50}{\metre\per\second}$. Moreover, the flows were converging into the active region, but the authors also detected outflows (called \textnormal{moat flows}) at further distances from the sunspots (beyond the so-called \textnormal{penumbra}).

Several papers confirmed these flows and investigated their properties independently with a different helioseismology method \citep{Haber2004, Hindman2004, Hindman2009}. These papers demonstrated that at larger depths there seem to be outflows from active regions rather than inflows. The authors also found smaller flow amplitudes of roughly $20$-$\SI{30}{\metre\per\second}$. The flows could be observed up to $\SI{10}{\degree}$ from the active region center.

As active regions can greatly vary in size, shape and lifetime, a solid statistical sample is crucial for studies of the flow patterns in their vicinity. \citet{Loeptien2017} confirmed the presence of active region inflows with local correlation tracking (Sect.~\ref{sect_intro_lct}). By averaging flow maps for many active regions, they found that the inflow is not symmetric, but rather converges toward the trailing polarity.

Finally, \citet{Braun2019} used a large sample of active regions and divided it into several bins of magnetic flux. They confirmed the prevalent inflows to the trailing polarity and demonstrated that they are not strongly dependent on the magnetic field strength. \citet{Braun2019} also observed a retrograde flow at the poleward side of the active regions (and weaker on the equatorward side), which had not been found in previous studies, and discussed how much the active region flows may contribute to time-varying larger-scale flows such as torsional oscillations or the residual meridional flow.

The dynamics of active regions is a topic of active research, because flows around active regions are thought to interact with, for example, the meridional flow. The latter is a key ingredient in \textnormal{flux transport models} \citep{Jouve2007}, where the poleward transport of magnetic flux plays a crucial role in the polarity reversal of the polar magnetic field, which itself is of importance for solar cycle predictions. The inflows might counteract the diffusion of the magnetic field in active regions by convection and thus could help in keeping the magnetic flux concentrated \citep{DeRosa2006, Martin-Belda2016}. Feedback mechanisms associated with active region flows could potentially also modulate the amplitude of the solar cycle \citep{Cameron2012}. We refer the reader to \citet{Charbonneau2010} for a review of various dynamo models.

\section{Motivation for the thesis}

The previous sections have shown the importance of various large-scale flows for our understanding of solar dynamics: The interplay of Rossby waves, convective flows and flows around active regions, in connection with differential rotation and meridional circulation, but also the magnetic field, has far-reaching consequences for basic solar physics models. This thesis therefore focuses on observations of large-scale solar flows.

While \citet{Loeptien2018} and \citet{Liang2019} successfully detected and identified solar Rossby waves and measured the wave frequencies, the question of how Rossby modes behave as a function of latitude and depth was only briefly addressed. However, it is crucial to understand the dependence of the waves on these spatial coordinates. A solid characterization of the mode sensitivity will allow us to understand which latitudes and depths can be probed with Rossby modes and, potentially serving as a test bed for different Rossby wave theories, may give us valuable information on mode physics in general. We therefore want to study the latitude and depth dependence of the Rossby waves.

As mentioned, the disagreement between various results regarding the strength of deep, large-scale convective flows, the convective conundrum, fundamentally puts our current view of solar turbulence in question. This motivates our study of deep, large-scale convection, where we show that the analysis that led to these former results contains various errors. While we will see that these errors are not the main cause of the discrepancy, the corrected curves presented by us together with new results will help build a solid foundation for future investigations.

\section{Data used in the thesis}

In this thesis, we use data from the \textnormal{Solar Dynamics Observatory} \citep[SDO,][]{Pesnell2012}. SDO is a satellite which has been launched into a geosynchronous orbit (following the Earth's rotation) in February 2010. It collects data since April 2010 via three instruments. Among these are the \textnormal{Atmospheric Imaging Assembly} (AIA) and the \textnormal{Extreme Ultraviolet Variability Experiment} (EVE). The data in this thesis, however, are from the \textnormal{Helioseismic and Magnetic Imager} \citep[HMI,][]{Schou2012, Scherrer2012}.

This instrument observes the full visible disk of the Sun with a high temporal cadence of $45$ or $720$~s and a high spatial resolution of $4096 \times 4096$~pixels. It was designed to study both solar oscillations and the solar magnetic field, both on the surface and in the interior. To this extent HMI obtains various kinds of raw data, which are then pre-processed and made publicly available in the form of different data products.

\begin{figure}
\centering
\begin{minipage}[t]{0.32\textwidth}
\includegraphics[width=\textwidth]{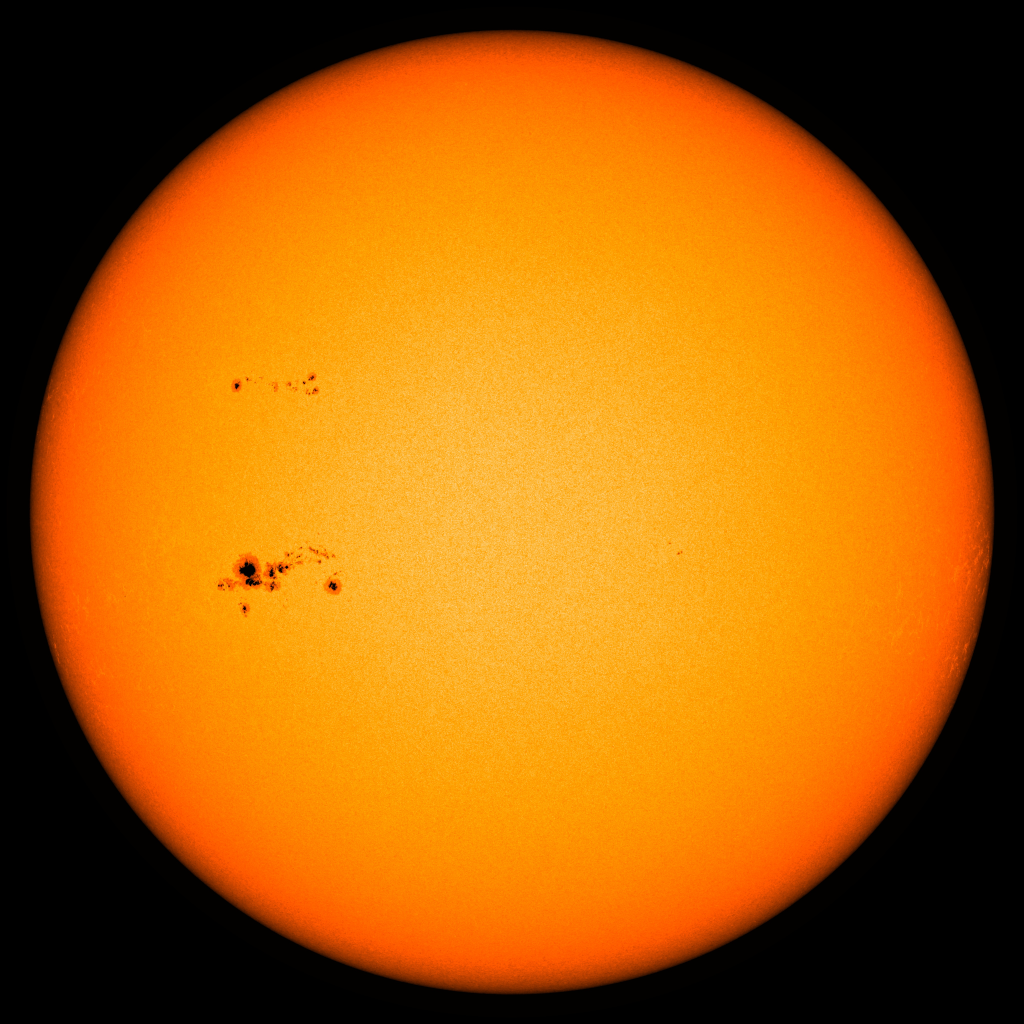}
\end{minipage}
\begin{minipage}[t]{0.32\textwidth}
\includegraphics[width=\textwidth]{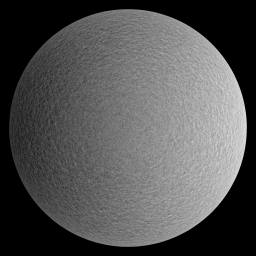}
\end{minipage}
\begin{minipage}[t]{0.32\textwidth}
\includegraphics[width=\textwidth]{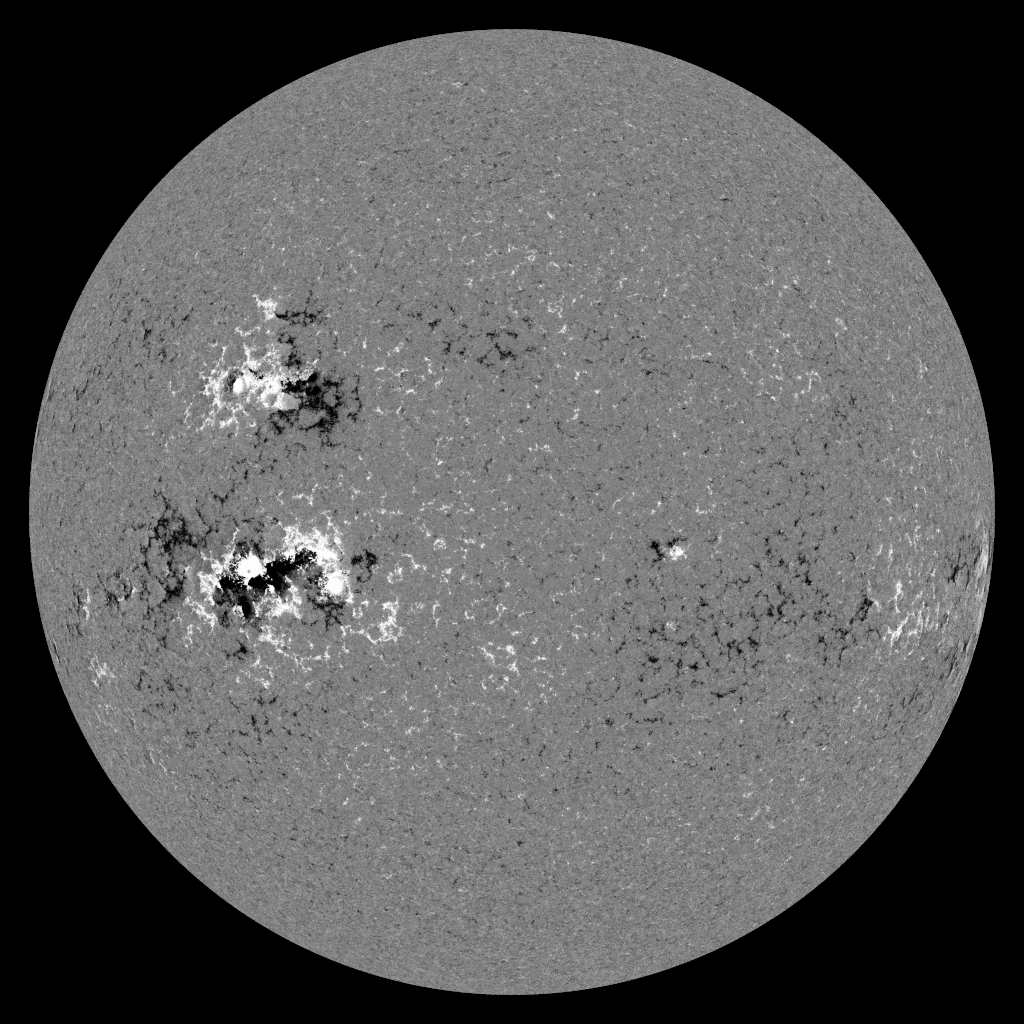}
\end{minipage}
\caption[Basic data products of the Helioseismic and Magnetic Imager]{
Basic HMI data products. Left: Continuum intensity on February 1, 2014. Several sunspots are clearly visible. Middle: Line-of-sight velocity on September 30, 2019. Black and white indicate plasma moving toward and away from the observer, respectively. Right: Line-of-sight magnetic field on February 1, 2014. Black and white indicate magnetic field lines pointing away from and toward the observer, respectively. Active regions (with high absolute field strengths) can be associated with the sunspots in the intensity image. From HMI quick-look data (\protect\textit{\url{http://jsoc.stanford.edu/data/hmi/images/latest/}}), courtesy of NASA/SDO and the HMI science team, with permission.
}
\label{fig_hmi_data}
\end{figure}

These include for example images of the vector magnetic field and of the line-of-sight magnetic field (\textnormal{magnetograms}), but also intensity images of the Sun, obtained in the continuum around the Fe I $\SI{6173}{\angstrom}$ line, and Dopplergrams, which give the velocity in the line-of-sight direction as measured from the wavelength shift of that Fe I line due to the \textnormal{Doppler effect} (Fig.~\ref{fig_hmi_data}).

\section{Processing methods used in the thesis}

While the aforementioned data are without doubt useful for various kinds of analysis, they often are not immediately usable for the study of solar flows, which typically requires knowledge about the horizontal velocities on the solar surface and in the solar interior. The basic question that this section wants to address is thus "How can we infer horizontal velocities from the basic HMI data products?". As we will see, there are multiple ways to obtain these velocities, such as local helioseismology. There are however also methods independent from local helioseismology such as local correlation tracking. Both local helioseismology and local correlation tracking are used for the analysis presented in this thesis. The following sections thus intend to illustrate these techniques in some detail.

\subsection{Local correlation tracking}
\label{sect_intro_lct}

Local correlation tracking (LCT) is was first used in the solar context by \citet{November1988}, who also coined the name of this technique. However, the basic principle behind this analysis method dates back further, since it was used in image processing for other fields before. Essentially it relies on tracking the motion of features and thus retrieving the related velocities.

\begin{figure}
\centering
\includegraphics[width=\textwidth]{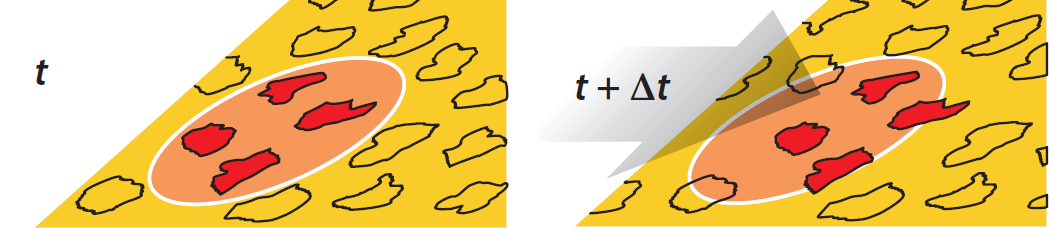}
\caption[Local correlation tracking (LCT, granulation tracking)]{
Local correlation tracking (granulation tracking). At a time step $t$, a small window (red shaded oval) containing several granules (outlined by solid black lines) is selected on an intensity image of the Sun. At a later time step $t + \Delta t$, the granules inside the window (dark red filled areas) have moved outside the window due to their proper motion. By following the granule motion via measuring the pixel shift between the two images, horizontal flows can be determined. From \citet{Hanasoge2016}, with permission.
}
\label{fig_hanasoge2016_lct}
\end{figure}

Suppose, for example, that we have a flow on the solar surface. Granules, the small convection cells visible in \textnormal{intensitygrams}, that are embedded in this flow field, will then be \textnormal{advected}. Consequently, if we take two intensity images at slightly different time steps, the positions of the granules will change (Fig.~\ref{fig_hanasoge2016_lct}). By measuring this change in position and dividing it by the known time difference between the two images, it is then possible to infer the horizontal velocity of the underlying flow field. 

An important caveat is that the time difference between the two images should be short compared to the lifetime of granules, since otherwise the evolution of the granules, for example changes in granule shape, may lead to a misdetermination of velocities. The typical granule lifetime of roughly $10$~min thus inherently limits the time lag between the intensitygrams to a few minutes or less. Also, since granules are very shallow, local correlation tracking is only sensitive to flows at the solar surface. Due to the usage of the granules as tracers of the flow the method is often referred to as \textnormal{granulation tracking}. However, \textnormal{supergranulation tracking} is possible as well, most easily using Dopplergrams, where the supergranules are best visible. Naturally, owing to the longer lifetime and the bigger spatial scale of supergranules (roughly $1$-$2$~days and $\SI{30}{\mega\metre}$), these features allow longer time lags between the input images, but with a worse spatial resolution.

In practice, a number of different implementations of granulation tracking exists. Among them are the \textnormal{coherent structure tracking} \citep[CST,][]{Rieutord2007} and the \textnormal{Fourier local correlation tracking} \citep[FLCT,][]{Fisher2008, Welsch2004}. The conceptual difference between the two algorithms is that FLCT, contrary to CST, also accounts for the intergranular lanes. We refer the reader to \citet{Tremblay2018} for a detailed comparison between these and other implementations of granulation tracking.

One of the two flow velocity datasets we will use in this thesis is based on the FLCT code. The code schematically works as follows: The code requires two input images as a function of pixel coordinates $x$ and $y$. For each individual reference pixel in either of the two images, sub-images are created by multiplying the corresponding image with a 2D Gaussian function (separable in $x$ and $y$), that drops with the distance from the reference pixels. This naturally decreases the weight of pixels far away from the reference. A crucial parameter for this windowing operation is $\sigma$, the standard deviation of the 2D Gaussian function, which sets the typical length scale of the structures for which the code will determine the pixel shifts. Too large windows will smear out the resulting velocities, such that spatial resolution is lost, while too small windows will lead to high noise. Once the sub-images have been created, the FLCT code computes the cross-covariance of all combinations of sub-images as a function of pixel shifts $\delta x$ and $\delta y$. This cross-covariance is computed via Fourier transforms. The pixel shifts for which the sub-images match best are then obtained by finding the maximum of a quadratic Taylor expansion to the absolute of the cross-covariance function. The output 2D pixel shift for each pixel is then converted to a 2D velocity vector through division by the time lag $\delta t$ between the two input images. For further details, we refer the reader to \citet{Fisher2008}.

\citet{Loeptien2017} have used the FLCT code to obtain maps of the horizontal velocity to study flows around active regions. For this they applied the FLCT code to pairs of continuum intensity images observed by HMI between May 19, 2010 and March 31, 2016. The two images in each pair are separated by $\delta t = 45$~s (thus much less than the granule lifetime) and the pairs are separated by $30$~min for computational reasons. The parameter $\sigma$ was chosen to be $6$~pixels, which for HMI corresponds to roughly $\SI{2}{\mega\metre}$ at disk center and thus roughly granule scales. Due to the presence of systematic effects in the output velocity maps, such as the \textnormal{shrinking-Sun effect} \citep{Lisle2004,Loeptien2016}, the data were then expanded into \textnormal{Zernike polynomials}, an orthogonal basis on the 2D disk. Temporal frequencies of one year and one day (and harmonics up to the \textnormal{Nyquist frequency}), which are related to the orbit of the SDO satellite, as well as the zero frequency were then removed via Fourier filtering of the Zernike coefficient time series. The filtered output velocities were converted from the CCD coordinate grid to \textnormal{heliographic} coordinates. Finally the \textnormal{mtrack} module was used to track the data at the \textnormal{sidereal Carrington rate} of $\SI{456.0}{\nano\hertz}$ (roughly $25.38$~days) and to map them onto a \textnormal{plate carr\'{e}e} (equirectangular) grid with a spatial sampling of $\SI{0.4}{\degree}$. The output data series of surface velocities as a function of time, latitude and longitude was also used by \citet{Loeptien2018} to study Rossby waves.

\subsection{Local helioseismology: ring-diagram analysis}
\label{sect_intro_local_helioseismology}

\begin{figure}
\centering
\includegraphics[width=\textwidth]{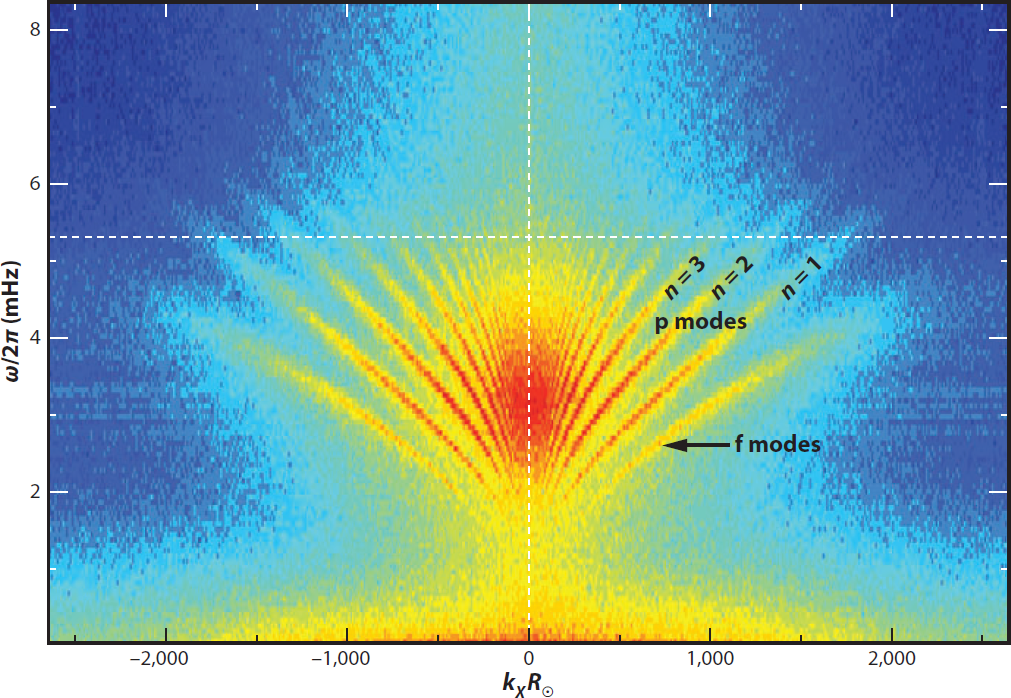}
\caption[Solar power spectrum of the Doppler velocity]{
Solar power spectrum. Power of Doppler velocity versus frequency and horizontal wavenumber $k_x$. The power is contained in distinct ridges, which belong to different wave modes (see text). From \citet{Gizon2010}, with permission.
}
\label{fig_gizon2010_ell_nu}
\end{figure}

Local helioseismology (see e.g. the review by \citealt{Gizon2005}) makes use of waves that are stochastically excited by convection. These waves can be observed in a power spectrum, where they appear as distinct ridges (Fig.~\ref{fig_gizon2010_ell_nu}). The waves are categorized into \textnormal{pressure/$p$-modes}, which are acoustic waves whose restoring force is pressure, \textnormal{internal gravity/$g$-modes}, which are driven by buoyancy, and \textnormal{fundamental/$f$-modes} (also called \textnormal{surface gravity waves}), which are similar to the deep ocean waves observed on Earth. These waves travel through and probe different regions within the solar interior (Fig.~\ref{fig_convection_zone_waves}). For example, $g$-modes are sensitive to the radiative core of the Sun, but they have not yet been convincingly observed, since their amplitude drops strongly with increasing distance from the Sun's core. The $f$-modes on the other hand probe only a very shallow region near the solar surface, whereas the $p$-modes are trapped within the convection zone. The ray paths of those waves are reflected at an \textnormal{upper turning point} near the solar surface (due to a strong decrease in density) and they become horizontal and then refracted at a \textnormal{lower turning point} (due to an increase in sound speed with depth). Additionally their frequency is modified by changes in the sound speed or the density of the matter they traverse, but also due to local flows.

\begin{figure}
\centering
\begin{minipage}[t]{0.30\textwidth}
\includegraphics[width=\textwidth]{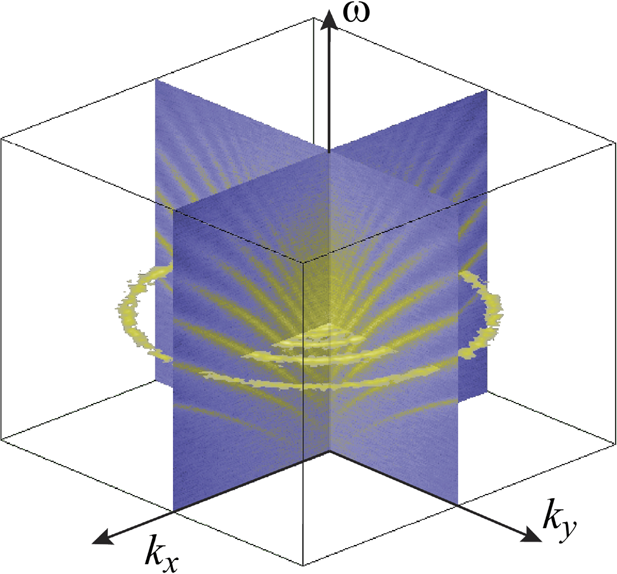}
\end{minipage}
\hspace*{0.1cm}
\begin{minipage}[t]{0.68\textwidth}
\includegraphics[width=\textwidth]{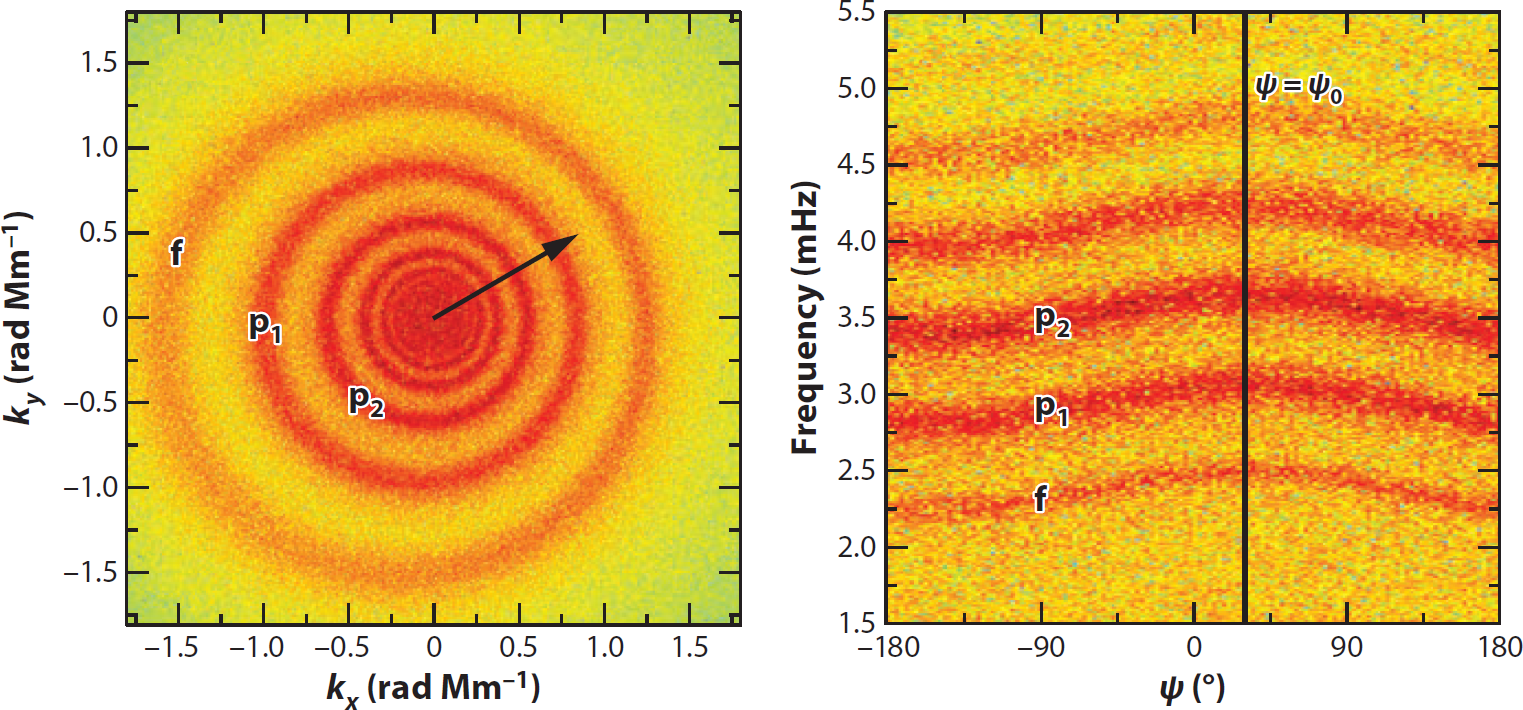}
\end{minipage}
\caption[Ring-diagram analysis (RDA)]{
Ring-diagram analysis. Left: Power of Doppler velocity versus frequency and horizontal wavenumbers $k_x$ and $k_y$. The trumpet-like structure is the 3D counterpart of Fig.~\ref{fig_gizon2010_ell_nu}. From \citet{Kosovichev2012}, with permission. Middle: Cut of 3D power at constant frequency. The power appears in the form of rings. A flow directed along $\SI{30}{\degree}$ north of the prograde direction (indicated by the arrow) causes a Doppler shift and deforms the rings along that direction. From \citet{Gizon2010}, with permission. Right: Cut of 3D power at constant $k$. The Doppler shift appears as a frequency shift along the direction indicated by the vertical black line. By fitting the individual ridges the horizontal velocities can be determined. From \citet{Gizon2010}, with permission.
}
\label{fig_rda_concept}
\end{figure}

An application of this is the local helioseismology technique of \textnormal{ring-diagram analysis} \citep[RDA,][]{Hill1988}, which determines horizontal velocities from distortions of the wave frequencies due to local flows. For this, for each time step, Dopplergrams of, for example, the full disk are split into small patches, which are called \textnormal{tiles}. For each of these tiles a local 3D power spectrum is computed, i.e. the power of the LOS velocity as a function of angular frequency $\omega$ and two wavenumber directions $k_x$ and $k_y$. These local power spectra contain the signature of the solar waves: The 2D power as a function of $\omega$ and the wavenumber $k = \sqrt{k_x^2 + k_y^2}$ (or an angular cut in the $k_x$-$k_y$ plane) appears in the form of distinct ridges (Fig.~\ref{fig_gizon2010_ell_nu}). The lowest of these ridges corresponds to the $f$-mode, above which there are the $p$-modes with an increasing number of radial nodes (ascending \textnormal{radial order} $n$). The 3D power spectrum resembles a trumpet-like structure, for each mode (Fig.~\ref{fig_rda_concept}, left panel).

In the absence of flows, this power should be isotropic, since there is no preference for any particular direction. Thus, when viewed at constant $\omega$, we would see concentric circles of power. However, if there is a flow, the Doppler effect will cause the frequency to be increased in the direction of the flow and decreased against it (Fig.~\ref{fig_rda_concept}, middle panel). This leads to a tilt of the power rings and therefore in the $k_x$-$k_y$ plane the circles are deformed into ellipses and additionally their center may be shifted. This explains why the method is called ring-diagram analysis: The distortion of the rings contains information about the flows through which the solar waves propagate. Therefore by fitting the shape of the rings it is possible to get the velocity for each individual mode (at each tile).

In practice, the fitting is often done by keeping the wavenumber $k$ constant and fitting the ridges in the $\omega$-$\psi$ plane, where $\psi$ is the azimuthal angle in the $k_x$-$k_y$ plane. The rings are basically unwrapped and the distortion caused by the Doppler effect due to a flow appears indeed as a change in the frequency (Fig.~\ref{fig_rda_concept}, right panel). Once the velocities have been determined for each mode, we can obtain the flows as a function of depth. This last step is referred to as an \textnormal{inversion}, of which prominent sub-classes are the \textnormal{regularized least squares} (RLS) and the \textnormal{optimally localized averages} \citep[OLA,][]{Backus1968,Pijpers1992} inversion. For this we can make use of the different depth sensitivity of the various modes. The sensitivity of ring-diagram \textnormal{kernels} to local flows was studied by \citet{Birch2007}. By combining the measurements (the ring fits) in a suitable way, it is possible to construct an \textnormal{averaging kernel}, which focuses the sensitivity at a particular target depth, while at the same time it suppresses unwanted side lobes present in individual mode sensitivity kernels. Different combinations of the ring fits therefore allow to obtain the flows as a function of depth.

For HMI data, a ring-diagram pipeline \citep{Bogart2011a, Bogart2011b} is in use. For this, Dopplergrams are mapped and tracked at the sidereal Carrington rate via the mtrack module, for various tile sizes ($5$, $15$ or $\SI{30}{\degree}$). Local power spectra are computed and the ring-fitting is done either via (a) a $6$-parameter model \citep{Haber2000} of a Lorentzian line profile in frequency or (b) a more complex $13$-parameter model \citep{Basu1999a}, which includes parameters to describe line asymmetry. Contrary to the former ring fits, however, the latter are not inverted. The inversion step for the \citet{Haber2000}-based ring fits is performed via a 1D OLA algorithm \citep{Basu1999a, Basu1999b}. The inverted flow velocities are available for the $15$ and $\SI{30}{\degree}$ tile sizes. For our analysis, we concentrate on the former tile size. The tile centers are separated by $\SI{7.5}{\degree}$ in latitude and in longitude, with the longitude spacing increasing toward the poles to keep the physical tile area constant. Adjacent tiles overlap by $\SI{50}{\percent}$ with each other. The RDA flow velocities are then post-processed before data from RDA and local correlation tracking (Sect.~\ref{sect_intro_lct}) are commonly analyzed further. More details on this are given in Chap.~\ref{chap_main_rossby}.

Apart from the standard HMI ring-diagram data, there is another ring-diagram pipeline by \citet{Greer2014}. One of the main differences is that the tiles are very densely spaced, i.e. tile centers are only $\SI{0.25}{\degree}$ apart. Also, instead of fitting individual modes independently, multiple ridges in the power spectrum are fit together. This is called \textnormal{multi-ridge fitting}. \citet{Nagashima2020} have made improvements to and fixed some bugs in the \citet{Greer2014} code and showed that the output ring fits are order-of-magnitude comparable with ring fits from the standard pipeline. Finally, contrary to the \citet{Bogart2011a} pipeline, the \citet{Greer2014} code employs a 3D inversion (including the horizontal dimensions). The effects of this kind of inversion on the output flow velocities are unknown and have not yet been analyzed in the literature.

\section{Structure of the thesis}

In Chap.~\ref{chap_main_rossby}, we investigate the latitudinal and radial dependence of Rossby wave eigenfunctions. The usage of two independent datasets allows us to compare the results for different methods to determine flows close to the surface of the Sun. Subsequently, in Chap.~\ref{chap_main_convection}, we look at the power spectrum of large-scale deep convection and re-evaluate the large discrepancy between existing results. Finally, Chap.~\ref{chap_discussion} gives a short discussion and extension of the results from the previous chapters and we try to illustrate how our observations might be part of a larger, common context. We will briefly address how Rossby waves appear in different observables and how solar activity may affect the energy spectrum of horizontal flows and we conclude the chapter with a short outlook.

%% file: main_rossby_waves.tex
\chapter{Exploring the latitude and depth dependence of solar Rossby waves using ring-diagram analysis}
\label{chap_main_rossby}

\blfootnote{This chapter reproduces the article \textit{Exploring the latitude and depth dependence of solar Rossby waves using ring-diagram analysis} by B.~Proxauf, L.~Gizon, B.~L{\"o}ptien, J.~Schou, A.~C.~Birch and R.~S.~Bogart, published in Astronomy and Astrophysics, 634, A44 (2020). Contributions: B.~Proxauf conducted the data analysis and contributed to the interpretation of the results and to writing the manuscript.}

\section{Abstract}

Global-scale equatorial Rossby waves have recently been unambiguously identified on the Sun. Like solar acoustic modes, Rossby waves are probes of the solar interior. We study the latitude and depth dependence of the Rossby wave eigenfunctions. By applying helioseismic ring-diagram analysis and granulation tracking to observations by HMI aboard SDO, we computed maps of the radial vorticity of flows in the upper solar convection zone (down to depths of more than $\SI{16}{\mega\metre}$). The horizontal sampling of the ring-diagram maps is approximately $\SI{90}{\mega\metre}$ ($\sim \SI{7.5}{\degree}$) and the temporal sampling is roughly $27$~hr. We used a Fourier transform in longitude to separate the different azimuthal orders $m$ in the range $3 \leq m \leq 15$. At each $m$ we obtained the phase and amplitude of the Rossby waves as functions of depth using the helioseismic data. At each $m$ we also measured the latitude dependence of the eigenfunctions by calculating the covariance between the equator and other latitudes. We conducted a study of the horizontal and radial dependences of the radial vorticity eigenfunctions. The horizontal eigenfunctions are complex. As observed previously, the real part peaks at the equator and switches sign near $\pm\SI{30}{\degree}$, thus the eigenfunctions show significant non-sectoral contributions. The imaginary part is smaller than the real part. The phase of the radial eigenfunctions varies by only $\pm\SI{5}{\degree}$ over the top $\SI{15}{\mega\metre}$. The amplitude of the radial eigenfunctions decreases by about $\SI{10}{\percent}$ from the surface down to $\SI{8}{\mega\metre}$ (the region in which ring-diagram analysis is most reliable, as seen by comparing with the rotation rate measured by global-mode seismology). The radial dependence of the radial vorticity eigenfunctions deduced from ring-diagram analysis is consistent with a power law down to $\SI{8}{\mega\metre}$ and is unreliable at larger depths. However, the observations provide only weak constraints on the power-law exponents. For the real part, the latitude dependence of the eigenfunctions is consistent with previous work (using granulation tracking). The imaginary part is smaller than the real part but significantly nonzero.

\section{Introduction}
\label{sect_introduction}

Recently, \citet[][hereafter \citetalias{Loeptien2018}]{Loeptien2018} discovered global-scale Rossby waves in maps of flows on the surface of the Sun. These waves are waves of radial vorticity that may exist in any rotating fluid body. Even though Rossby waves were predicted to exist in stars more than $40$~years ago \citep{Papaloizou1978, Saio1982}, solar Rossby waves were difficult to detect because of their small amplitudes ($\sim \SI{1}{\metre\per\second}$) and long periods of several months. Solar Rossby waves contain almost as much vorticity as large-scale solar convection. The dispersion relation of solar Rossby waves is close to the standard relation for sectoral modes, $\omega = -2\Omega/(m + 1)$, where $\Omega$ is the rotation rate of a rigidly rotating star and $m$ is the azimuthal order \citep{Saio1982}. Rossby waves have a retrograde phase speed and a prograde group speed. In \citetalias{Loeptien2018}, the authors also measured the horizontal eigenfunctions, which peak at the equator.

The detection of solar Rossby waves was confirmed by \citet[][hereafter \citetalias{Liang2019}]{Liang2019} with time-distance helioseismology \citep{Duvall1993} using data covering more than $20$~years, obtained from the Solar and Heliospheric Observatory (SOHO) and from the Solar Dynamics Observatory \citep[SDO;][]{Pesnell2012}. \citet{Alshehhi2019}, in an effort to speed up ring-diagram analysis \citep[RDA;][]{Hill1988} via machine learning, also saw global-scale Rossby waves. \citet{Hanasoge2019} and \citet{Mandal2019} provide another recent Rossby wave confirmation using a different technique of helioseismology known as normal-mode coupling \citep{Woodard1989, Hanasoge2017}.

Knowledge about the latitude dependence of Rossby wave eigenfunctions is incomplete, as \citetalias{Loeptien2018} studied only their real parts. In a differentially rotating star, the horizontal eigenfunctions are not necessarily spherical harmonics (and may not even separate in latitude and depth). Also, little is known observationally about the depth dependence of the Rossby waves. It would be well worth distinguishing between the few existing theoretical models of the depth dependence \citep{Provost1981, Smeyers1981, Saio1982, Wolff1986}.

In this paper, we explore the latitude dependence of the eigenfunctions, as well as the phase and amplitude of solar Rossby waves as functions of depth from the surface down to more than $\SI{16}{\mega\metre}$ using helioseismology. We use observations from the Helioseismic and Magnetic Imager \citep[HMI;][]{Schou2012} on board SDO, processed with RDA. From these we attempt to measure the eigenfunctions of the Rossby waves in the solar interior. For comparison near the surface, we also use data from local correlation tracking of granulation \citep[LCT;][]{November1988}.

\section{Data and methods}
\label{sect_data_methods}

We used maps of the horizontal velocity, derived from two different techniques applied to SDO/HMI observations. The first dataset consists of LCT (granulation tracking) flow maps at the surface \citep{Loeptien2017} and covers almost six years from May 20, 2010 to March 30, 2016. The second dataset comprises RDA flow maps from the HMI ring-diagram pipeline (\citealt{Bogart2011a, Bogart2011b}; see also \citealt{Bogart2015}). For comparisons with LCT, we took a period as close to the LCT period as possible, i.e., May 19, 2010 to March 31, 2016, while for all other results we used a longer period of more than seven years from May 19, 2010 to December 29, 2017; this corresponds to 102 Carrington rotations (CRs), i.e., CR $2097$ - $2198$.

\subsection{Overview of LCT data}

The LCT flow maps are obtained from and processed as described in \citet{Loeptien2017}. They are created by applying the Fourier LCT code \citep[FLCT;][]{Welsch2004, Fisher2008} to track the solar granulation in pairs of consecutive HMI intensity images. The image pairs are separated by $30$~min. Several known systematic effects such as the shrinking-Sun effect \citep{Lisle2004, Loeptien2016} and effects related to the SDO orbit are present in the LCT maps. Therefore the maps are decomposed into Zernike polynomials, a basis of 2D orthogonal functions on the unit disk, and the time series of the coefficient amplitudes for the lowest few Zernike polynomials are filtered to remove frequencies of one day and one year (associated with the SDO orbit) as well as all harmonics up to the Nyquist frequency. The zero frequency is also removed. The filtered maps are then tracked at the sidereal Carrington rate and remapped onto an equi-spaced longitude-latitude grid with a step size of $\SI{0.4}{\degree}$ in both directions.

\subsection{Overview of ring-diagram data}
\label{sect_data_methods_rda_overview}

The ring-diagram pipeline \citep{Bogart2011a, Bogart2011b} takes HMI Dopplergrams as input and remaps them onto tiles spanning $182 \times \SI{182}{\mega\metre}$ (i.e., $\SI{15}{\degree}$ each in latitude $\lambda$ and longitude $\varphi$ at the equator). The tiles overlap each other by roughly $\SI{50}{\percent}$ in each direction such that the tile borders fall onto the centers of adjacent tiles. Both the latitude and longitude sampling are half the tile size. The latitude grid is linear and includes the equator, while the longitude grid is also linear, but is latitude-dependent. Each tile is tracked for $1728$~min ($28.8$~hr) at the sidereal Carrington rate. The temporal grid spacing is, on average, $1/24$ of the synodic Carrington rotation period of $27.2753$~days.

In the pipeline, for each tile a 3D local power spectrum is computed from the tracked Dopplergrams. The velocity fit parameters $U_{x,\,n \ell}$ (prograde) and $U_{y,\,n \ell}$ (northward) are extracted via a ring-fit algorithm \citep{Haber2000} for different solar oscillation modes, which are indexed by their radial order $n$ and angular degree $\ell$. The flow velocities $u_x$ and $u_y$ are inferred for various target depths via a 1D optimally localized averages (OLA) inversion. The inversion results for the six-parameter fits of the $\SI{15}{\degree}$ tiles sample a range of target depths from $\SI{0.97}{R_\odot}$ to $\SI{1}{R_\odot}$ (step size $\SI{0.001}{R_\odot}$), corresponding to a nonlinear grid of measurement depths (median of the ring-diagram averaging kernels) from $\SI{0.976}{R_\odot}$ to $\SI{1}{R_\odot}$. In this paper, the term depth always refers to measurement depth and not to target depth.

The inversion results are stored in the Joint Science Operations Center (JSOC) data series hmi.V\_rdvflows\_fd15\_frame. However, up to inversion module rdvinv v.0.91, the inversion results depended on the input tile processing order due to an array initialization bug. This caused significantly lower velocity uncertainties for tiles near latitude $\SI{7.5}{\degree}$ and Stonyhurst longitude $\SI{37.5}{\degree}$, even when averaged over seven years, but also slightly affected the velocities. At the same disk locations the bug caused a correlation of $u_x$ with the $B_0$ angle. Since rdvinv v.0.92 is officially only applied since March 2018, we re-inverted the entire dataset ourselves for the analysis shown in this work.

Apart from the array initialization bug, we found several other issues with the default HMI ring-diagram pipeline that have not yet been solved. Among these are under-regularization in the inversion for some individual tiles, leading to relatively narrow averaging kernels and anomalously high noise. Finally, the number of ring fits used for the inversion depends strongly on disk position. This may lead to systematic effects and additional noise.

The ring-diagram velocities $u_x$ reported at a certain measurement depth $r$ at the equator for an angular rotation rate $\Omega(r)$ are equal to $\Omega(r) R_\odot$ instead of the local velocity $\Omega(r) r$. Since we are interested in the latter, we multiplied $u_x$ by $r/R_\odot$. By analogy, we also applied this factor to $u_y$ and to all other latitudes. Additionally, the inversion does not account for the quantity $\beta_{n \ell}$, defined, for example, in Eq.~3.357 of \citet{Aerts2010}. The quantity $\beta_{n \ell}$ is related to the effect of the Coriolis force on the mode frequency splitting. For uniform rotation in particular, at fixed $m$, $\beta_{n \ell}$ completely describes the effect of the rotation on the mode frequency splitting. Both issues are described in more detail in App.~\ref{app_rda_inversion_issues}.

\subsection{Post-processing of ring-diagram data}

The ring-diagram data are organized in CRs, which undergo several processing steps, including the removal of systematic effects, an interpolation in longitude, an interpolation in time, and the removal of limb data.

Several systematic effects are present in the ring-diagram velocities, such as center-to-limb effects that depend on the disk position of the tile \citep{Baldner2012, Zhao2012}. There are time-independent effects and systematics with a one-year period, which are probably related to the $B_0$ angle. To remove the systematics, we fit the time series at each position on the disk (in Stonyhurst coordinates) with sinusoids
\begin{equation}
\begin{aligned}
& u_x(t) = a_x \sin(2\pi t/(\SI{1}{yr})) + b_x \cos(2\pi t/(\SI{1}{yr})) + c_x, \\
& u_y(t) = a_y \sin(2\pi t/(\SI{1}{yr})) + b_y \cos(2\pi t/(\SI{1}{yr})) + c_y,
\end{aligned}
\end{equation}
and subtract the fits from the flow velocities. We used all available CRs to determine the fit parameters.

Because of the specific tile coordinate selection used by the ring-diagram pipeline \citep{Bogart2011a}, which seeks to optimally cover the visible disk, tile centers at different latitudes have Stonyhurst longitudes that are offset by multiples of $\SI{2.5}{\degree}$ from each other. To obtain a latitude-independent longitude grid, we interpolated the flow maps in Stonyhurst longitude using splines (App.~\ref{app_rda_processing_steps}).

We also interpolated the ring-diagram flows in time similarly with splines to fill missing time steps due to instrumental issues (only $12$ out of $2448$ time steps), which cause a too low observational duty cycle ($\leq \SI{70}{\percent}$). We interpolated the data in the Carrington reference frame so as to use always roughly the same physical locations on the Sun. This mixes different systematics, which are primarily dependent on disk position, but we should already have removed the dominant contributions at this stage. We interpolated every missing time step from roughly the same number of data points (all available time steps within the corresponding disk passage) using splines (App.~\ref{app_rda_processing_steps}).

The output uncertainties from the ring-diagram pipeline increase strongly toward the limb, in particular beyond an angular great-circle distance of roughly $\SI{65}{\degree}$ to the crossing of the central meridian with the equator ($\lambda = \SI{0}{\degree}, \varphi = \SI{0}{\degree}$). We thus only used ring-diagram data within $\SI{65}{\degree}$ of ($\lambda = \SI{0}{\degree}, \varphi = \SI{0}{\degree}$).

\subsection{From velocity maps to power spectra of radial vorticity}

From this stage onward ring-diagram and LCT data are processed similarly. The processing steps include a shift to the equatorial rotation rate $\nu_{\text{eq}} = \Omega_{\text{eq}}/2\pi = \SI{453.1}{\nano\hertz}$, the subtraction of the longitude mean, the calculation of the radial vorticity, a spherical harmonic transform (SHT), and a Fourier transform of the SHT coefficient time series.

The flow maps are shifted from the tracking rate (sidereal Carrington rate) to the surface sidereal equatorial rotation rate of $\SI{453.1}{\nano\hertz}$, an average of zonal flows inferred from global-mode analysis of SDO/HMI observations \citep{Larson2018}. We shifted the LCT data in Fourier space via a time-dependent phase factor, applying the same convention for the Fourier transform as \citetalias{Loeptien2018}. The ring-diagram data are first apodized by a raised cosine in angular great-circle distance to the point ($\lambda = \SI{0}{\degree}, \varphi = \SI{0}{\degree}$) to suppress near-limb data and are shifted via spline-interpolation (App.~\ref{app_rda_processing_steps}).

We next subtracted the longitude mean from the data to remove any remaining large-scale flows. Differential rotation and meridional circulation should have already been subtracted in the RDA or LCT post-processing, but any possible longitude-independent flows still in the data are removed in this step.

Subsequently, we calculated the radial vorticity (via second-order central finite differences) as follows:
\begin{equation}
\label{eq_definition_radial_vorticity}
\begin{aligned}
\zeta(t,r,\lambda,\varphi) =& - \frac{1}{r \cos\lambda} \frac{\partial (u_x(t,r,\lambda,\varphi) \cos\lambda)}{\partial\lambda} + \frac{1}{r \cos\lambda} \frac{\partial u_y(t,r,\lambda,\varphi)}{\partial \varphi},
\end{aligned}
\end{equation}
where $r$ is the measurement depth. We decomposed the resulting maps into spherical harmonics and performed a temporal Fourier transform of the spherical harmonic coefficients. Last, we calculated the power and the phase (where the phase range is the half-open interval ($\SI{-180}{\degree},\SI{180}{\degree}$]). The sign convention is such that waves with positive $m$ and negative frequency $\nu$ have a retrograde phase speed.

If not stated otherwise, the terms power spectrum or Fourier transform used in this paper always refer to the power spectrum or Fourier transform of the radial vorticity. Similarly, we discuss eigenfunctions of radial vorticity. These eigenfunctions are not spherical harmonics, however \citepalias{Loeptien2018}. In particular, while the modes can be meaningfully indexed by $m$, the angular degree $\ell$ is not observable. Throughout the paper $\ell$ thus only refers to the projection of the Rossby wave modes onto the corresponding spherical harmonic and not to the Rossby wave eigenfunction itself. We also use the terms latitudinal and radial eigenfunctions, which assumes separability in the $r$ and $\lambda$ coordinates. This assumption is addressed in more detail in Sect.~\ref{sect_summary}.

\section{Results}
\label{sect_results}

\subsection{Radial vorticity maps}

\begin{figure}
\centering
\includegraphics[width=\hsize]{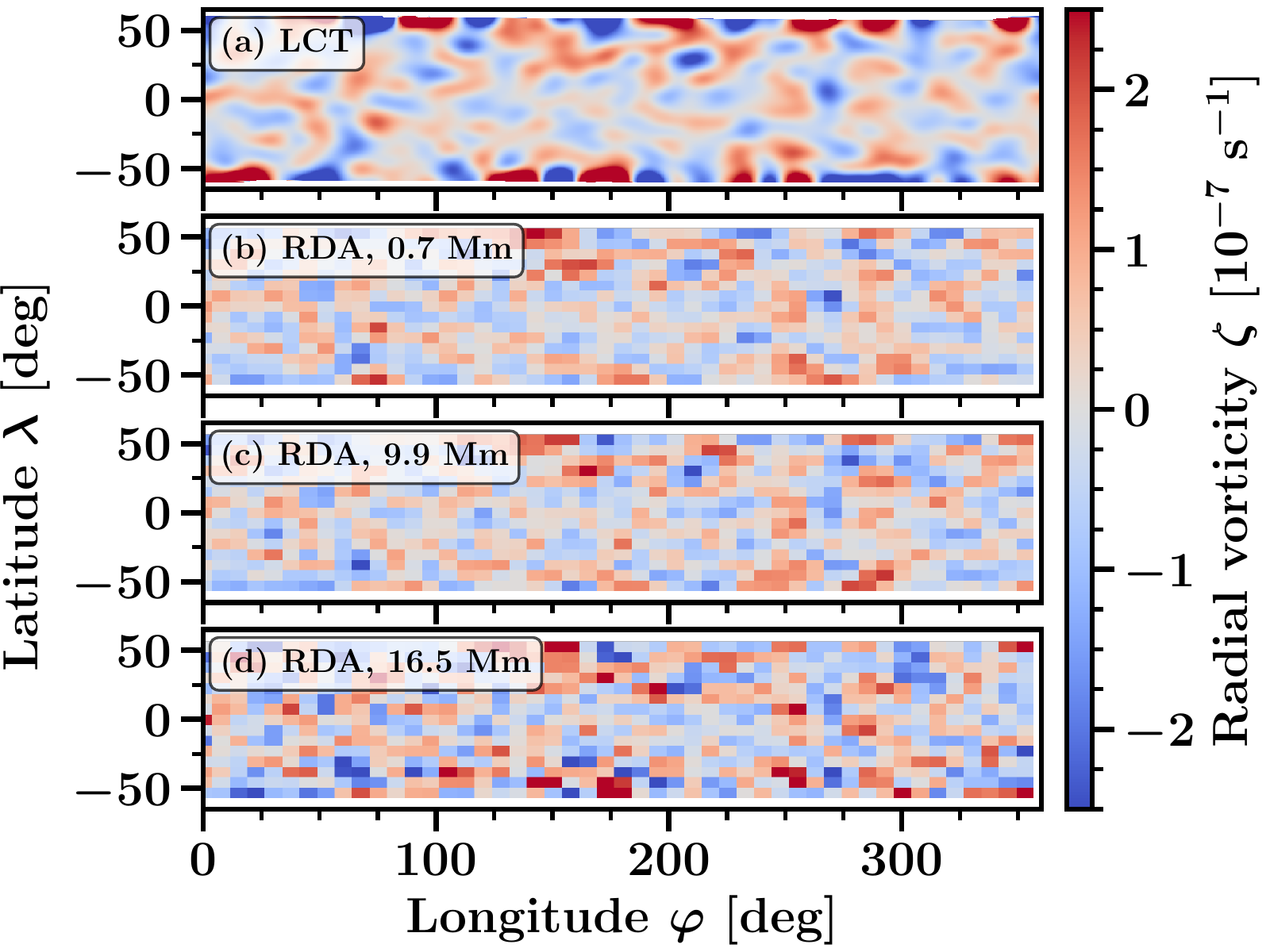}
\caption[Radial vorticity maps from LCT and RDA]{
Radial vorticity maps from LCT at the surface and from RDA at depths $0.7$, $9.9$, and $\SI{16.5}{\mega\metre}$. The radial vorticity is averaged over one rotation (from May 20, 2010 to June 16, 2010). The LCT map is smoothed in latitude and longitude with a Gaussian filter ($\sigma = \SI{6}{\degree}$) to filter out small-scale convection.
}
\label{fig_vorticity_maps}
\end{figure}

Figure~\ref{fig_vorticity_maps} shows example vorticity maps from LCT surface flows and from RDA flows near the surface, at intermediate, and at large depths ($0.7$, $9.9$, and $\SI{16.5}{\mega\metre}$), averaged over the first rotation in the dataset (May 20, 2010 to June 16, 2010). The LCT data have a much better horizontal resolution than the ring-diagram data and thus pick up small-scale convective contributions. To be able to compare LCT with RDA, we thus smooth the LCT vorticity with 1D Gaussian filters of width $\sigma = \SI{6}{\degree}$ both in latitude and longitude.

We do not expect perfect agreement of the two methods because of their different sensitivities to horizontal scales and to different depths. Nonetheless, the LCT map shows similar features as the near-surface ($\SI{0.7}{\mega\metre}$) ring-diagram map. While large absolute radial vorticities are visible at high latitudes (beyond $\pm\SI{50}{\degree}$) in the LCT but not in the ring-diagram data, the vorticities near the equator agree. As a test, we interpolate the LCT data to the RDA grid using a 2D bicubic spline. The correlation coefficient between the interpolated LCT and the ring-diagram maps decreases with the latitude width of the strip of pixels considered and there is a steep decrease beyond $\pm\SI{45}{\degree}$. The correlation is 0.92 when including only equatorial pixels, 0.79 for pixels within $\pm\SI{45}{\degree}$, and 0.59 for all pixels, i.e., within $\pm\SI{52.5}{\degree}$. The noise increases strongly with depth (see lower panels of Fig.~\ref{fig_vorticity_maps}), but the main vorticity features are still visible.

\subsection{Power spectra of radial vorticity}
\label{sect_power_spectra_rossby}

\begin{figure}
\centering
\includegraphics[width=\hsize]{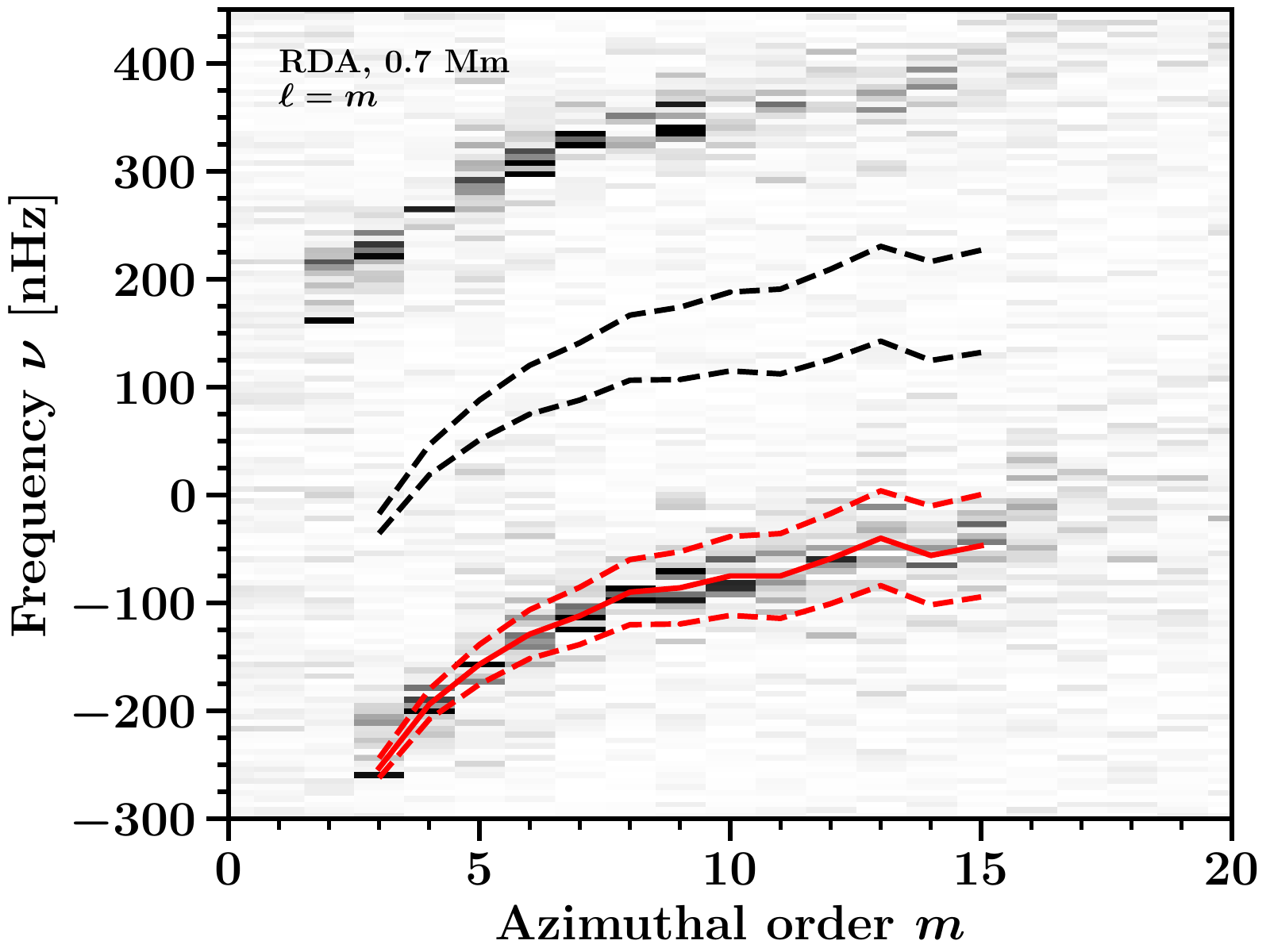}
\caption[Sectoral power spectrum of the radial vorticity from RDA (2D)]{
Sectoral power spectrum ($\ell = m$) of the radial vorticity for RDA data at depth $\SI{0.7}{\mega\metre}$. The solid red line indicates the Rossby wave frequencies from \citetalias{Liang2019} for $m = 3$ and from \citetalias{Loeptien2018} for $m \geq 4$. Frequency intervals for the Rossby wave region and the background region are indicated by the red and black dashed lines, respectively. The ridge of power at positive frequencies is due to the first side lobe of the window function. For better visibility of the Rossby waves at low $m$, the power is normalized at each $m$ by the frequency average over [$-300,100$]~nHz. The color scale is truncated at $\SI{50}{\percent}$ of the maximum value (black).
}
\label{fig_power_spectrum_2d}
\end{figure}

Figure~\ref{fig_power_spectrum_2d} shows the Rossby wave power of the $\ell = m$ component for the ring-diagram data near the surface ($\SI{0.7}{\mega\metre}$) versus frequency and azimuthal order $m$ (\citetalias{Loeptien2018} detected only the sectoral component of the Rossby waves). We divide the power, at each $m$, by the frequency average over [$-300,100$]~nHz near the surface ($\SI{0.7}{\mega\metre}$). The visible power ridge corresponds to the Rossby wave signal. The mode frequency increases with $m$ roughly according to the textbook dispersion relation for sectoral waves, $\omega = -2\Omega_{\text{eq}}/(m + 1)$, as seen earlier by \citetalias{Loeptien2018}.

Besides the Rossby wave signal there are other ridges, that, at fixed $\Delta m = m - m'$, are shifted from the Rossby waves by roughly $\Delta m \left(\nu_{\text{eq}} - 1/(\SI{1}{yr})\right)$, where $\nu_{\text{eq}} - 1/(\SI{1}{yr}) \sim \SI{421.4}{\nano\hertz}$. The main contribution to these side lobes comes from a temporal window function, which is introduced by solar rotation and not by time gaps; the time coverage is very good (see Sect.~\ref{sect_data_methods}). This leads to side lobes of wave power from modes at $m'$ to modes at $m$. We only show the side lobes for $\Delta m = +1$, but we typically observe the side lobes above the noise between $\Delta m = -2$ and $\Delta m = +3$. In \citetalias{Liang2019}, the authors use $21$~years of time-distance data from a combined sample of observations from the Michelson Doppler Imager (MDI) on board SOHO and from SDO/HMI and they discuss the window function in detail.

\begin{figure*}
\centering
\includegraphics[width=\hsize]{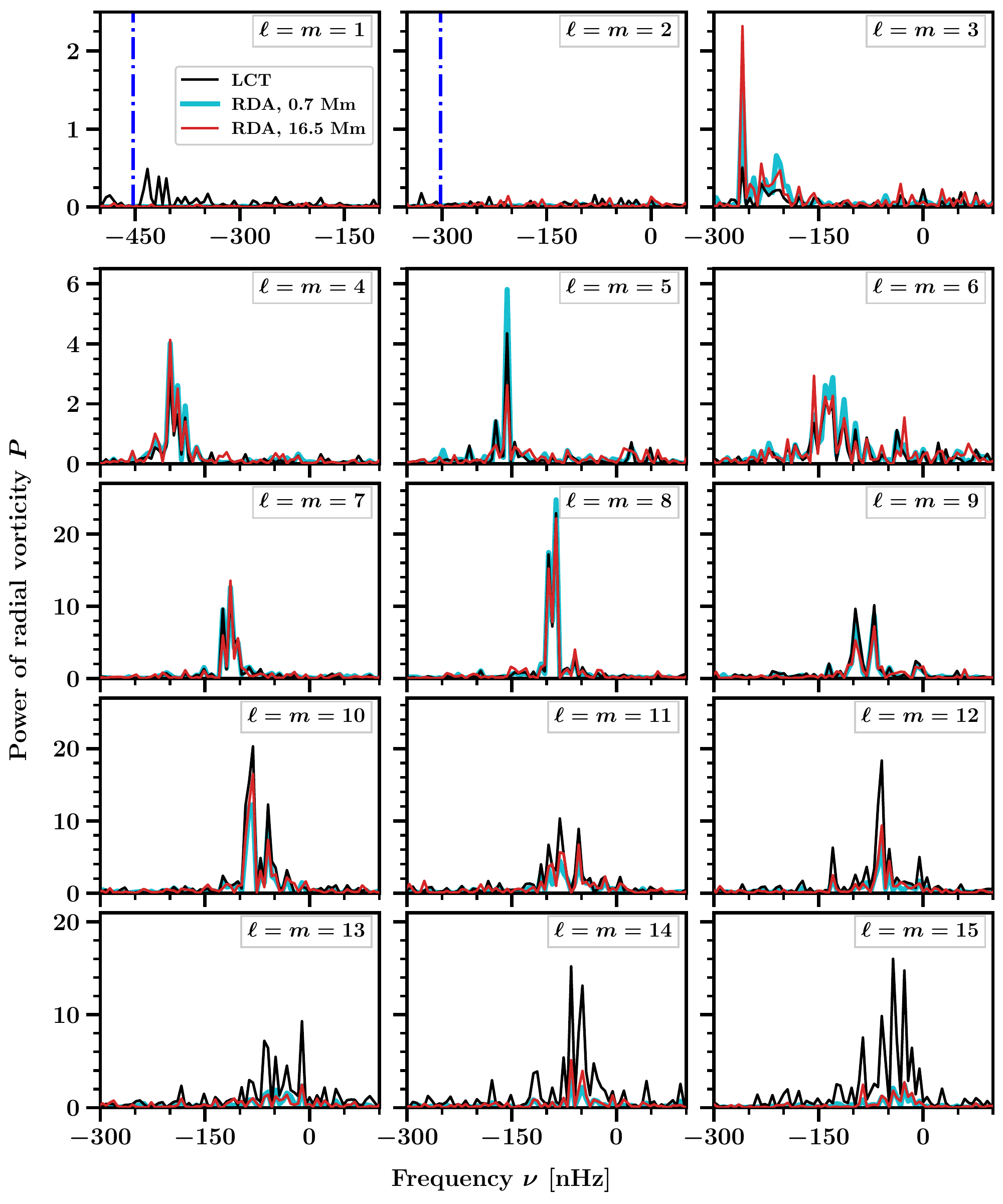}
\caption[Sectoral power spectrum of the radial vorticity from RDA (1D cuts)]{
Sectoral power spectra ($\ell = m$) showing the Rossby wave power in the LCT data (black line) and the RDA data at depths $0.7$ and $\SI{16.5}{\mega\metre}$ (cyan and red lines). The power is normalized by the average of the $m = 8$ power in the range [$-300,100$]~nHz. The dash-dotted vertical lines in the $m = 1$ and $m = 2$ panels indicate the frequencies $\omega = -2\Omega_{\text{eq}}/(m + 1)$. The frequency axes of the $m = 1$ and $m = 2$ panels are shifted with respect to the other panels.
}
\label{fig_power_spectrum_all_mvals}
\end{figure*}

Figure~\ref{fig_power_spectrum_all_mvals} shows the power versus frequency for different azimuthal orders $m$. We divide the power, at each $m$, by the frequency average of the $\ell = m = 8$ mode over [$-300,100$]~nHz near the surface ($\SI{0.7}{\mega\metre}$). The power decreases from $0.7$ to $\SI{9.9}{\mega\metre}$, then increases toward $\SI{16.5}{\mega\metre}$, but the depth dependence is modest ($\leq \SI{20}{\percent}$). We also see that the wave power decreases with $m$ faster for RDA than for LCT owing to the different sensitivity kernels, as found by \citetalias{Loeptien2018}. The $\ell = m = 3$ signal has a multi-peak structure and is thus difficult to measure. We do not observe Rossby waves for $\ell = m \leq 2$; the dash-dotted blue lines for $\ell = m = 1$ and $\ell = m = 2$ indicate the expected mode frequencies from the textbook dispersion relation.

The wave power side lobes due to the window function explain why the $\ell = m = 6$ side lobe in Fig.~\ref{fig_power_spectrum_2d} even exceeds the main signal: the adjacent $\ell = m = 7$ mode has a higher relative power (see Fig.~\ref{fig_power_spectrum_all_mvals}). Systematic effects that are fixed in the Stonyhurst reference frame can be easily misinterpreted as an $\ell = m = 1$ Rossby wave signal (see the LCT curve in Fig.~\ref{fig_power_spectrum_all_mvals}), as their frequency (the rotation rate) is equal to the $\ell = m = 1$ Rossby wave frequency.

We assume that there is background power contributing to the observed power at the Rossby peak, but measuring its contribution directly at the peak is impossible. Since we are limited by the side lobes, we use a region halfway between the peak and the next side lobe, i.e., shifted from the peak by half the rotation rate. We checked that the shift direction does not matter much, so for the central background frequency, we use the Rossby wave frequencies $\nu_0^{\text{ref}}$ from \citetalias{Liang2019} and \citetalias{Loeptien2018} for $m = 3$ and $m \geq 4$, respectively, plus half the rotation frequency $\nu_{\text{eq}}$. We use the full widths at half maximum $\gamma^{\text{ref}} = \Gamma^{\text{ref}}/2\pi$ from \citetalias{Liang2019} and \citetalias{Loeptien2018} for $m = 3$ and $m \geq 4$, respectively, and perform a least-squares second-order polynomial fit in $m$ to obtain a smoothed linewidth $\gamma_{\text{smooth}}$. We use $\gamma_{\text{smooth}}$ for the width of the peak and background frequency intervals. Thus our peak and background frequency intervals at each $m$ are
\begin{equation}
\label{eq_freqintvals}
\begin{aligned}
&\text{peak interval:} \quad &&\nu_0^{\text{ref}} &&\pm \gamma_{\text{smooth}}, \\
&\text{background interval:} \quad &&(\nu_0^{\text{ref}} + \nu_{\text{eq}}/2) &&\pm \gamma_{\text{smooth}}.
\end{aligned}
\end{equation}
These definitions are used in the analysis of latitudinal eigenfunctions in Sect.~\ref{sect_latfunc}.

In Fig.~\ref{fig_power_spectrum_2d}, we see that the peak interval (dashed red lines) typically captures the main wave power well. The 1D power spectra, however, reveal that the background interval (not shown in Fig.~\ref{fig_power_spectrum_all_mvals}, but see dashed black lines in Fig.~\ref{fig_power_spectrum_2d}), however, is potentially contaminated by scattered signal power, for example, for $\ell = m = 5$, $6$, and $14$. To check how the frequency interval definition affects our results, we performed our analysis for several different peak and background intervals. The results are consistent, thus we adopt Eq.~\ref{eq_freqintvals} for the peak and background intervals.

Unlike \citetalias{Loeptien2018}, we see evidence for non-sectoral components of the Rossby waves. For $\ell = m + 2$ the 2D power spectrum shows, for $6 \leq m \leq 13$, a ridge of power at very similar frequencies to those of the $\ell = m$ Rossby waves seen in Fig.~\ref{fig_power_spectrum_2d}, apart from a higher relative noise level and side lobes. This is confirmed by the 1D cuts at fixed values of $m$. We do not see structure in the power spectra for $\ell = m + k$ other than for $k = 0$ and $k = 2$. In Sect.~\ref{sect_latfunc_results}, we indeed show that the latitudinal eigenfunctions of Rossby waves are not sectoral spherical harmonic functions (in agreement with \citetalias{Loeptien2018}).

\subsection{Latitudinal eigenfunctions of Rossby waves}
\label{sect_latfunc}

To estimate the latitudinal eigenfunctions, we first remove small-scale convection from the LCT maps via smoothing with a $\SI{6}{\degree}$ Gaussian in latitude. Next we compute the Fourier transform of the radial vorticity maps $\zeta(t,r,\lambda,\varphi)$ from LCT and RDA in time and longitude as follows:
\begin{equation}
\hat{\zeta}_m(\nu,r,\lambda) = \sum_{t} \sum_{\varphi} \zeta(t,r,\lambda,\varphi) e^{i (2\pi \nu t - m\varphi)}.
\end{equation}
The variables are discrete and take values at time steps $t_j = j T/N_t$ (integer $0 \leq j < N_t$), longitudes $\varphi_k = 2\pi k/N_\varphi$ (integer $0 \leq k < N_\varphi$), frequencies $\nu_s = s/T$ (integer $s$, with $-N_t/2 \leq s \leq N_t/2 - 1$ for even $N_t$), and azimuthal orders $m$ (integer, with $-N_\varphi/2 \leq m \leq N_\varphi/2 - 1$ for even $N_\varphi$). In this case, $T$, $N_t$, and $N_\varphi$ are the observation period and the number of data points in time and longitude, respectively. We apply a filter to select the Rossby waves one $m$ at a time, i.e.,
\begin{equation}
\bar{\zeta}_m(\nu,r,\lambda) = \hat{\zeta}_m(\nu,r,\lambda) F_m(\nu).
\end{equation}
The filter $F_m(\nu)$ is equal to one within the Rossby wave ridge and zero elsewhere. Since $\zeta(t,r,\lambda,\varphi)$ is real, the symmetry $\bar{\zeta}_m(\nu,r,\lambda) = \bar{\zeta}_{-m}^*(-\nu,r,\lambda)$ applies. We then transform back to time to obtain
\begin{equation}
\tilde{\zeta}_m(t,r,\lambda) = \frac{1}{N_t} \sum_{\nu} \bar{\zeta}_m(\nu,r,\lambda) e^{-i 2\pi \nu t}.
\end{equation}
In this way we obtain filtered time-latitude vorticity maps for every $m$. Because there is no symmetry $\bar{\zeta}_m(\nu,r,\lambda) = \bar{\zeta}_m^*(-\nu,r,\lambda)$, the filtered vorticity maps $\tilde{\zeta}_m(t,r,\lambda)$ are in general complex.

\citetalias{Loeptien2018} do a similar analysis for LCT data, in particular for rotation-averaged maps and filtering within $\pm\SI{30}{\nano\hertz}$ around the central mode frequencies. We do the entire latitudinal eigenfunction analysis for LCT and RDA, for full time-resolution maps and maps averaged in time within individual solar rotations, and for a $\pm\SI{27}{\nano\hertz}$ (five frequency pixels) and a linewidth filter (Eq.~\ref{eq_freqintvals}) around the central mode frequencies. The different time-resolution and filtering cases yield consistent results; we thus show only the outcome for the full time-resolution and linewidth filtering. However, \citetalias{Loeptien2018} take the real part of the complex $\tilde{\zeta}_m(t,r,\lambda)$. This is equivalent to assuming that the phase of the eigenfunction is independent of latitude. We address the implications of this in Sect.~\ref{sect_latfunc_results} in more detail.

To estimate uncertainties for all results in this paper, we split the data into equal-size time intervals, apply our analysis to each chunk, and calculate the standard deviation over the results (for complex quantities separately for the real and imaginary part). Appendix~\ref{app_error_estimation_validation} gives more details on error estimation and validation. Because of the small number of chunks, low-number statistics are an issue and the reported error bars are relatively uncertain. 

For the sake of clarity, for the simple case of a single $m$ Rossby wave with a frequency $\nu_m$ and an eigenfunction $C_m(r,\lambda)$, the vorticity field for that mode, $\zeta_m(t,r,\lambda,\varphi)$, would be given by
\begin{equation}
\zeta_m(t,r,\lambda,\varphi) \propto \textrm{Re}\left( C_m(r,\lambda) e^{i (m \varphi - 2\pi \nu_m t)} \right).
\end{equation}
We apply two different methods to obtain the eigenfunctions $C_m(r,\lambda)$ near the surface, the covariance method (Sect.~\ref{sect_latfunc_cov}), and the SVD method (Sect.~\ref{sect_latfunc_svd}). The former is used also by \citetalias{Loeptien2018}.

\subsubsection{Covariance}
\label{sect_latfunc_cov}

We calculate, at each $m$, the temporal covariance of the vorticity $\tilde{\zeta}$ between the equator near the surface (target depth $R = R_\odot - \SI{0.7}{\mega\metre}$ for RDA) and all other latitudes and depths, normalized by the variance at the equator near the surface
\begin{equation}
\label{eq_eigenfunc_cov_covnorm}
C_m(r,\lambda) = \frac{\langle \tilde{\zeta}^{'}_{m}(t,r,\lambda) \tilde{\zeta}_m^{'*}(t,r = R,\lambda = \SI{0}{\degree}) \rangle_t}{\langle \vert \tilde{\zeta}_m^{'}(t,r = R,\lambda = \SI{0}{\degree}) \vert^2 \rangle_t}, 
\end{equation}
where the angle brackets $\langle \cdot \rangle_t$ denote a temporal average and $\tilde{\zeta}^{'} = \tilde{\zeta} - \langle \tilde{\zeta} \rangle_t$ is the centered vorticity. The function $C_m(r,\lambda)$ is complex-valued since $\tilde{\zeta}_m$ is in general complex. By construction $C_m(r = R,\lambda = \SI{0}{\degree})$ is unity. Appendix~\ref{app_latitudinal_eigenfunction_methods} shows that $C_m$ can also be obtained by a linear fit to the vorticity. The same covariance can be computed with the LCT data.

\subsubsection{Singular value decomposition}
\label{sect_latfunc_svd}

We present a second, new method to obtain latitudinal eigenfunctions. We want to separate the filtered vorticity at each azimuthal order $m$ and depth $r$, i.e., a 2D matrix, into a latitude and a time dependence, i.e.,
\begin{equation}
\label{eq_svd_concept}
\tilde{\zeta}_m(t,r,\lambda) \propto f_m(t) C_m(r,\lambda).
\end{equation}
Applying a singular value decomposition (SVD), we can decompose the vorticity as
\begin{equation}
\tilde{\zeta}_m(t,r,\lambda) = \sum_{j = 0}^{k-1} s_{(r,\,m),\,j} U_{(r,\,m),\,j}(t) V_{(r,\,m),\,j}(\lambda),
\end{equation}
where $s_{(r,\,m),\,j}$ is the singular value of index $j$ with left and right singular vectors $U_{(r,\,m),\,j}$ and $V_{(r,\,m),\,j}$ and $k$ is the minimum between the number of grid points in time and latitude. The square of $s_{(r,\,m),\,j}$ measures the variance captured by its singular vectors. By convention the singular values are sorted in descending order, thus the first singular vector contains more variance than any other individual singular vector.

Assuming that there is only one nonzero singular value, $s_{(r,\,m),\,0}$, the SVD gives the desired decomposition of the vorticity into one time and one latitude function. This assumption is valid if there is only one excited mode in our filtered vorticity maps. Our observations indeed have one clearly dominant singular value: The first singular value, $s_{(r,\,m),\,0}$, is typically two to three times larger than the second singular value, $s_{(r,\,m),\,1}$. In addition, by looking at the latitude singular vectors from different time chunks, we noticed that the first singular vector, $V_{(r,\,m),\,0}$, always had a similar shape, whereas the second singular vector, $V_{(r,\,m),\,1}$, had different shapes for different chunks. This is another indication that there is only one significant mode.

Given that the noise at high latitudes increases steeply, we crop our vorticity maps for the SVD to latitudes within $\pm\SI{50}{\degree}$ of the equator. Also, the SVD does not account for the varying noise of the remaining latitudes. To ensure that latitudes with larger uncertainties are given less weight, we filter the original vorticity maps once more in Fourier space for the noise, calculate the temporal standard deviation $\sigma_m$ of the noise-filtered maps, and compute $\tilde{\zeta}_{\text{nw},\,m}(t,r,\lambda) = \tilde{\zeta}_m(t,r,\lambda)/\sigma_m(r,\lambda)$. We filter for the noise by taking either all frequencies except for five pixels around the peak or all frequencies within the background interval (see Eq.~\ref{eq_freqintvals}). The two different filters give consistent results. At each $m$, the SVD is performed on the weighted maps $\tilde{\zeta}_{\text{nw},\,m}$ and the resulting latitude vectors are multiplied by $\sigma_m$ again to undo the weighting. We apply the weighting only to LCT, since the ring-diagram data are already apodized (see Sect.~\ref{sect_data_methods}). We select the first latitude singular vector near the surface and normalize it by its value at the equator.

\subsubsection{Results for the latitudinal eigenfunctions}
\label{sect_latfunc_results}

\begin{figure*}
\centering
\includegraphics[width=\hsize]{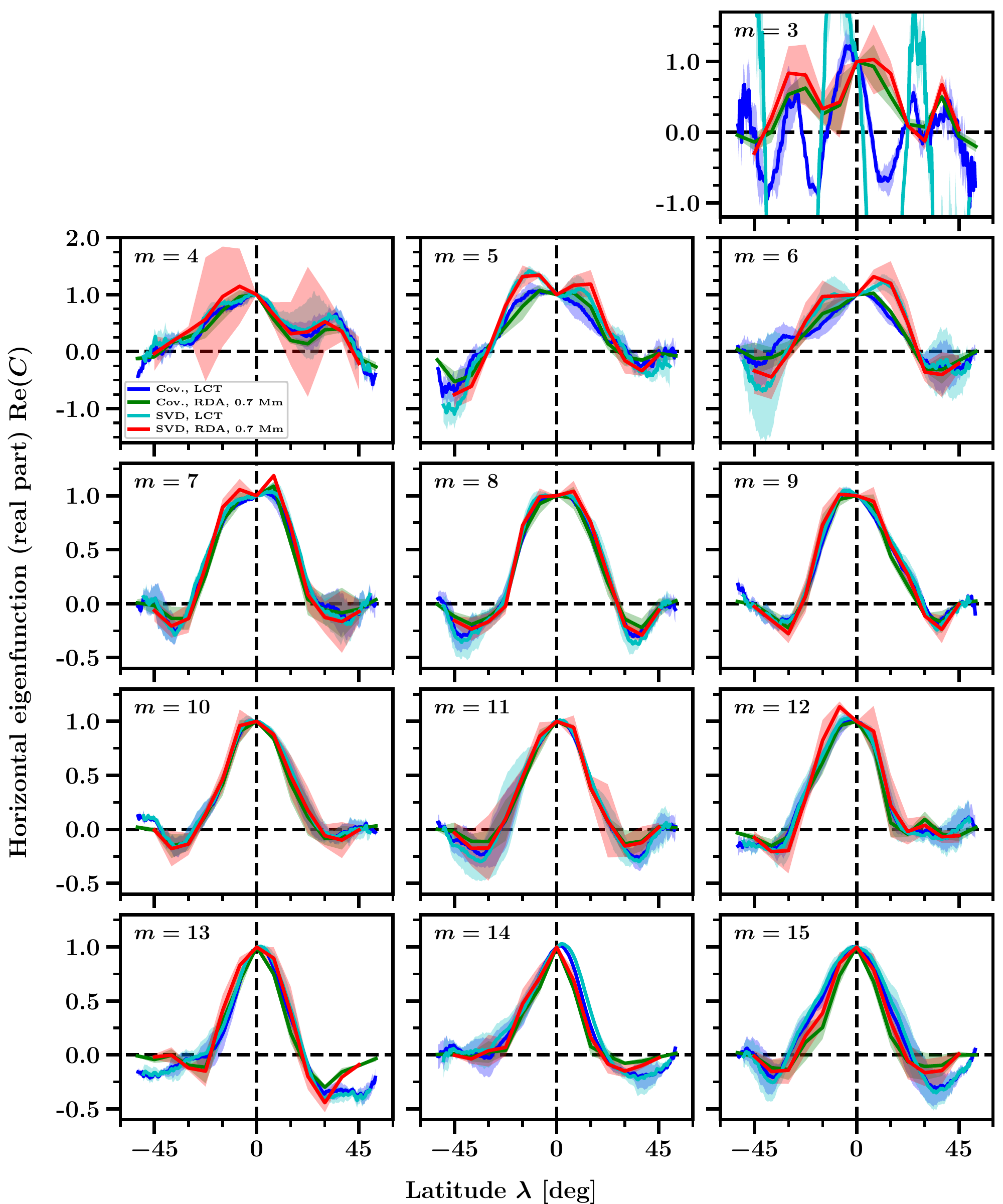}
\caption[Latitudinal eigenfunctions of Rossby waves, real part]{
Real part of $C_m(\lambda)$ for different azimuthal orders $m$ and four different methods (see legend in panel $m = 4$). The shaded areas indicate the 1$\sigma$ error estimates. 
}
\vspace{1cm}
\label{fig_horiz_eigenfunction_all_mvals_real}
\end{figure*}

\begin{figure*}
\centering
\includegraphics[width=\hsize]{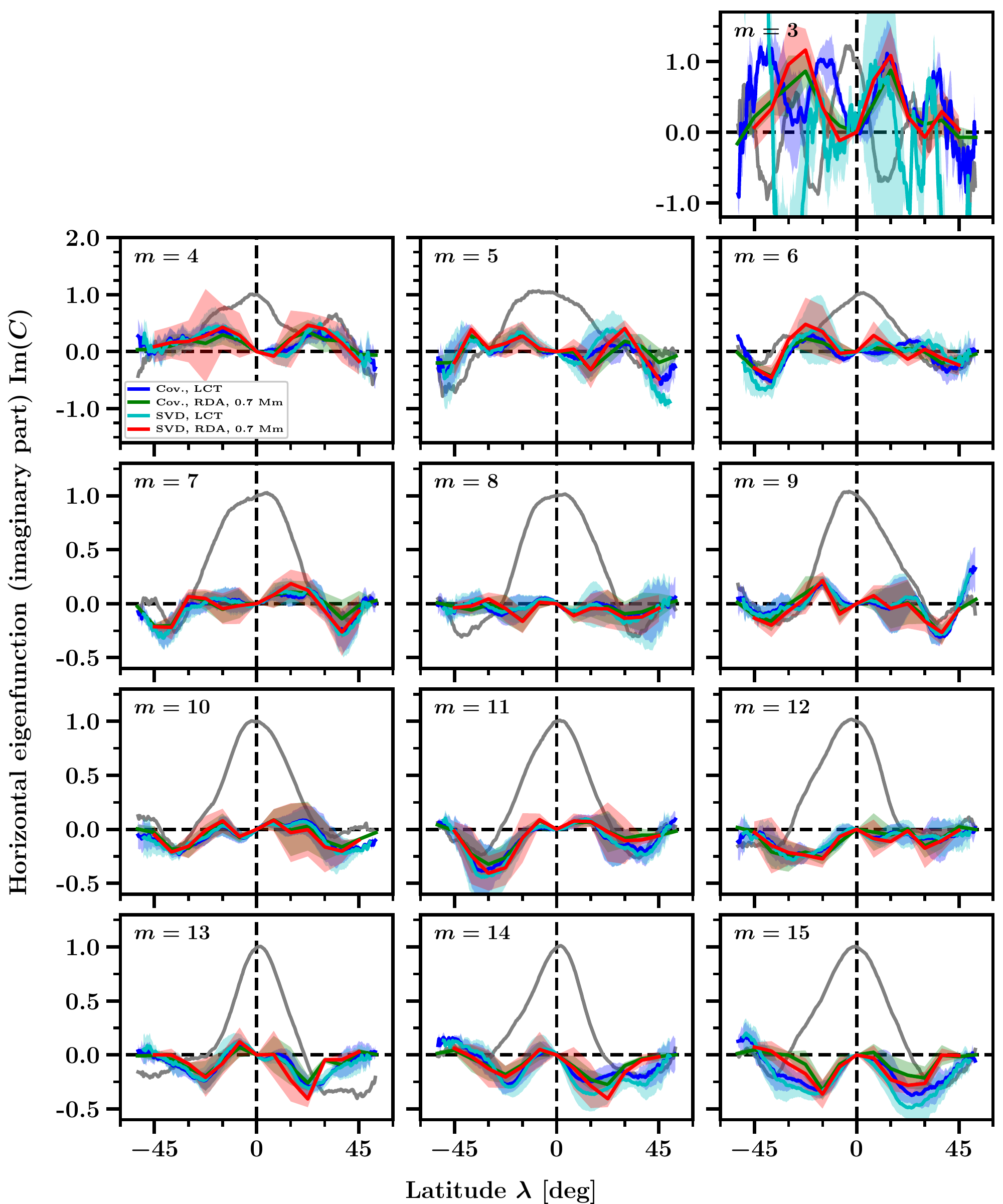}
\caption[Latitudinal eigenfunctions of Rossby waves, imaginary part]{
Imaginary part of $C_m(\lambda)$ for different azimuthal orders $m$ and four different methods (see legend in panel $m = 4$). The shaded areas indicate the 1$\sigma$ error estimates. For comparison, the solid gray curves show the real part of $C_m$ for the LCT covariance-based data. The plotting ranges are the same as in Fig.~\ref{fig_horiz_eigenfunction_all_mvals_real}.
}
\vspace{1cm}
\label{fig_horiz_eigenfunction_all_mvals_imag}
\end{figure*}

Figures~\ref{fig_horiz_eigenfunction_all_mvals_real} and \ref{fig_horiz_eigenfunction_all_mvals_imag} show the real and imaginary parts of the horizontal eigenfunctions of Rossby waves versus latitude for different $m$. The real part is consistent with the findings from \citetalias{Loeptien2018}. The imaginary part, however, was not discussed by \citetalias{Loeptien2018}.

In the current paper, we find that the LCT and the RDA results are mostly consistent for the near-surface layers. Also, almost all $m$ show agreement between the covariance and SVD results. This in particular holds for the modes with the largest amplitudes, i.e., for $7 \leq m \leq 10$. On the other hand, the modes $m = 4$ and to a lesser extent $m = 15$, where Rossby wave measurements become difficult, display larger errors but nonetheless consistent results. The $m = 3$ results for the different techniques disagree and are noisy. The $m = 5$ and $m = 6$ results for the real part differ slightly between the covariance and SVD methods. While the covariance yields a real part of the eigenfunction quite similar to those of other modes, the SVD-based results show maxima around latitudes of $\pm 10$-$\SI{15}{\degree}$. Apparently, there the SVD picks up some variance that is uncorrelated with the equator. It is unclear whether it is just noise, or a real signal of a different kind of latitudinal eigenfunctions.

The eigenfunction shape is similar for different modes. The real part decreases away from the equator, flips sign, and approaches zero after going through a local minimum. The imaginary part is much noisier than the real part, as indicated by the error estimates. For most $m$, it is close to zero and flat near the equator, but reaches minima at high latitudes. The latitude of the minima appears to move toward the equator with increasing $m$.

As can be seen from, for example, the red curves in Fig.~\ref{fig_horiz_eigenfunction_all_mvals_imag}, the imaginary part appears to be mostly positive for $3 \leq m \leq 6$. For $7 \leq m \leq 9$ the sign of the imaginary part is unclear. For $10 \leq m \leq 15$, the imaginary part is predominantly negative. The presence of an imaginary part induces a phase for the latitudinal eigenfunctions that can be interpreted as a latitude-dependent shift of the sinusoid in longitude. A positive sign of the imaginary part means that the horizontal eigenfunctions at high latitudes are leading in the retrograde direction with respect to the equator. Conversely, a negative sign would indicate that the eigenfunctions at high latitudes are trailing with respect to the equator. This may provide important constraints on the theory of latitudinal eigenfunctions of Rossby waves.

\begin{figure}
\centering
\includegraphics[width=\hsize]{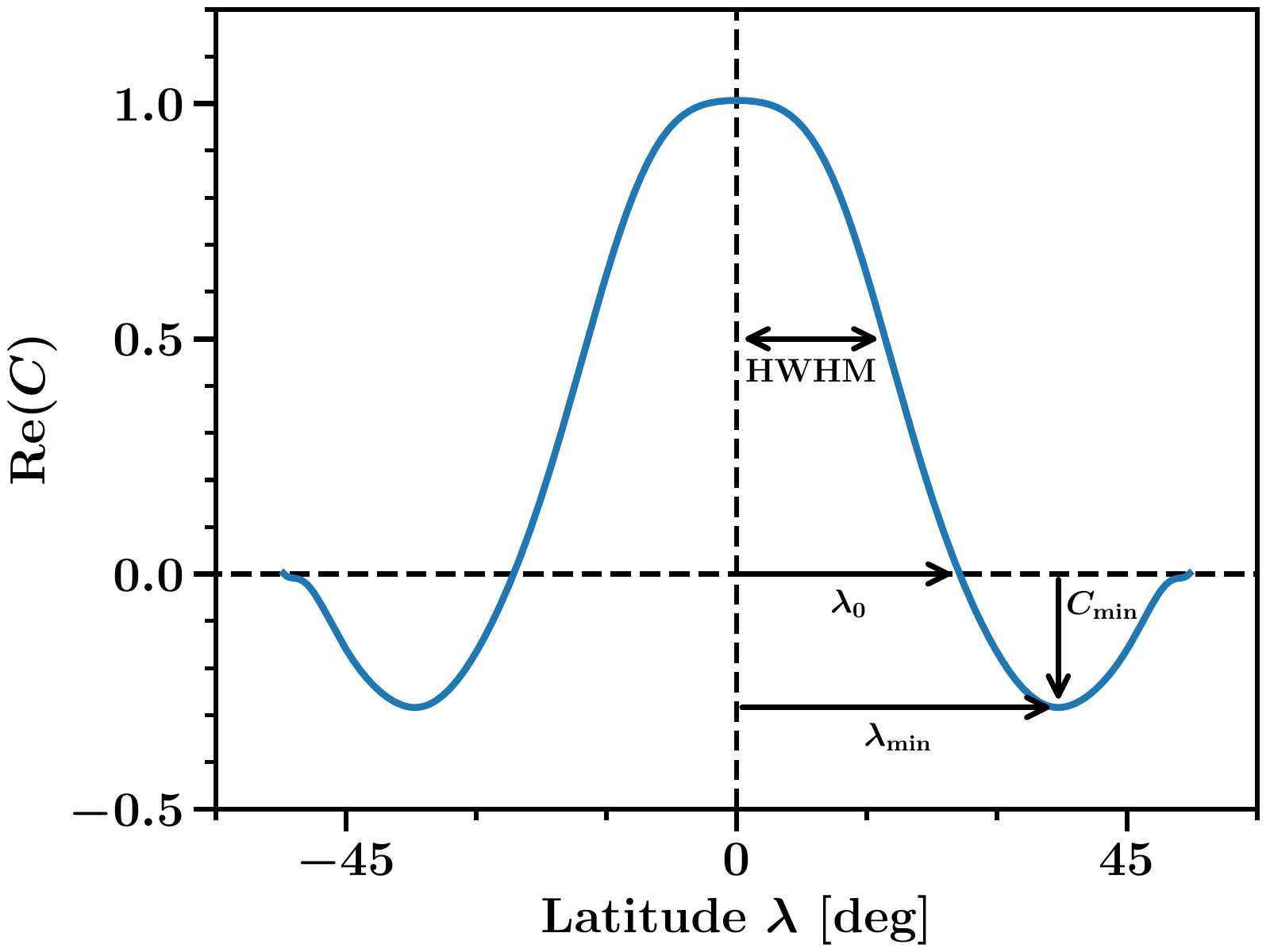}
\caption[Latitudinal eigenfunctions of Rossby waves, schematic with parameters]{
Schematic description of the real part of $C_m(\lambda)$ for a given $m$. The various parameters that describe the curve are the HWHM, the latitude at zero crossing ($\lambda_0$), the latitude at minimum ($\lambda_{\text{min}}$), and the minimum value ($C_{\text{min}}$).
}
\label{fig_horiz_eigenfunction_symm_m8}
\end{figure}

Figure~\ref{fig_horiz_eigenfunction_all_mvals_real} suggests that the real part of the eigenfunctions is more confined to low latitudes for higher values of $m$. We study the $m$-dependence of several characteristic parameters illustrated in Fig.~\ref{fig_horiz_eigenfunction_symm_m8}, i.e., the width at an eigenfunction real part of $\text{Re}(C) = 0.5$ (a half width at half maximum; HWHM), the latitude of the eigenfunction real part sign reversal, $\lambda_0$, and the latitude and value of the minimum, $\lambda_{\text{min}}$ and $C_{\text{min}}$, respectively. To reduce the noise level we derive the eigenfunctions from maps symmetrized in latitude before measuring these parameters.

To obtain the latitude widths at $\text{Re}(C) = 0.5$ and $\text{Re}(C) = 0$, we linearly fit the two points closest to these $\text{Re}(C)$ values. The latitude and value of the minimum are obtained by quadratically fitting three points around the minimum derived without fitting. We do this to avoid oscillating RDA results due to the coarse $\SI{7.5}{\degree}$ latitude sampling. For LCT, the effects of fitting the minimum (or not) are minimal. There are no results for $m = 3$ and $m = 4$ because of the poor quality and different shape of the eigenfunctions. We already stated the difficulties in characterizing these modes. As described at the beginning of Sect.~\ref{sect_latfunc} and in App.~\ref{app_error_estimation_validation}, to derive uncertainties, we compute the standard deviation over the results for different time chunks, separately for the real and the imaginary part.

\begin{table}
\caption[Latitudinal eigenfunctions of Rossby waves, measured parameters for the real part]{Parameters of the real part of $C_m(\lambda)$ for the LCT covariance-based data; see Fig.~\ref{fig_horiz_eigenfunction_symm_m8}. The parameters for $m = 3$ and $m = 4$ are not given owing to the large uncertainties.}
\label{table_params_eigenfunction}
\centering
\begin{tabular}{c c c c c}
\hline
\hline
$m$ & HWHM & $\lambda_0$ & $\lambda_{\text{min}}$ & $C_{\text{min}}$ \\
& [deg] & [deg] & [deg] & \\
\hline
5 & $20.7 \pm 2.8$ & $31.5 \pm 4.0$ & $46.5 \pm 3.6$ & $-0.38 \pm 0.17$ \\
6 & $16.1 \pm 2.8$ & $31.5 \pm 2.2$ & $44.0 \pm 3.3$ & $-0.24 \pm 0.13$ \\
7 & $18.7 \pm 0.4$ & $28.7 \pm 1.2$ & $37.0 \pm 4.2$ & $-0.17 \pm 0.07$ \\
8 & $17.1 \pm 0.8$ & $25.8 \pm 0.2$ & $36.5 \pm 1.8$ & $-0.28 \pm 0.06$ \\
9 & $16.0 \pm 1.3$ & $27.4 \pm 1.4$ & $35.2 \pm 0.4$ & $-0.17 \pm 0.02$ \\
10 & $14.7 \pm 1.1$ & $27.8 \pm 1.6$ & $34.5 \pm 2.1$ & $-0.11 \pm 0.06$ \\
11 & $14.3 \pm 1.2$ & $24.8 \pm 2.3$ & $34.9 \pm 2.4$ & $-0.25 \pm 0.09$ \\
12 & $13.8 \pm 1.4$ & $28.9 \pm 1.0$ & $39.6 \pm 1.8$ & $-0.14 \pm 0.01$ \\
13 & $11.3 \pm 2.0$ & $21.1 \pm 1.3$ & $47.1 \pm 2.2$ & $-0.31 \pm 0.02$ \\
14 & $12.0 \pm 0.8$ & $24.6 \pm 9.0$ & $35.5 \pm 5.5$ & $-0.11 \pm 0.07$ \\
15 & $14.7 \pm 2.0$ & $27.0 \pm 3.1$ & $37.7 \pm 2.0$ & $-0.26 \pm 0.04$ \\
\hline
\end{tabular}
\end{table}

Table~\ref{table_params_eigenfunction} shows how these parameters, measured for the LCT data from the covariance method, depend on $m$. Although not given in the table, we also measure the parameters for the RDA and the SVD results. We thus also discuss the $m$-dependence for those measurements; this dependence is mostly consistent with that of the LCT covariance-based parameters.

The latitude width at $\text{Re}(C) = 0.5$ indeed decreases with $m$, quasi-linearly between $m = 7$ and $m = 13$. The slope is roughly $\SI{-1}{\degree}$ per $m$. The decrease might flatten off at high $m$, but this could also be caused by noise. We observe slightly different latitude widths between the covariance and SVD eigenfunctions at low $m$ for $\text{Re}(C) = 0.5$, but similar widths at $\text{Re}(C) = 0$. Toward higher $m$, $\lambda_0$ is consistent with a flat profile, until around $m = 13$ the eigenfunction widths become smaller. The latitude of the minimum, $\lambda_{\text{min}}$, shows an $m$-(in)dependence similar to $\lambda_0$. There is a strong discrepancy for $m = 13$ between LCT and RDA, indicating that this mode is not trivial to characterize. This could be caused by noise. To some extent we could already see this in the power spectrum in Fig.~\ref{fig_power_spectrum_all_mvals}, where the $m = 13$ linewidth is large compared to all other $m$. The error on $\lambda_{\text{min}}$ might be underestimated here, since as seen in the asymmetric eigenfunctions in Fig.~\ref{fig_horiz_eigenfunction_all_mvals_real} the minimum of the LCT data is more poorly defined for $m = 13$ than for other modes. Finally, the value of the minimum, $C_{\text{min}}$, is different between the different analysis methods at $m = 5$ and $m = 6$, as seen before. Otherwise, it is quasi $m$-independent and has at most a very mild increase with $m$ from $m = 7$ onward, which is likely covered by noise, however.

As mentioned before, the latitudinal eigenfunctions appear to have two nodes (zero crossings) at latitudes $\pm\lambda_0$. This in combination with Fig.~\ref{fig_power_spectrum_2d} and the subsequent discussion indicates that the eigenfunctions have significant contributions from $\ell = m$ and $\ell = m + 2$ components. To quantify these contributions, we project the symmetric eigenfunctions $C_m(\lambda)$ onto associated Legendre polynomials $P_{\ell}^{m}(\sin\lambda)$, to obtain the coefficients
\begin{equation}
c_{\ell m} = \frac{\pi}{2 N_\lambda} \sum_{\lambda} C_m(r = R,\lambda) P_{\ell}^{m}(\sin\lambda) \cos\lambda.
\end{equation}
The sum goes over all latitudes $\lambda = k \pi/N_\lambda$ (integer $-N_\lambda/2 \leq k < N_\lambda/2$), where $N_\lambda$ is the number of data points in latitude. The $P_{\ell}^{m}(\sin\lambda)$ are normalized such that
\begin{equation}
\pi/N_\lambda \sum_\lambda (P_{\ell}^{m}(\sin\lambda))^2 \cos\lambda = 2.
\end{equation}
The associated Legendre polynomials used in the decomposition are not orthogonal over the limited observed latitude range. However, we do not expect this to be a problem since we see later in this section that the near-sectoral associated Legendre polynomials, whose amplitude is concentrated toward the equator, are the dominant contributions to the latitudinal eigenfunctions. Because of the symmetry of the eigenfunctions, only $c_{\ell m}$ with even $\ell - m \geq 0$ are nonzero. We find that the real and the imaginary parts of the eigenfunctions for almost all $m$ can be approximated well (within 1$\sigma$) when using only the contributions from $c_{\ell m}$ for $m \leq \ell \leq m + 6$, except for $m = 3$, which is very noisy. The approximation also does not work well at the high latitudes (beyond $\pm\SI{40}{\degree}$) for the real part (for some modes) and at the near-equatorial latitudes for the imaginary part (for almost all modes).

\begin{sidewaystable*}
\caption[Latitudinal eigenfunctions of Rossby waves, coefficients of associated Legendre polynomial components]{Coefficients $c_{\ell m}$ for the LCT covariance-based data. Each bracketed pair of numbers refers to the real and imaginary parts of $c_{\ell m}$. The numbers in italics are not significantly different from zero (zero within 1$\sigma$).}
\label{table_params_eigenfunc_legendre_coeffs}
\footnotesize
\centering
\begin{tabular}{c c c c c}
\hline
\hline
$m$ & $c_{m m}$ & $c_{m+2,\,m}$ & $c_{m+4,\,m}$ & $c_{m+6,\,m}$ \\
\hline
3 & $(+0.026,+0.417) \pm (0.018,0.105)$ & $(-0.155,\mathit{+0.089}) \pm (0.032,0.103)$ & $(+0.118,-0.108) \pm (0.075,0.074)$ & $(-0.117,\mathit{+0.004}) \pm (0.015,0.029)$ \\
4 & $(+0.457,+0.136) \pm (0.016,0.060)$ & $(-0.097,+0.061) \pm (0.027,0.017)$ & $(\mathit{-0.010},-0.066) \pm (0.036,0.030)$ & $(-0.071,\mathit{+0.019}) \pm (0.014,0.026)$ \\
5 & $(+0.478,\mathit{+0.037}) \pm (0.031,0.070)$ & $(-0.263,\mathit{-0.021}) \pm (0.086,0.067)$ & $(-0.059,\mathit{-0.106}) \pm (0.052,0.112)$ & $(\mathit{+0.041},\mathit{-0.011}) \pm (0.045,0.040)$ \\
6 & $(+0.411,+0.021) \pm (0.055,0.019)$ & $(-0.202,\mathit{-0.050}) \pm (0.025,0.057)$ & $(\mathit{-0.005},-0.118) \pm (0.040,0.032)$ & $(\mathit{-0.006},+0.059) \pm (0.007,0.018)$ \\
7 & $(+0.473,\mathit{+0.026}) \pm (0.024,0.028)$ & $(-0.183,-0.055) \pm (0.015,0.024)$ & $(\mathit{-0.012},-0.073) \pm (0.012,0.023)$ & $(+0.043,+0.018) \pm (0.006,0.012)$ \\
8 & $(+0.441,\mathit{-0.013}) \pm (0.015,0.022)$ & $(-0.231,-0.038) \pm (0.021,0.027)$ & $(-0.047,\mathit{-0.009}) \pm (0.031,0.068)$ & $(+0.038,\mathit{+0.001}) \pm (0.017,0.027)$ \\
9 & $(+0.434,\mathit{+0.028}) \pm (0.013,0.042)$ & $(-0.162,-0.032) \pm (0.023,0.014)$ & $(\mathit{-0.010},-0.053) \pm (0.018,0.044)$ & $(+0.025,\mathit{+0.022}) \pm (0.005,0.030)$ \\
10 & $(+0.423,\mathit{+0.008}) \pm (0.024,0.012)$ & $(-0.139,-0.057) \pm (0.017,0.041)$ & $(\mathit{+0.011},-0.076) \pm (0.034,0.028)$ & $(+0.020,-0.013) \pm (0.001,0.010)$ \\
11 & $(+0.400,-0.010) \pm (0.027,0.003)$ & $(-0.173,-0.117) \pm (0.049,0.043)$ & $(-0.043,-0.064) \pm (0.027,0.058)$ & $(\mathit{+0.002},+0.024) \pm (0.018,0.024)$ \\
12 & $(+0.419,-0.040) \pm (0.029,0.023)$ & $(-0.117,-0.078) \pm (0.006,0.011)$ & $(-0.016,\mathit{-0.021}) \pm (0.010,0.032)$ & $(\mathit{-0.027},\mathit{-0.007}) \pm (0.042,0.030)$ \\
13 & $(+0.345,\mathit{-0.016}) \pm (0.036,0.023)$ & $(-0.195,-0.112) \pm (0.018,0.020)$ & $(-0.048,-0.028) \pm (0.029,0.011)$ & $(-0.071,+0.026) \pm (0.013,0.023)$ \\
14 & $(+0.380,-0.057) \pm (0.026,0.025)$ & $(-0.128,-0.102) \pm (0.038,0.044)$ & $(\mathit{-0.007},\mathit{+0.006}) \pm (0.027,0.022)$ & $(-0.019,\mathit{+0.001}) \pm (0.015,0.017)$ \\
15 & $(+0.431,-0.074) \pm (0.017,0.013)$ & $(-0.085,-0.151) \pm (0.054,0.029)$ & $(-0.072,\mathit{-0.041}) \pm (0.020,0.057)$ & $(-0.053,\mathit{-0.012}) \pm (0.036,0.020)$ \\
\hline
\end{tabular}
\end{sidewaystable*}

Table~\ref{table_params_eigenfunc_legendre_coeffs} shows the coefficients $c_{\ell m}$ for $m \leq \ell \leq m + 6$ for the LCT covariance-based latitudinal eigenfunctions. As usual the uncertainties are calculated from the standard deviation over the coefficients for different time chunks (App.~\ref{app_error_estimation_validation}), separately for the real and the imaginary part. The real part of the eigenfunctions is clearly dominated by the $\ell = m$ component. The contribution from the $\ell = m + 2$ component is significant as well and has a relative strength of $30$-$\SI{50}{\percent}$. This is consistent with the observations from the 2D and 1D power spectra in Sect.~\ref{sect_power_spectra_rossby}. The real part of the $c_{m m}$ and $c_{m+2,\,m}$ each depend weakly on $m$. The real part of several of the coefficients with larger $\ell$ is insignificant. The imaginary part, on the other hand, has significant, dominant contributions at $\ell = m + 4$ for $m \leq 10$ and at $\ell = m + 2$ for $m \geq 11$, whereas the $\ell = m$ and $\ell = m + 6$ components are often insignificant. The term insignificant refers to an absolute value of $c_{\ell m}$ of less than 1$\sigma$. Nonetheless, independent of the estimated error bars, 11 out of 12 modes within $4 \leq m \leq 15$ have the same sign for $c_{m+4,\,m}$, suggesting that the $\ell = m + 4$ contribution to the real part is significant, despite being within $1\sigma$ from zero. A similar argument holds for the imaginary part of the latitude dependence of the Rossby wave eigenfunctions.

For the latitudinal eigenfunctions of Rossby waves there are so far only a few theoretical studies such as \citet{Lee1997} and \citet{Townsend2003}. These studies typically gave either analytic (asymptotic) expressions and/or numeric calculations, but the former expressions do not agree well with the latter calculations for Rossby waves \citep{Townsend2003}. Although both studies indicate that the latitudinal eigenfunctions are not concentrated near the equator, we cannot sensibly compare their findings to our measurements. In particular these models assume a uniform rotation rate. Also these authors used the traditional approximation, i.e., they neglected the horizontal component of the rotation vector. This approximation requires the squared Brunt-V{\"a}is{\"a}l{\"a} frequency $N^2$ to be much higher than both the squared oscillation frequency $\omega^2$ and the squared rotation rate $\Omega^2$. The validity of the traditional approximation thus has to be critically examined within the convection zone of the Sun.

\subsection{Radial eigenfunctions of Rossby waves}

\subsubsection{Depth-dependent ring-diagram systematics}

To study the Rossby wave depth dependence, we must check to which depths RDA is reliable. For this we compare the solar rotation profile from ring-diagram velocities with the results from SDO/HMI global modes from the JSOC data series hmi.V\_sht\_2drls \citep{Larson2018}. The global modes have a $72$-day time sampling from April 30, 2010 to June 4, 2017, a $\SI{1.875}{\degree}$ latitude sampling, and a nonlinear depth grid with many more points near the surface than at larger depths. Global modes are expected to give a precise and accurate solar rotation profile. We interpolate the global mode results to the ring-diagram latitude-depth grid via 2D bicubic splines, which is reasonable because the global-mode inversions do not vary on scales of their original grid; we then average the $72$-day chunks over time. The chunk scatter of the rotation rate is used to estimate the uncertainty. We divide the ring-diagram data into five intervals of length $480$ time steps ($20$ rotations), average the chunks over time and estimate the error from the scatter, convert the velocities into rotation rates, and add the sidereal Carrington rate to correct for the ring-diagram tracking.

\begin{figure}
\centering
\includegraphics[width=\hsize]{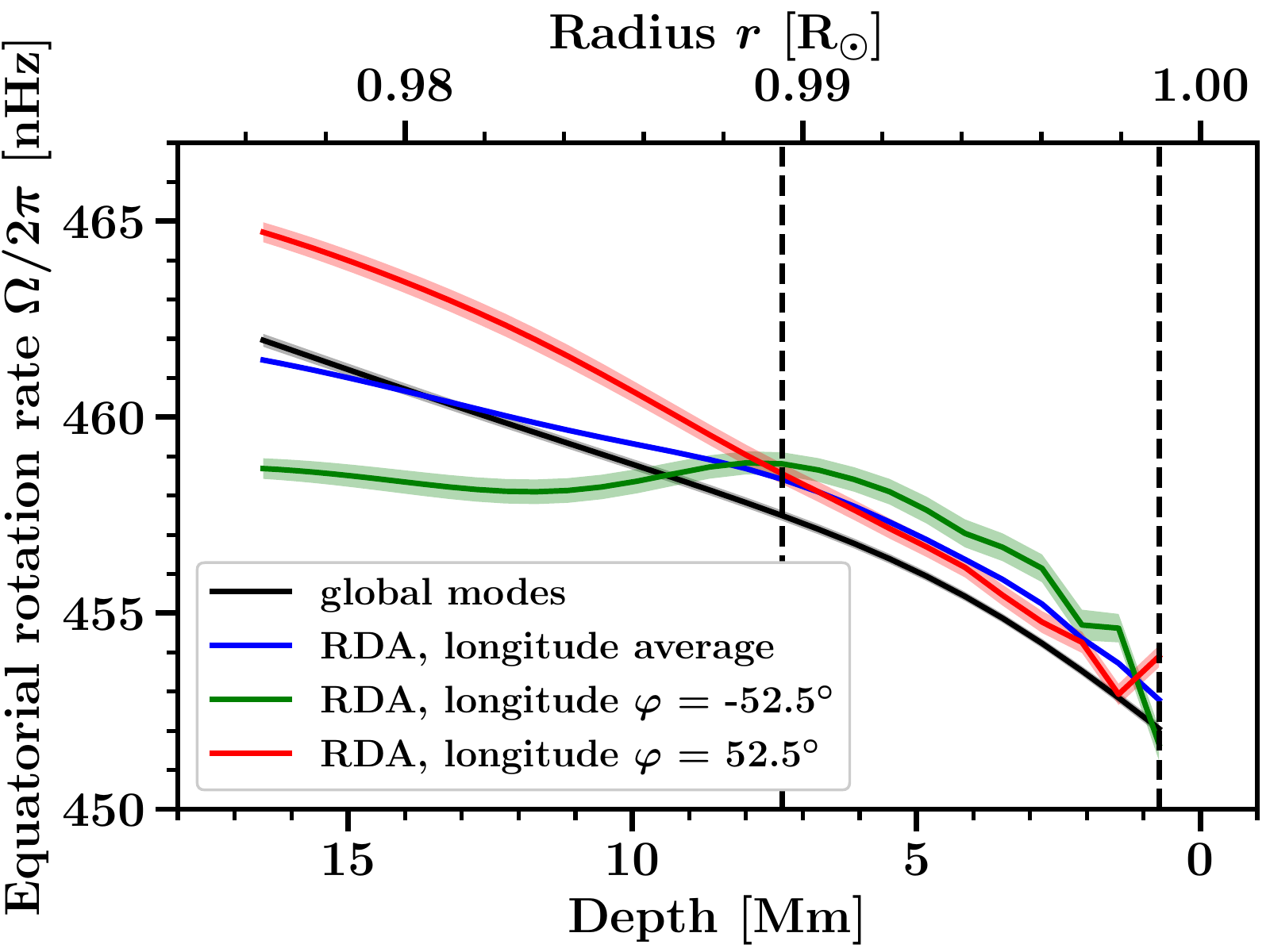}
\caption[Solar equatorial rotation, RDA versus global-mode helioseismology]{
Solar equatorial rotation rate as a function of depth. The global-mode helioseismology result is given by the black curve. The blue curve is for RDA after averaging over all longitudes. The green and red curves show the ring-diagram results at Stonyhurst longitudes $\pm\SI{52.5}{\degree}$.The shaded areas give the 1$\sigma$ error estimates. The observations cover more than seven years. The dashed black lines indicate the depth range within which the ring-diagram results are in best agreement with each other. 
}
\label{fig_rotation_systematics}
\end{figure}

Figure~\ref{fig_rotation_systematics} shows the equatorial rotation rate versus depth from global modes and ring-diagram velocities, both averaged over longitude and at Stonyhurst longitudes of $\pm\SI{52.5}{\degree}$ (the outermost longitudes in our vorticity maps). The global modes yield a smooth profile with extremely small errors. The ring-diagram data show a small offset at small depths, but, more importantly, inconsistency with the global modes at large depths. Of course, it is difficult to judge how well the results should agree because of the different kernels of the datasets and thus different depth (and latitude) sensitivities. The $\SI{-52.5}{\degree}$ longitude curve has a small local maximum around $\SI{8}{\mega\metre}$. For longitudes even further east (not shown), the rotation rate has an even stronger excess (a bump) there.

The most worrisome point is the disagreement between different ring-diagram longitudes themselves and also with the longitude average, below roughly $\SI{8}{\mega\metre}$ (indicated by the left dashed line in Fig.~\ref{fig_rotation_systematics}). Because we averaged the data over more than seven years, any short-lived flows and even longer-lived structures should be filtered away and the longitude gradient from east to west should thus not exist. This points to a deeper problem with the ring fits and the pipeline processing that generated these fits. The presence of systematic effects in HMI ring-diagram data has also been extensively discussed in \citet{Komm2015}.

Finally, we note that Fig.~\ref{fig_rotation_systematics} is affected by an issue related to the ring-diagram inversion, since the inversion does not account for the quantity $\beta_{n \ell}$. A discussion of this issue and a brief check of the magnitude of the effect is given in App.~\ref{app_rda_inversion_issues}. The latter showed that the main effect is a depth-independent underestimation of the ring-diagram velocities by $1$-$\SI{2}{\metre\per\second}$ or equivalently of the rotation rate by less than $\SI{0.5}{\nano\hertz}$. This does not affect our main conclusions. The small, but significant difference between the rotation rates from global modes and ring diagrams cannot be caused by the $\beta_{n \ell}$ issue (it has the wrong sign), but may possibly instead be due to different averaging kernel widths, systematics, or other unknown effects.

\subsubsection{Determining the Rossby wave depth dependence}

In this section, we discuss only the sectoral ($\ell = m$) component of the power spectrum of radial vorticity. The Rossby wave power $P_m(\nu,r)$ and phase $\Phi_m(\nu,r)$ thus depend on frequency, depth, and azimuthal order. Based on the assumption of damped oscillations and stochastic wave excitation, we perform a maximum-likelihood Lorentzian fit \citep{Anderson1990} to the power spectra for the longer ring-diagram period, separately at each $m$. We use the functional form
\begin{equation}
\label{eq_power_fit_lorentz}
P_{\text{fit},\,m}(\nu,r) = \frac{A_m(r)}{4(\nu - \nu_{0,\,m})^2/\gamma_m^2 + 1} + B_m(r).
\end{equation}
We fit all the depths (except for the surface, i.e., $r = \SI{0.0}{\mega\metre}$, where the ring-diagram data are unreliable) at once, with a common central frequency $\nu_{0,\,m}$ and linewidth $\gamma_m$, but with individual amplitudes $A_m(r)$ and backgrounds $B_m(r)$. The Lorentzian fit of the power spectra, in most cases, fits well to the observations. As seen in Fig.~\ref{fig_power_spectrum_all_mvals}, the $\ell = m = 6$ and $\ell = m = 13$ modes have large linewidths and their power spectra show fine structure. The $\ell = m = 3$ mode has been fit by \citetalias{Liang2019}, but not by \citetalias{Loeptien2018}.

To determine error bars for the amplitudes and backgrounds via chunked data (App.~\ref{app_error_estimation_validation}), we also perform the Lorentzian fit for each chunk separately, fitting again all depths together, but keeping the central frequency and linewidth fixed at the fit results of $\nu_{0,\,m}$ and $\gamma_m$ from Eq.~\ref{eq_power_fit_lorentz} for the full time period (to prevent unstable fits). Because we keep these parameters fixed for each chunk, we cannot derive their uncertainties based on the standard deviation over the chunks. We thus do a Monte Carlo simulation and generate $1000$ realizations of synthetic power spectra according to Eq.~\ref{eq_monte_carlo_model} (App.~\ref{app_error_estimation_validation}) and perform the Lorentzian fit for each realization analogously to the fit for the observations. The median of the parameters over the Monte Carlo realizations is consistent with the fit parameters for the observations. While the error based on the Monte Carlo simulation contains realization noise, the model we use (Lorentzian and stochastic excitation) does not include all features of the observed power spectra. The chunk-based error likely describes the physical system more accurately, by also including other variance contributions, such as from temporal effects on the Rossby waves; we could imagine, for example, solar cycle effects. This may also explain the discrepancy between the two types of errors of order $\SI{30}{\percent}$ (App.~\ref{app_error_estimation_validation}). This disagreement is, however, small enough to not affect the significance of the results for the radial eigenfunctions. We thus use the uncertainties based on the Monte Carlo simulation for the central frequency and the linewidth.

\setlength{\extrarowheight}{2pt}
\begin{table}
\caption[Measured frequencies and linewidths of solar Rossby waves from RDA]{Measured frequencies and linewidths of the Rossby waves from RDA sectoral power spectra with azimuthal orders in the range $3\leq m \leq 15$. Previous measurements (with superscript 'ref') are also listed for comparison.}
\label{table_params_lorentz_fit}
\centering
\begin{tabular}{c r@{}l r@{}l r@{}l r@{}l c}
\hline
\hline
 & \multicolumn{4}{c}{This work} & \multicolumn{5}{c}{Previous work} \\
\hline
$m$ & \multicolumn{2}{c}{$\nu_{0,\,m}$} & \multicolumn{2}{c}{$\gamma_m$} & \multicolumn{2}{c}{$\nu_{0,\,m}^{\text{ref}}$} & \multicolumn{2}{c}{$\gamma_{m}^{\text{ref}}$} & Ref. \\
& \multicolumn{2}{c}{[$\SI{}{\nano\hertz}$]} & \multicolumn{2}{c}{[$\SI{}{\nano\hertz}$]} & \multicolumn{2}{c}{[$\SI{}{\nano\hertz}$]} & \multicolumn{2}{c}{[$\SI{}{\nano\hertz}$]} & \\
\hline
$3$ & $-230$ & $^{+5}_{-4}$ & $40$ & $^{+13}_{-11}$ & $-253$ & $~\pm~2$ & $7$ & $^{+4}_{-3}$ & \citetalias{Liang2019} \\
$4$ & $-195$ & $~\pm~3$ & $16$ & $^{+7}_{-5}$ & $-194$ & $^{+5}_{-4}$ & $18$ & $^{+14}_{-7}$ & \citetalias{Loeptien2018}\\
$5$ & $-159$ & $^{+3}_{-2}$ & $12$ & $^{+6}_{-5}$ & $-157$ & $~\pm~4$ & $11$ & $^{+14}_{-6}$ & \citetalias{Loeptien2018}\\
$6$ & $-119$ & $~\pm~6$ & $84$ & $^{+22}_{-19}$ & $-129$ & $~\pm~8$ & $47$ & $^{+28}_{-16}$ & \citetalias{Loeptien2018} \\
$7$ & $-111$ & $~\pm~3$ & $20$ & $^{+7}_{-5}$ & $-112$ & $~\pm~4$ & $17$ & $^{+10}_{-7}$ & \citetalias{Loeptien2018} \\
$8$ & $-89$ & $~\pm~3$ & $19$ & $^{+7}_{-6}$ & $-90$ & $~\pm~3$ & $12$ & $^{+7}_{-5}$ & \citetalias{Loeptien2018} \\
$9$ & $-77$ & $~\pm~4$ & $40$ & $~\pm$ 11 & $-86$ & $~\pm~6$ & $37$ & $^{+21}_{-11}$ & \citetalias{Loeptien2018} \\
$10$ & $-77$ & $^{+4}_{-3}$ & $29$ & $^{+10}_{-7}$ & $-75$ & $~\pm~5$ & $28$ & $^{+12}_{-10}$ & \citetalias{Loeptien2018} \\
$11$ & $-64$ & $^{+4}_{-5}$ & $47$ & $^{+13}_{-12}$ & $-75$ & $~\pm~7$ & $43$ & $^{+23}_{-13}$ & \citetalias{Loeptien2018} \\
$12$ & $-59$ & $~\pm~4$ & $35$ & $^{+11}_{-9}$ & $-59$ & $~\pm~6$ & $42$ & $^{+20}_{-12}$ & \citetalias{Loeptien2018} \\
$13$ & $-45$ & $~\pm~6$ & $76$ & $^{+22}_{-20}$ & $-40$ & $~\pm~10$ & $71$ & $^{+38}_{-22}$ & \citetalias{Loeptien2018} \\
$14$ & $-47$ & $~\pm~5$ & $40$ & $^{+13}_{-11}$ & $-56$ & $^{+6}_{-7}$ & $36$ & $^{+20}_{-13}$ & \citetalias{Loeptien2018} \\
$15$ & $-39$ & $^{+5}_{-4}$ & $41$ & $^{+12}_{-11}$ & $-47$ & $^{+7}_{-6}$ & $40$ & $^{+21}_{-12}$ & \citetalias{Loeptien2018} \\
 \hline
\end{tabular}
\end{table}
\setlength{\extrarowheight}{0pt}

Table~\ref{table_params_lorentz_fit} compares the fit parameters from this study with the results from \citetalias{Liang2019} and \citetalias{Loeptien2018}. As in \citetalias{Liang2019} and \citetalias{Loeptien2018}, the upper and lower errors give the difference between the quantiles comprising the central $\SI{68.3}{\percent}$ (the distributions are non-Gaussian) and the fit parameters for the observations. We also calculate the uncertainties on the central frequency following \citet{Libbrecht1992} and find that those (symmetric) errors typically underestimate the Monte Carlo quantile errors by roughly $\SI{1}{\nano\hertz}$. A possible reason for this could be that we use a finite frequency fitting interval.

The fit parameters for the observations and those from \citetalias{Liang2019} and \citetalias{Loeptien2018} typically agree within 1$\sigma$ or better. The central frequencies and the linewidths for the $\ell = m = 6$ mode differ by $10$ and $\SI{37}{\nano\hertz}$, but the fit is sensitive to the fitting range. The $\ell = m = 3$ fit parameters do not agree. In \citetalias{Liang2019}, the authors, using $21$~years of data, observed that the multi-peak structure of the $\ell = m = 3$ power spectrum (Fig.~\ref{fig_power_spectrum_all_mvals}) seen in data with shorter periods collapses to a narrow single peak, which indicates that the discrepancy of the fit parameters is explained by stochastic excitation of the Rossby waves and vanishes when fitting data with a longer time period. Our errors are typically more symmetric and often smaller than those of \citetalias{Liang2019} and \citetalias{Loeptien2018}. The lower errors for the linewidth often agree better than the upper errors, indicating a tail of high values (skewness) in the \citetalias{Liang2019} and \citetalias{Loeptien2018} estimate distributions. Reasons for the differences in the error estimates may lie in the simultaneous fitting of all depths at once or in the different observation periods of our datasets and those of \citetalias{Loeptien2018}.

To determine the power depth dependence, we use the amplitude $A$ derived from the Lorentzian fit (see Eq.~\ref{eq_power_fit_lorentz}) and we define the normalized power of the signal as
\begin{equation}
\label{eq_power_normalized_lorentz}
\mathcal{P}_{\text{signal},\,m}(r) = \frac{A_m(r)}{\langle A_m(r) \rangle_r}.
\end{equation}
We thus normalize by the depth average of the amplitude of the Lorentzian. The normalized power is independent of temporal amplitude variations due to Rossby wave excitation.

\subsubsection{Results for the radial eigenfunctions}

\begin{figure}
\centering
\includegraphics[width=\hsize]{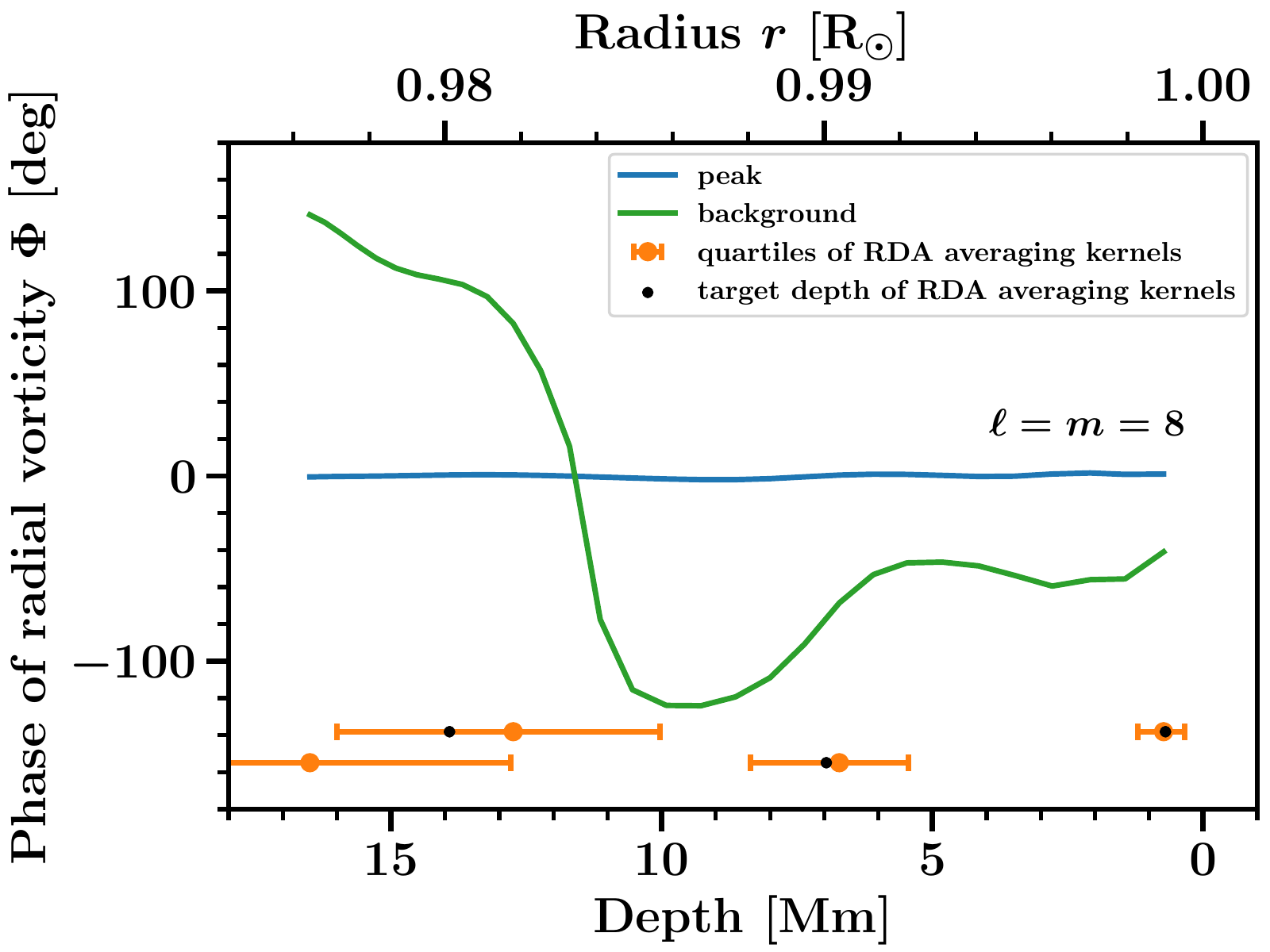}
\caption[Depth dependence of Rossby wave phase and power, for $\ell = m = 8$]{
Phase at a frequency of roughly $\SI{-87.4}{\nano\hertz}$ corresponding to the peak of power for the Rossby mode $\ell = m = 8$, as a function of depth (blue curve). The green line shows the phase of the background at the center of the background interval (see Eq.~\ref{eq_freqintvals}). The depth refers to the median of the ring-diagram averaging kernels (orange dots), which corresponds to certain target depths (black dots).
}
\label{fig_phase_depth_m8}
\end{figure}

Figure~\ref{fig_phase_depth_m8} shows the depth dependence of the $\ell = m = 8$ phase, but the behavior is similar for other $m$. For easier comparison, we remove phase jumps of $\SI{360}{\degree}$ and move the depth average to zero. The phase at the frequency of maximum power is almost constant with depth, within roughly $\pm\SI{3}{\degree}$. The phase at the background, at the center of the background interval, varies much more strongly with depth, within roughly $\pm\SI{100}{\degree}$, although the phases at other background frequencies sometimes show much less variation. The background phase is not random in depth (see App.~\ref{app_error_estimation_validation}). In particular, phase changes are gradual and smooth; the depths are correlated. This could indicate a significant contribution from scattered signal power to the background. Nonetheless, peak and background display distinctly different depth dependences. We also find that phases at different frequencies across the peak and background are different. Frequency averages of phases are thus not useful. However, as seen, for single frequencies the phase at the peak is nearly constant with depth, while the background phase varies with depth.

Figure~\ref{fig_phase_depth_m8} also shows the main parameters of the ring-diagram averaging kernels for a few target depths, i.e., the first, second (median) and third quartiles and the width (interquartile range). The flow measurements are well-localized near the surface, but smeared out over a broad depth range at large depths. The ring-diagram depth covariance matrix (not shown) indicates a similar behavior and shows that different depths are mostly independent near the surface, while at the largest depths there is high correlation and they thus do not give independent results. This could maybe also explain why the background phase is not random in depth. At large depths, the center of the averaging kernels (second quartile) moves away from the target depth, but the averaging kernels are relatively symmetric.

The background power for different $m$ (not shown) generally increases with depth and at least for some modes there could be a minimum at $8$-$\SI{9}{\mega\metre}$, albeit with little significance given the large errors.

\begin{figure*}
\centering
\includegraphics[width=\hsize]{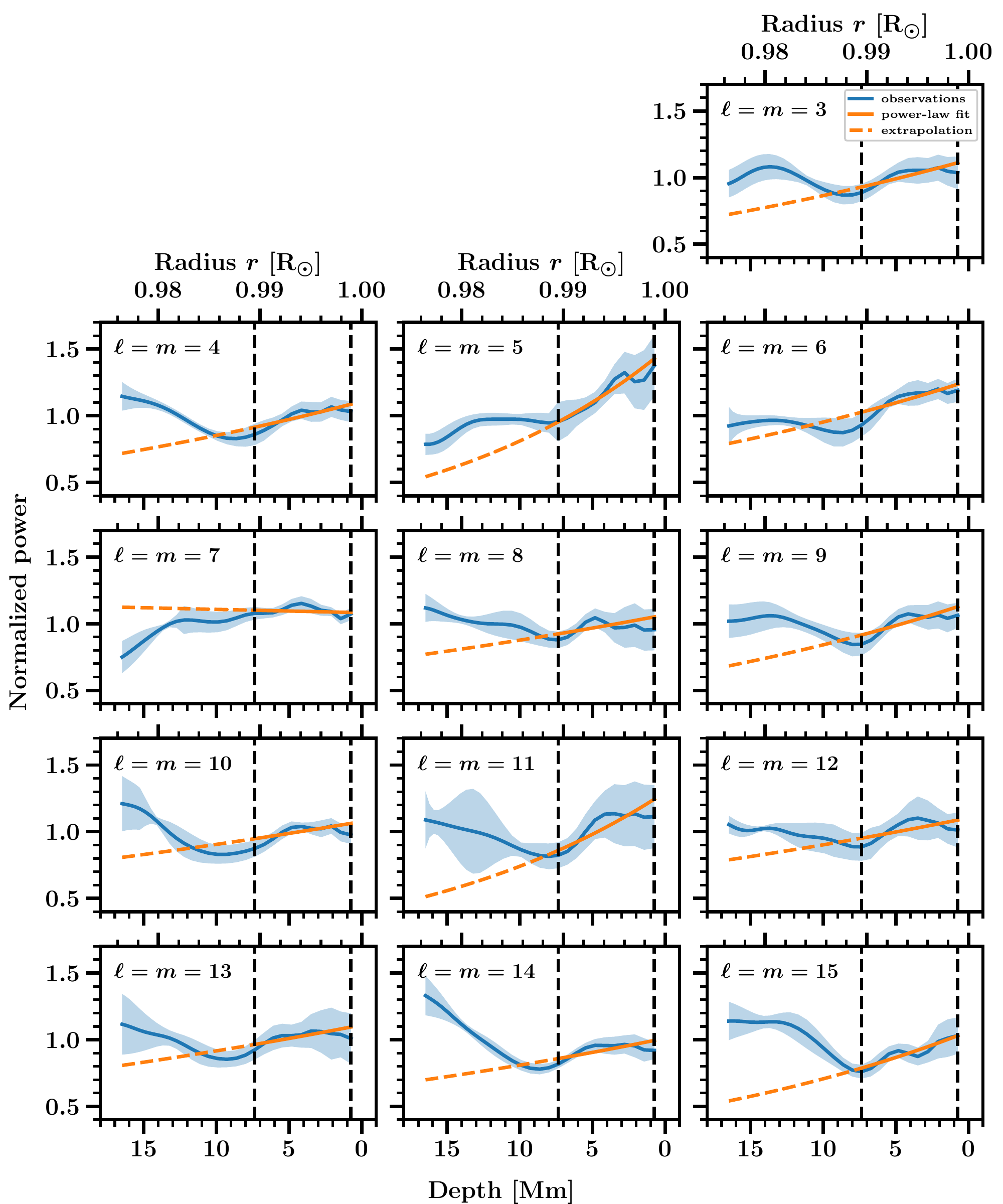}
\caption[Depth dependence of Rossby wave power, all measured $\ell = m$]{
Blue lines show the Rossby wave power $\mathcal{P}_{\text{signal}}$ as a function of depth for different values of $m$. The blue shaded areas indicate the 1$\sigma$ errors. The orange curves are fits of the form $const. \times r^{2\alpha}$ over depths between $0.7$ and $\SI{7.4}{\mega\metre}$ (between the vertical dashed lines). The orange dashed curves are extrapolations to larger depths.
}
\vspace{1cm}
\label{fig_power_peakminusbackground_depth_all_mvals}
\end{figure*}

Figure~\ref{fig_power_peakminusbackground_depth_all_mvals} shows the signal power (Eq.~\ref{eq_power_normalized_lorentz}). The quantity $\mathcal{P}_{\text{signal}}$ typically decreases from the surface toward a depth of $\SI{8}{\mega\metre}$, significantly as shown by the errors. Even further inside the Sun the power often increases again and reaches near-surface or even higher values. The 1$\sigma$ errors shown in this plot give the standard deviation, but they do not indicate $\SI{68.3}{\percent}$ probability intervals, since the power distribution is non-Gaussian (power cannot be negative). More information about error estimation can be found in App.~\ref{app_error_estimation_validation}.

\citet{Provost1981} presented a theoretical argument that the Rossby wave eigenfunctions for the horizontal displacement are proportional to $r^m$ under the assumption that the modes are incompressible. Thus, under this theory, the radial vorticity is expected to be proportional to $r^{m - 1}$. To compare this to our observations, we perform a fit of the form $const. \times r^{2\alpha}$ to $\mathcal{P}_{\text{signal}}$ within the dashed black lines ($0.7$ to $\SI{7.4}{\mega\metre}$) where the RDA is more reliable (see Fig.~\ref{fig_rotation_systematics}). We assume that the data points are uncorrelated in depth. Obviously, the fit does not reproduce the increase of power at large depths. 

\begin{figure}
\centering
\includegraphics[width=\hsize]{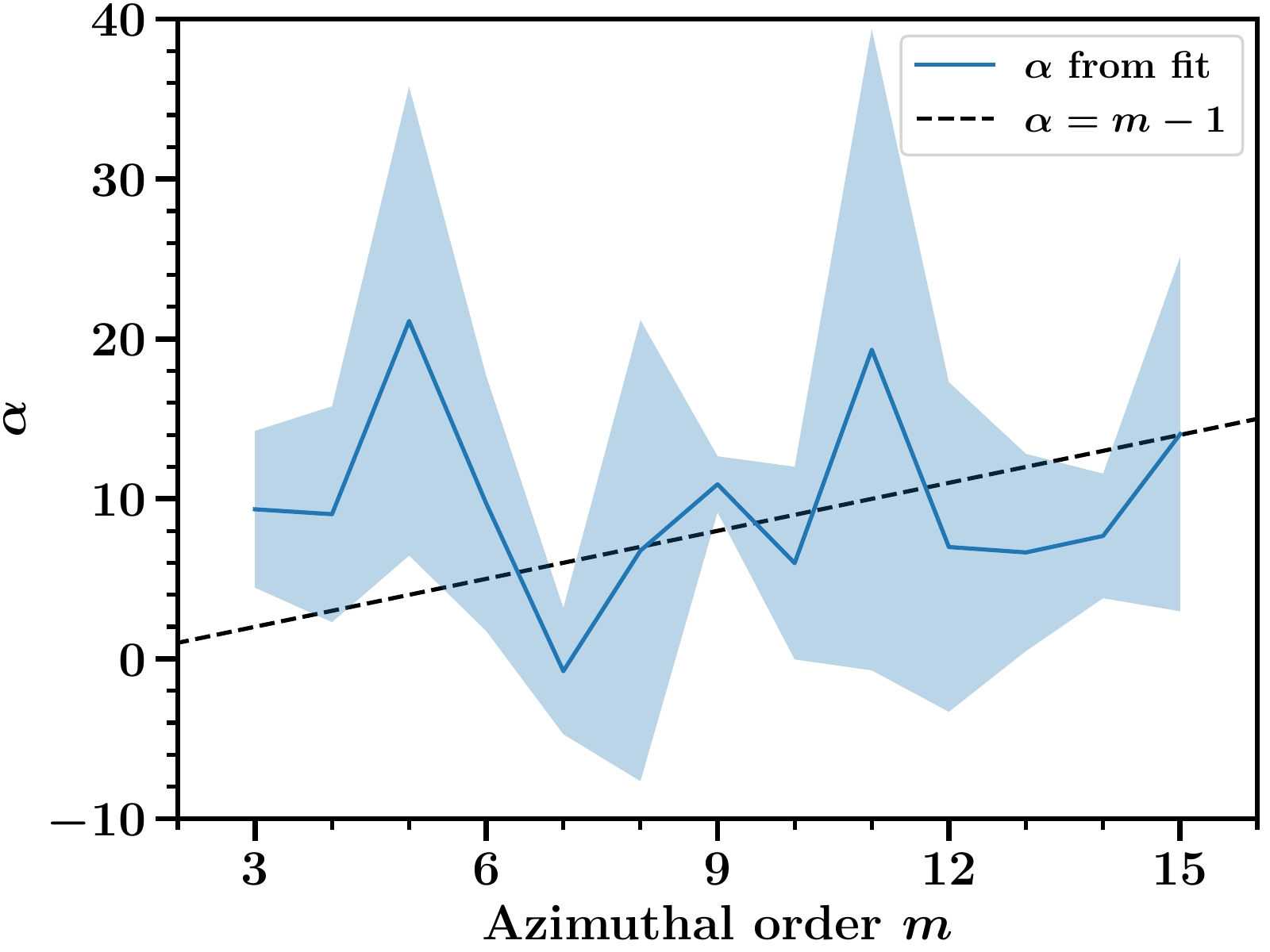}
\caption[Exponents for depth dependence of Rossby wave power]{
Exponent $\alpha$ as a function of $m$, measured in the top $\SI{7.4}{\mega\metre}$ (blue line) and 1$\sigma$ error (blue shaded area). The dashed black line corresponds to the model $\alpha = m - 1$, obtained under the assumption of non-divergent motions \citep{Provost1981}.
}
\label{fig_power_slope_fit}
\end{figure}

Figure~\ref{fig_power_slope_fit} compares the observed and theoretical exponent $\alpha$. The fitted exponent has very large error bars. It is consistent with the theoretical model from \citet{Provost1981}, but also with the absence of any trend with $m$. Although the exponent depends strongly on the fit range because of the kink at roughly $\SI{5}{\mega\metre}$ in Fig.~\ref{fig_power_peakminusbackground_depth_all_mvals}, we also do not find inconsistency with a flat dependence on $m$ within other fit intervals. Thus the current error estimates do not allow a definitive statement on the radial dependence of Rossby waves.

\section{Summary}
\label{sect_summary}

We build on \citetalias{Loeptien2018}, who investigated Rossby waves mostly using granulation tracking, by studying several Rossby wave properties via the analysis of radial vorticities computed from RDA at different depths and LCT at the surface. We obtained several new results: independently the latitudinal eigenfunctions with RDA (including a more complete, complex-valued description of the eigenfunctions), and the Rossby wave power and phase depth dependence.

We calculated latitudinal eigenfunctions of Rossby waves from the radial vorticity maps via the covariance between the equator and different latitudes and from the singular vectors of an SVD. We confirmed the shape of the real part of the eigenfunction from \citetalias{Loeptien2018}, who used the covariance method on symmetrized LCT data. We also saw consistency between covariance and SVD results, except for $m = 5$ and $m = 6$, where the SVD eigenfunctions had maxima around $\pm 10$-$\SI{15}{\degree}$ instead of at the equator. The shape of the real part of the latitudinal eigenfunctions seen for most $m$ indicates that the Rossby waves have maximum amplitudes near the equator, as found by \citetalias{Loeptien2018}. The imaginary part appears to be mostly positive for low $m$ ($3 \leq m \leq 6$); this part varies around zero for intermediate $m$ ($7 \leq m \leq 9$) and is mostly negative for high $m$ ($10 \leq m \leq 15$). A nonzero imaginary part may be due to attenuation and to the interaction of the waves with large-scale flows. In particular, the interaction of viscous Rossby waves with latitudinal differential rotation leads to the formation of critical layers at intermediate latitudes \citep[][submitted]{Gizon2020}.

We defined and measured characteristic parameters for the real part of the eigenfunctions and we found that the width at an eigenfunction value of $0.5$ (the HWHM) decreased with $m$, in contrast to the $m$-independent width at a value of $0$ and the latitude and value of the eigenfunction minimum. We also decomposed the eigenfunctions into associated Legendre polynomials and saw that the real part is dominated by $\ell = m$ and $\ell = m + 2$ contributions, while the imaginary part consists mostly of $\ell = m + 4$ and $\ell = m + 2$ contributions for low and high $m$, respectively.

We compared rotation rates from ring-diagram data and global modes as functions of depth and saw a small offset at small depths and disagreement at large depths, but most importantly inconsistency of different ring-diagram longitudes. This indicated systematic effects in the ring-diagram pipeline (see also \citealt{Komm2015}).

We studied the Rossby wave power and phase depth dependence in detail for the first time. The phase at the peak is stable with depth, in contrast to the phase of the background. The background power almost monotonically increases with depth, while the signal power decreases toward a depth of $8$-$\SI{9}{\mega\metre}$ and then increases again. The radial eigenfunctions of the Rossby waves are (at small depths) consistent with a power-law decrease, in particular both with the theoretical \citet{Provost1981} model (exponent $m - 1$) and an $m$-independent exponent. However, the \citet{Provost1981} model is based on assumptions that are not exactly correct for the Sun (e.g., uniform rotation). We can constrain the radial dependence of the eigenfunctions only very weakly owing to the high uncertainties on the observed exponents.

The analysis presented in this paper implicitly makes the assumption that the Rossby wave eigenfunctions are separable in depth and latitude. Our data show a similar latitude dependence for the different depths, separability thus appears to be a reasonable assumption. The results shown in this work motivate further research on Rossby wave eigenfunctions, which is a necessary condition for the interpretation of the measured mode frequencies.

\paragraph{Acknowledgements}
B.~Proxauf is a member of the International Max Planck Research School for Solar System Science at the University of G{\"o}ttingen; he conducted the data analysis and contributed to the interpretation of the results and to the writing of the manuscript. We thank \mbox{Z.-C.}~Liang for providing help with the fitting of the modes in frequency space and V.~B{\"o}ning for providing beta values for App.~\ref{app_rda_inversion_issues}. The HMI data are courtesy of NASA/SDO and the HMI Science Team. The data were processed at the German Data Center for SDO funded by the German Aerospace Center DLR. We acknowledge partial support from the European Research Council Synergy Grant WHOLE SUN \#810218.

\section{Appendix}

\subsection{Issues of the ring-diagram inversions}
\label{app_rda_inversion_issues}

In this appendix, we discuss two issues regarding the ring-diagram pipeline inversions. In order to obtain local velocities at a certain measurement depth $r$, the reported pipeline velocities must be multiplied by $r/R_\odot$. Additionally the pipeline inversion does not take the quantity $\beta_{n \ell}$ (see, e.g., \citet{Aerts2010}, Eq.~3.357) into account and thus the reported inversion velocities $u_x$ are slightly incorrect.

To see this, we study a simple case. For now, let us assume that the ring diagrams are not tracked. The frequency perturbation $\delta \omega_{n \ell m}$ of the mode indexed by radial order $n$, angular degree $\ell$, and azimuthal order $m$ due to a radial differential rotation rate $\Omega(r)$ is
\begin{equation}
\label{eq_freq_splitting_rotation}
\delta \omega_{n \ell m} = m \beta_{n \ell} \int_0^{R_\odot} K_{n \ell}(r) \Omega(r) dr,
\end{equation}
where $K_{n \ell}$ is the normalized rotation kernel for that mode, i.e., $\int_0^{R_\odot} K_{n \ell}(r) dr = 1$ (see, e.g., \citet{Aerts2010}, Eq.~3.358). On the other hand, ring diagrams assume that the velocity mode fits $U_{x,\,n \ell}$ are equal to a radial integral over the true velocity flow field $u_x(r)$ weighted by flow sensitivity kernels. Based on inspection of the pipeline, we think that the used HMI kernels are normalized rotation kernels $K_{n \ell}$ from Eq.~\ref{eq_freq_splitting_rotation}. Thus
\begin{equation}
\label{eq_ring_fits_kernel}
U_{x,\,n \ell} = \int_0^{R_\odot} K_{n \ell}(r) u_x(r) dr.
\end{equation}
To connect the two equations in a simple case, consider the Doppler shift of a sectoral ($\ell = m$) mode as seen by a ring diagram at the equator, i.e.,
\begin{equation}
U_{x,\,n \ell} k_x = \delta \omega_{n \ell m} = m \beta_{n \ell} \int_0^{R_\odot} K_{n \ell}(r) \Omega(r) dr.
\end{equation}
In this equation, $k_x$ is the wavenumber in the prograde direction, which is related to $m$ via $k_x = m/R_\odot$. We conclude that
\begin{equation}
\label{eq_ring_fits_kernel_beta}
U_{x,\,n \ell} = \beta_{n \ell} \int_0^{R_\odot} K_{n\ell}(r) R_\odot \Omega(r) dr.
\end{equation}
This is not consistent with Eq.~\ref{eq_ring_fits_kernel} since $\beta_{n \ell}$ is missing from Eq.~\ref{eq_ring_fits_kernel}. Additionally we see that $u_x(r)$ should be interpreted as $R_\odot \Omega(r)$ and not as the local linear velocity $r \Omega(r)$.

To see what happens if the tracking rate is not zero, we now suppose that we track at rotation rate $\Omega_{T}$. Equation~\ref{eq_ring_fits_kernel_beta} then becomes
\begin{equation}
\tilde{U}_{x,\,n \ell} = \left[ \beta_{n \ell} \int_0^{R_\odot} K_{n \ell}(r) R_\odot \Omega(r) dr \right] - R_\odot \Omega_{T},
\end{equation}
where $\tilde{U}_{x,\,n \ell}$ is the ring measurement in the rotating frame. We now define the local deviation from the tracking rate $\delta \Omega(r) = \Omega(r) - \Omega_{T}$. Then we obtain
\begin{equation}
\label{eq_ring_fits_kernel_beta_tracking}
\tilde{U}_{x,\,n \ell} = \left[ \beta_{n \ell} \int_0^{R_\odot} K_{n \ell}(r) R_\odot \delta \Omega(r) dr \right] + (\beta_{n \ell} - 1) R_\odot \Omega_{T}.
\end{equation}
The first term is the same form as in Eq.~\ref{eq_ring_fits_kernel_beta}, while the second term is an offset that depends on $n$ and $\ell$.

The conversion factor $r/R_\odot$ is multiplied onto the data before any analysis is performed for this paper; see Sect.~\ref{sect_data_methods_rda_overview}. The offset due to $\beta_{n \ell}$ depends on the set of mode ring fits, but it should be mostly time-independent, since the ring-diagram mode set does not vary much with time. Thus the time-dependent Rossby waves should not be sensitive to this effect and the only affected result in this paper should be the comparison of rotation rates in Fig.~\ref{fig_rotation_systematics}.

To estimate the effect of $\beta_{n \ell}$ on the inversion result, for a given input flow $u_x$ we generate artificial ring fits $\tilde{U}_{x,\,n \ell}$ via Eq.~\ref{eq_ring_fits_kernel_beta_tracking}, on which we run the ring-diagram inversion module to retrieve the output velocities. To compute $\tilde{U}_{x,\,n \ell}$, we assume a depth-independent flow equivalent to the tracking rate (sidereal Carrington rate), thus $\delta \Omega(r) = 0$. We thus check only the second term of Eq.~\ref{eq_ring_fits_kernel_beta_tracking} and neglect that $\beta_{n \ell}$ also appears in the first term as a scaling factor. However, the effect due to the second term should be much larger than that due to the first term, as $\Omega_T$ is much larger than $\delta\Omega(r)$ for the ring diagrams.

We use $\beta_{n \ell}$ values provided by V.~B{\"o}ning (priv. comm.). These were computed from eigenfunctions obtained from the Aarhus adiabatic oscillation package (ADIPLS; \citealt{Christensen-Dalsgaard2008,Christensen-Dalsgaard2011}). We lose roughly $\SI{25}{\percent}$ of the original ring-fit modes, as we only have $\beta_{n \ell}$ values up to frequencies of $\SI{5}{\milli\hertz}$. However, this does not critically change the mode set used during the inversion. We replace the actual pipeline ring fits with the artificial data. We leave all other data, including uncertainties on mode-fit velocities, as is and perform the inversion. The aforementioned conversion factor of $r/R_\odot$ is multiplied onto the output velocities $u_x$.

We see that the effect of $\beta_{n \ell}$ does not depend much on depth and that the retrieved $u_x$ are on the order of only $\SI{1.5}{\metre\per\second}$ (equivalent to roughly $\SI{0.1}{\percent}$ of the tracking rate). The reason for this is that the inversion gives much more weight to the high $\ell$ modes for which the uncertainties are comparatively small. These modes typically have $\beta_{n \ell}$ values around $0.999$, thus $1 - \beta_{n \ell} \sim \SI{0.1}{\percent}$. We performed this check exemplarily for a ring-diagram tile at the first time step in our dataset (May, 20, 2010) at the point ($\lambda = \SI{0}{\degree}, \varphi = \SI{0}{\degree}$). However, tests using different tiles show that this result does not depend much on time or disk position.

The effect of the pipeline inversion not accounting for $\beta_{n \ell}$ is thus an underestimation of the true velocity fields by roughly $1$-$\SI{2}{\metre\per\second}$, or equivalently approximately $\SI{0.4}{\nano\hertz}$. This difference would be visible in Fig.~\ref{fig_rotation_systematics}, but does not change our main conclusions.

\subsection{Interpolation and apodization of ring-diagram velocities}
\label{app_rda_processing_steps}

We interpolate the ring-diagram velocities separately in time and longitude (see Sect.~\ref{sect_data_methods}) with different functions, depending on the number of available data points:
\begin{alignat*}{3}
&\text{-- $\geq 4$ data points:} \quad &&\text{cubic splines} \\
&\text{-- $3$ data points:} \quad &&\text{quadratic splines} \\
&\text{-- $2$ data points:} \quad &&\text{linear splines}
\end{alignat*}
Before we interpolate the ring-diagram velocities to the surface equatorial rotation rate, we apodize these velocities with a raised cosine in angular great-circle distance $\rho$ to the point ($\lambda = \SI{0}{\degree}, \varphi = \SI{0}{\degree}$), see Sect.~\ref{sect_data_methods}), as follows:
\begin{equation}
H(\rho) =
\begin{cases}
1 & \text{if } |\rho| \leq \frac{1 - \beta}{2T}, \\
\frac{1}{2} \left\{ 1 + \cos \left[ \frac{\pi T}{\beta} \left( |\rho| - \frac{1 - \beta}{2T} \right) \right] \right\} & \text{if } \frac{1 - \beta}{2T} < |\rho| \leq \frac{1 + \beta}{2T}, \\
0 & \text{else},
\end{cases}
\end{equation}
where $\beta$ defines the steepness of the raised cosine flanks. We choose $\beta = 0.3$. The quantity $T$ defines the central position of the flanks. We choose $T$ such that zero is reached at $\rho = \SI{67.5}{\degree}$ (where there are no more valid pixels). Apodizing the ring-diagram velocities (with different $\beta$), or not, gives consistent results.

\subsection{Error estimation and error validation}
\label{app_error_estimation_validation}

\subsubsection{Error estimation via chunked data}

The uncertainties on all results are derived by dividing the time series of vorticity maps (in total $2448$ time steps, i.e., $102$ rotations, for RDA) into equal-size chunks and calculating the scatter over the results for the chunks. We find that for chunks longer than a few months (roughly six rotations), the Rossby wave signature is visible in the power spectra. We make a compromise between noise level and chunk statistics and divide the dataset into five chunks that are $480$ time steps long each ($20$ rotations).

For the latitudinal eigenfunctions, where we only study the shorter LCT period (i.e., $78$ rotations), we first used four chunks of length $19$ rotations (rotation-averaged maps) and $470$ time steps (full time resolution, for RDA), but we obtained very large errors for the SVD method for different single $m$, where single chunks gave singular vectors different from the usual eigenfunction shape. We think that the noise in our filtered maps might have been detected as the dominant term in the decomposition. We thus use a chunking with three chunks of length $26$ rotations and $625$ time steps, where all chunks have the expected first singular vectors.

\subsubsection{Error validation via Monte Carlo simulation}

We validate the chunking approach for the depth dependence via a Monte Carlo simulation. As a plausible physical model for the Rossby wave power spectrum, we assume a Lorentzian profile and a background (constant in frequency), each with a $\chi^2$-distributed random variable (stochastic excitation). In analogy to Eq.~\ref{eq_power_fit_lorentz} we generate $1000$ realizations of synthetic data for the Fourier transform of the radial vorticity $\mathcal{F}_{\text{syn}}$ (not the power $P$), at each $m$, as
\begin{equation}
\label{eq_monte_carlo_model}
\mathcal{F}_{\text{syn},\,m}(\nu,r) = \sqrt{\frac{A_m(r)}{4(\nu - \nu_{0,\,m})^2/\gamma_m^2 + 1}} \mathcal{N}_{A,\,m}(\nu) + \sqrt{B_m(r)} \mathcal{N}_{B,\,m}(\nu,r).
\end{equation}
We fix the amplitude $A_m(r)$, background $B_m(r)$, central frequency $\nu_{0,\,m}$, and full width at half maximum $\gamma_m$ via the fit parameters for the observations. Furthermore we assume that the random variable $\mathcal{N}_{A,\,m}(\nu)$ is constant with depth, i.e., the signal is fully correlated in depth, while for the background, we take a random variable $\mathcal{N}_{B,\,m}(\nu,r)$ that is uncorrelated in depth. The random variables are complex Gaussian variables (with independent real and imaginary parts) with zero mean and unit variance, independent for each frequency. We analyze the realizations in the same way as the observations.

We observe inconsistency between two Monte Carlo estimates (by roughly $\SI{30}{\percent}$): one from chunking (like the observations), averaged over the realizations, and the other from the scatter of the power over the realizations. We find that the discrepancy is due to the temporal correlation of the different chunks; the temporal correlation matrix has values around $0.1$ on the first off-diagonal. The two quantities agree when using a weighted average with weights based on the temporal covariance matrix.

Both Monte Carlo estimates disagree with the error for the observed data. This could be due to the depth correlation of the observations. We determine a (noisy) estimate of the depth covariance and correlation matrix of the observed background power from the different background frequencies and find strong correlations between different depths, even between the largest depth and the surface, with values above $0.25$ in the off-diagonal corners. We use the observed depth covariance matrix for the Fourier transform (not the power) to correlate the Gaussian background random variable of our realizations, via a Cholesky decomposition.

Correlating the background in depth noticeably improves the agreement between Monte Carlo and observed errors. However, there is still a remaining discrepancy that can be attributed to our model, which is missing some features of the observed power spectrum. We do not account for the window function nor a background decreasing with frequency present in the observations. A detailed analysis goes beyond the scope of this paper.

\subsection{Relation of covariance to linear fit}
\label{app_latitudinal_eigenfunction_methods}

The covariance method (Sect.~\ref{sect_latfunc_cov}) is conceptually equivalent to a linear fit of the vorticity at each depth $r$ and latitude $\lambda$, i.e.,
\begin{equation}
\label{eq_eigenfunc_cov_linfit}
\tilde{\zeta}_m^{'}(t,r,\lambda) = a_m(r,\lambda) f_m(t).
\end{equation}
Let us for simplicity assume that the vorticity $\tilde{\zeta}_m^{'}(t,r,\lambda)$ is real. The latitude and depth dependence in this vorticity separation ansatz is contained in the fit parameter $a_m(r,\lambda)$. Let us assume that only $\tilde{\zeta}^{'}$ is uncertain. For zero-mean quantities (such as our vorticity maps $\tilde{\zeta}^{'}$), the slope of a linear fit without intercept, i.e., $a_m(r,\lambda)$, is given as
\begin{equation}
a_m(r,\lambda) = \frac{\langle \tilde{\zeta}_m^{'}(t,r,\lambda) f_m^*(t) \rangle_t}{\langle \vert f_m(t) \vert^2 \rangle_t}.
\end{equation}
Assuming the time dependence is given by the surface equatorial vorticity time series, i.e., $f_m(t) = \tilde{\zeta}_m^{'}(t,r = R, \lambda = \SI{0}{\degree})$, in Eq.~\ref{eq_eigenfunc_cov_covnorm} we can identify $a_m(r,\lambda)$ with $C_m(r,\lambda)$. Equation~\ref{eq_eigenfunc_cov_linfit} implies that $a_m(r = R, \lambda = \SI{0}{\degree})$ is unity.

The main disadvantage of the covariance method is the assumption of a noise-free vorticity at the equator, $\tilde{\zeta}_m^{'}(t,r = R, \lambda = \SI{0}{\degree})$, required so that the time-dependence $f_m(t)$ is noise-free and the vorticity $\tilde{\zeta}_m^{'}(t,r,\lambda)$ is the only uncertain quantity of the fit.

%% file: main_convection_power.tex
\chapter{Revisiting helioseismic constraints on solar convection}
\label{chap_main_convection}

\blfootnote{This chapter reproduces the manuscript \textit{Revisiting helioseismic constraints on solar convection} by A.~C.~Birch, T.~L.~Duvall, L.~Gizon, S.~Hanasoge, B.~W.~Hindman, B.~Proxauf and K.~R.~Sreenivasan, in prep. Contributions: B.~Proxauf carried out the granulation tracking and ring-diagram analysis shown in Sect.~\ref{sec.new} and contributed to the interpretation of the results and to writing the manuscript.}

\section{Abstract}

There is substantial disagreement between the helioseismic upper limits on the amplitude of east-west subsurface convective flows from \citet{Hanasoge2012} and the inferences of the strength of these flows by \citet{Greer2015}. In addition, the upper limit obtained by \citet{Hanasoge2012} disagrees with simulations of solar convection \citep{Miesch2008}. Additional observational and theoretical work on the topic of solar subsurface convection is crucial. Motivated by the need to establish a clear baseline for future work, we describe and remedy several inconsistencies in the figures shown in \citet{Hanasoge2012}, \citet{Gizon2012}, \citet{Greer2015}, and \citet{Hanasoge2016}. After these corrections, all of the previous disagreements remain. To provide a larger context for these measurements, we provide new estimates of the strength of surface convection from correlation tracking of SDO/HMI continuum images and of subsurface convection from the SDO/HMI ring-diagram pipeline. These new measurements are both above, but qualitatively similar to, the upper limit on convection from \citet{Hanasoge2012}, but well below the inferences of \citet{Greer2015}.

\section{Introduction}

Solar subsurface convection is a crucial ingredient in solar global-scale dynamics as it determines the Reynolds stresses that maintain differential rotation \citep[e.g.][]{Miesch2005}. Subsurface convection may also play a key role in the dynamics of the subsurface magnetic flux concentrations that emerge through the solar surface and form active regions \citep[e.g.][]{Cheung2014}.

\citet[hereafter \shravan]{Hanasoge2012} used local helioseismology \citep[see][for a review]{Gizon2010} to obtain an upper limit on the strength of horizontal convective flows in the depth range of $20$-$30$~Mm below the photosphere. This result suggested that simulations of solar convection \citep[e.g.][hereafter \miesch]{Miesch2008} overestimate the strength of convective flows in this depth range by roughly two orders of magnitude at large scales. Using three-dimensional inversions of helioseismic measurements \citet[hereafter \greer]{Greer2015}, however, found that the amplitudes of the flows in this depth range are (order-of-magnitude) compatible with the simulations. The stark differences between \shravan and \greer measurements and also between the result of \shravan and simulations call for a detailed look at the current state of understanding of solar subsurface convection \citep[see also][]{Gizon2012,Hanasoge2016}.

Here, we revisit the existing comparisons of various estimates of, or limits on, flow amplitudes. In the course of this work we have found a number of ``apples and oranges'' comparisons in the literature (in particular \shravan; \citealt{Gizon2012}, hereafter \gizon; \greer; and \citealt{Hanasoge2016}, hereafter \arfm) as well as a few cases where the analysis that was performed was not correctly described in the publication. We build on these previous results and present consistent comparisons of inferences of the amplitudes of solar convection. In addition, we provide new measurements of the strength of convection at the surface from the correlation tracking of granulation \citep{Loeptien2017} and below the surface using ring-diagram analysis \citep{Hill1988} from the SDO/HMI ring-diagram pipeline \citep{Bogart2011a, Bogart2011b}. Our overarching goal is to a provide a clearly described baseline for further investigations.

\section{Revisiting previous work}
\label{sec.old_work}

In this section we review the datasets and methods that led to the comparison figures of \shravan, \gizon, and \greer. In addition we describe a problem introduced by the journal in Fig.~5 of \arfm. As described in the introduction, the goal of this work is to provide a baseline for future work. We emphasize that our current aim is {\it not} to test the assumptions that led to the conclusions of these studies.

\subsection{HDS2012}
\label{sec.pnas}

Here we elaborate on the descriptions from \shravan for a few steps that were not clearly described in the original publication. The maps of east-west travel-time differences were generated using a phase-speed filter and deep-focusing geometry as described in \shravan. The spherical harmonic power spectrum $\tilde{P}_{\ell,m}$ (the tilde is to show that the \shravan normalization of the SHT is used; see Eq.~\ref{eq.shravan_power}; $\ell$ is the angular degree and $m$ is the azimuthal order) was computed from the Carrington map of east-west travel-time differences. The power spectrum was divided by $0.27162$, with the aim of correcting for the effect of missing data in the Carrington map (latitudes above about $\pm{}58.3^\circ$ and the missing data in longitude $\phi$\footnote{In contrast to Chap.~\ref{chap_main_rossby}, here we denote the longitude with $\phi$ instead of $\varphi$. In Chap.~\ref{chap_discussion}, both symbols are used. This ambiguity arises from different conventions in the papers forming the basis of these chapters.} due to using only $17$~days of data instead of a full rotation of about $27$~days). This factor should have been
\begin{equation}
\alpha = \frac{1}{4\pi} \int_0^{2\pi}\!\!\!\int_0^\pi w(\theta,\phi) \sin{\theta} \rmd\theta \rmd\phi = 0.529\; ,
\end{equation}
where $\theta$ is co-latitude, $\phi$ is longitude, and $w(\theta,\phi)$ is the window function ($w = 1$ where data is available and $w = 0$ otherwise). The factor of $\sin{\theta}$ was neglected in the original calculation.

The rms travel time {\it per mode} was then estimated from the $m$-averaged power spectrum as $\delta \tau_\ell = \sqrt{\langle \tilde{P}_{\ell,m} \rangle_{m > 0} / \alpha}$; the angle brackets $\langle \cdot \rangle$ denote the average. The $m = 0$ modes are not included in this calculation (this removes the contribution of the differential rotation and torsional oscillations). The rms travel time per mode is then converted to a rms flow per mode $v_\ell$ using:
\begin{equation}
\label{eq.calibration}
v_\ell = \delta \tau_\ell / \left[ c_\ell D N/S \right] \; ,
\end{equation}
where $c_\ell$ denotes the calibration curve (see Fig.~\ref{fig.calib}), $D$ is the ratio of the assumed radial extent of the flow field to the radial extent of the test functions used to compute $c_\ell$, and the factor $N/S$ corrects for the contribution of noise to the $\delta \tau_\ell$ (see Sect.~\ref{sec.fitting}). For the target depth $r = 0.96~R_\odot$, \shravan used $D = 9.64$ based on the assumption that the flows cover a radial extent given by the mixing length, assumed to be given by $1.8$ pressure scale heights. For the target depth $r = 0.96~R_\odot$, \shravan used $N/S = 2$. Section~\ref{sec.fitting} shows that re-analysis of the data (using the original method) suggests that this factor should have been $N/S = 4.7$.

\subsection{Figure~5 of \shravan}

Figure~5 of \shravan is not consistent as it compares the upper limit rms {\it per mode} from helioseismology with the rms {\it per multiplet} from \miesch. In addition, as discussed in Sect.~\ref{sec.pnas}, the area correction and $N/S$ correction factors are not as described in the text.

\subsection{Figure~1 of \gizon}
\label{sec.gizon}

Figure~1 of \gizon contains several errors: (1)~The calculation of the $E_\phi$ (Sect.~\ref{sec.sht_pnas}) for the helioseismology upper limit of \shravan was based on the misunderstanding that the upper limit was an rms {\it per multiplet} rather than an upper limit {\it per mode}. In addition a factor of $1/4\pi$ was missing due to different conventions for the spherical harmonics transform (see Sect.~\ref{sec.sht_pnas}). (2)~The calculation of the $E_\phi$ for the \citet[hereafter \roudier]{Roudier2012} correlation tracking is missing a factor of $1/2$.

\subsection{Figure~5 of \greer}
\label{sec.greer}

Figure~5 of \greer is not consistent as it compares the rms {\it per multiplet} from the ASH simulations, with the rms {\it per bin} in horizontal wavenumber from the ring-diagram inversions. The bin size is about $9.2$ in angular degree. Thus the helioseismic measurements should have been scaled down a factor of $\sqrt{9.2}$ for comparison with the simulations. In addition, the upper limit from \shravan is rms {\it per mode} and should have been converted to rms {\it per multiplet} for the sake of comparison.

\subsection{Figure~5 of \arfm}

Figure~5 of \arfm is based on Fig.~1 of \gizon and contains the same issues. In addition, the $y$-axis labels are not correct. The $y$-axis in the figure provided to the journal extends from $10^{-3}$ to $3 \times 10^{3}$.

\subsection{Summary of the revisions}
\label{sec.revised_figures}

\begin{figure}
\includegraphics[width=\linewidth]{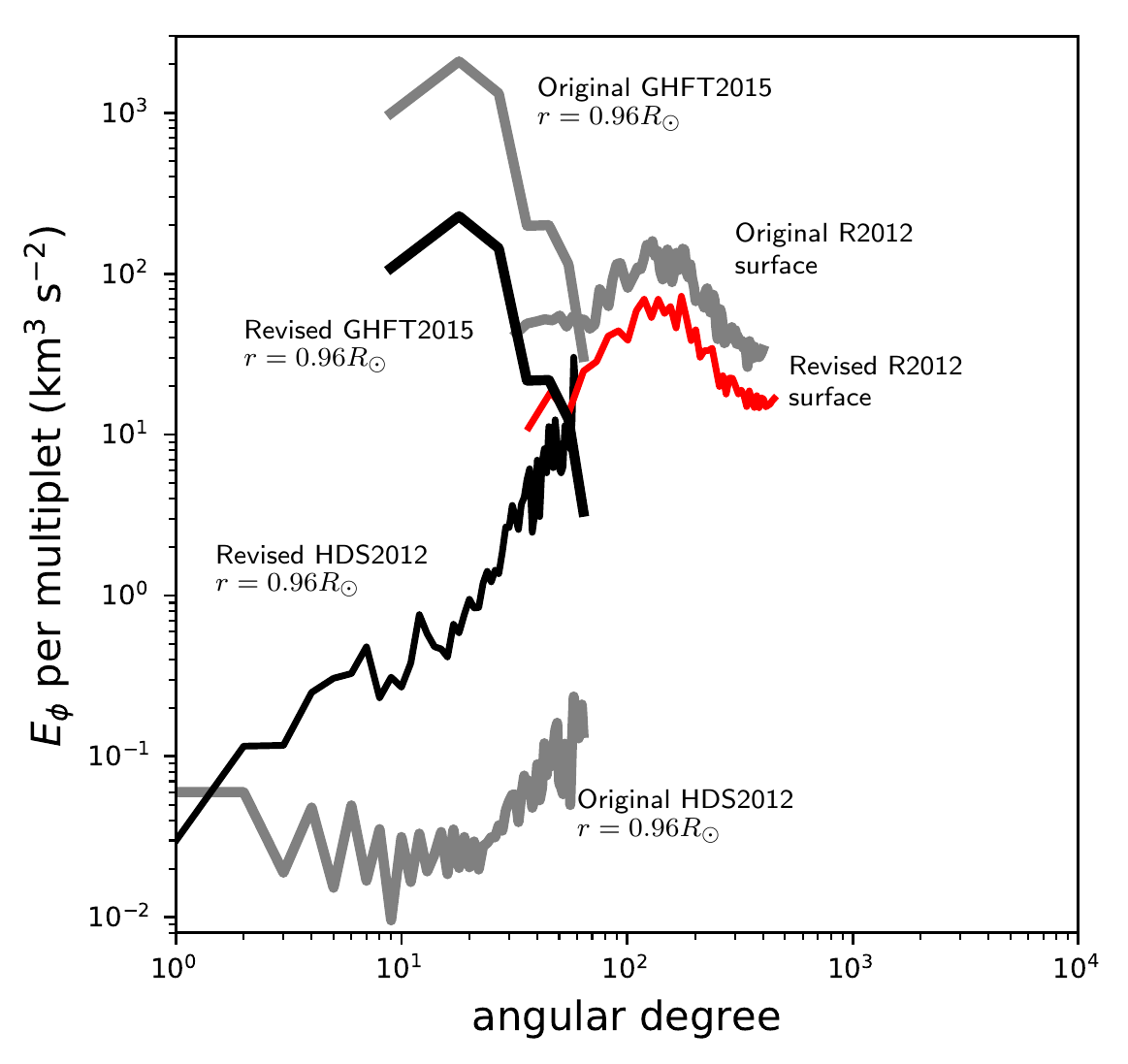}
\caption[Energy spectra of prograde flows, revised and original estimates]{
Revised (black or red) and original (grey) estimates of $E_\phi(\ell)$ for the \shravan, \greer, and \roudier results. In this work we have not found any issues with the original (\gizon) calculations for the stagger or ASH results and so no additional curves are plotted for these two cases. The discrepancy between the upper limit of \shravan and the simulations and the ring-diagram result (\greer) remains for $\ell \lesssim 50$ but is reduced in amplitude. The small deviation from a pure scaling between the original and revised \roudier curve is due to a change in the spatial apodization.
}
\label{fig.E_of_k_full}
\end{figure}

Figure~\ref{fig.E_of_k_full} shows the original and revised versions of \shravan, \greer, and \roudier. As discussed above, the corrections are multiplicative factors for \greer and \roudier. The $E_\phi$ for \greer is reduced by a factor of the bin size in angular degree (about $9.2$) and the \roudier curve is corrected downward by a factor of two. The situation is more complicated for the \shravan curve. The change in the slope of the curve comes from the correction from computing the $m$-averaged power to the $m$-summed power. This multiplies the original curve by a factor of $2\ell + 1$. The other corrections (new $N/S$ correction and new area correction) are scalings.

\section{New measurements}
\label{sec.new}

To establish a larger context for the flow amplitudes shown in the previous section, we here use two additional existing measurements of horizontal flows to estimate the spectrum of convective velocities. \citet{Loeptien2017} measured horizontal flows at the solar surface by tracking the granulation pattern seen in the SDO/HMI continuum images. The SDO/HMI ring-diagram pipeline \citep{Bogart2011a,Bogart2011b} applies ring-diagram analysis to Doppler images to infer horizontal flows in the depth range from just below the surface down to about $16$~Mm.

\subsection{Surface flows from granulation tracking}
\label{sec.new_surface_flows}

The correlation tracking described by \citet{Loeptien2017} is based on applying the FLCT code \citep{Welsch2004, Fisher2008} to pairs of SDO/HMI continuum images separated by $45$~s. A flow map with a pixel size of $0.4^\circ$ (about $4.8$~Mm at disk center) was computed every $30$~min. Working in Stonyhurst coordinates, large-scale systematic effects with periods of one day and one year (and harmonics) and steady flows were removed from a time-series of maps using an expansion in Zernike polynomials followed by Fourier filtering in time \citep[see][for details]{Loeptien2017}. The resulting $30$~min cadence flow maps were used by \citet{Loeptien2017} to measure the inflows into active regions. These maps have also been used to study equatorial Rossby waves \citep{Loeptien2018, Proxauf2020}.

There is a wide range of possible approaches to estimating the spectrum of east-west velocities from these maps. Here we explore three options:
\begin{itemize}
\item Construct Carrington maps of the east-west velocity by concatenating strips that cover $15.12^\circ$ in longitude, each strip is a $27.5$-hr average. This is essentially the approach of \shravan but with slightly wider strips and slightly longer averaging in time.
\item Extract a $60^\circ \times 60^\circ$ patch near disk center from a $27.5$-hr time average of east-west velocities. This is the approach of \greer but with a slightly different time averaging (\greer used $25.6$~hr).
\item Extract a $650" \times 650"$ patch near disk center ($\sim 40^\circ \times 40^\circ$) from a $2$-hr average map. This is the approach of \roudier.
\end{itemize}
For the ring diagrams, the patch widths refer to latitude or longitude ranges measured between the centers of the ring diagram tiles. For the local correlation tracking they are measured from pixel centers. The final step in all cases was to compute the SHT and then the $m$-summed power. In all cases, to reduce noise we averaged the resulting power spectra over the $25\%$ of Carrington rotations with the smallest sunspot numbers.

Over the range of $\ell$ covered by all three methods, the \shravan and \greer averaging schemes give similar results for $E_\phi$. The power measured by the \roudier averaging scheme is somewhat is higher because of the shorter averaging time ($2$~hr vs. about one day). From here on, we choose to only show the result from the \shravan-like analysis as this averaging provides full Carrington maps (and thus full resolution in $\ell$).

\subsection{Subsurface flows from the SDO/HMI ring-diagram pipeline}
\label{sec.new_ring_flows}

The SDO/HMI ring-diagram pipeline produces maps of horizontal flows in the depth range from the surface down to about $16$~Mm with an average cadence of about $27$~hr. Each ring tile covers a patch of about $180$~Mm~$\times$~$180$~Mm on the solar surface and this is expected to set the horizontal resolution of the inferred flows \citep{Birch2007}. The flow maps produced by the pipeline oversample this resolution by a factor of two. The depth dependence of the flow in each tile is obtained by a one-dimensional inversion of the ring-fit parameters for that tile \citep{Basu1999a, Basu1999b}. This approach of tile-by-tile one-dimensional depth inversions is very different than the three-dimensional (horizontal and vertical) inversion of many highly-overlapped tiles used by \greer.

The pipeline ring-diagram data undergo a sequence of post-processing steps. \citet{Proxauf2020} describe these steps in detail. Here we give only a brief summary. The inverted ring-diagram flows from the pipeline are first multiplied by a scaling factor $r/R_\odot$. This is used to convert the surface-equivalent velocities $\Omega(r) R_\odot$ to linear velocities $\Omega(r) r$. We then subtract fits of sinusoids with a period of one year and of time-independent flows (differential rotation) at each Stonyhurst coordinate in order to reduce the systematic effects related to the orbit of SDO. Next, we interpolate the data onto a regular longitude grid, since the ring-diagram data originally have different longitude grids at different latitudes. Finally, we apply either of the approaches indicated in Sect.~\ref{sec.new_surface_flows} to extract regions or to compute maps. We subsequently compute the SHT coefficients for the extracted data and obtain the $m$-summed power.

We then use the east-west velocities obtained in this method to compute $E_\phi$ using the \shravan method of constructing Carrington maps described above, except with a strip width of $15^\circ$ instead of $13^\circ$ and $27.2753$-hr instead of $1$-day time-averaging. We then compute $E_\phi$ averaged over the $25\%$ of the Carrington rotations with the lowest activity.

\subsection{Summary of the new results}
\label{sec.results}

\begin{figure}
\includegraphics[width=\linewidth]{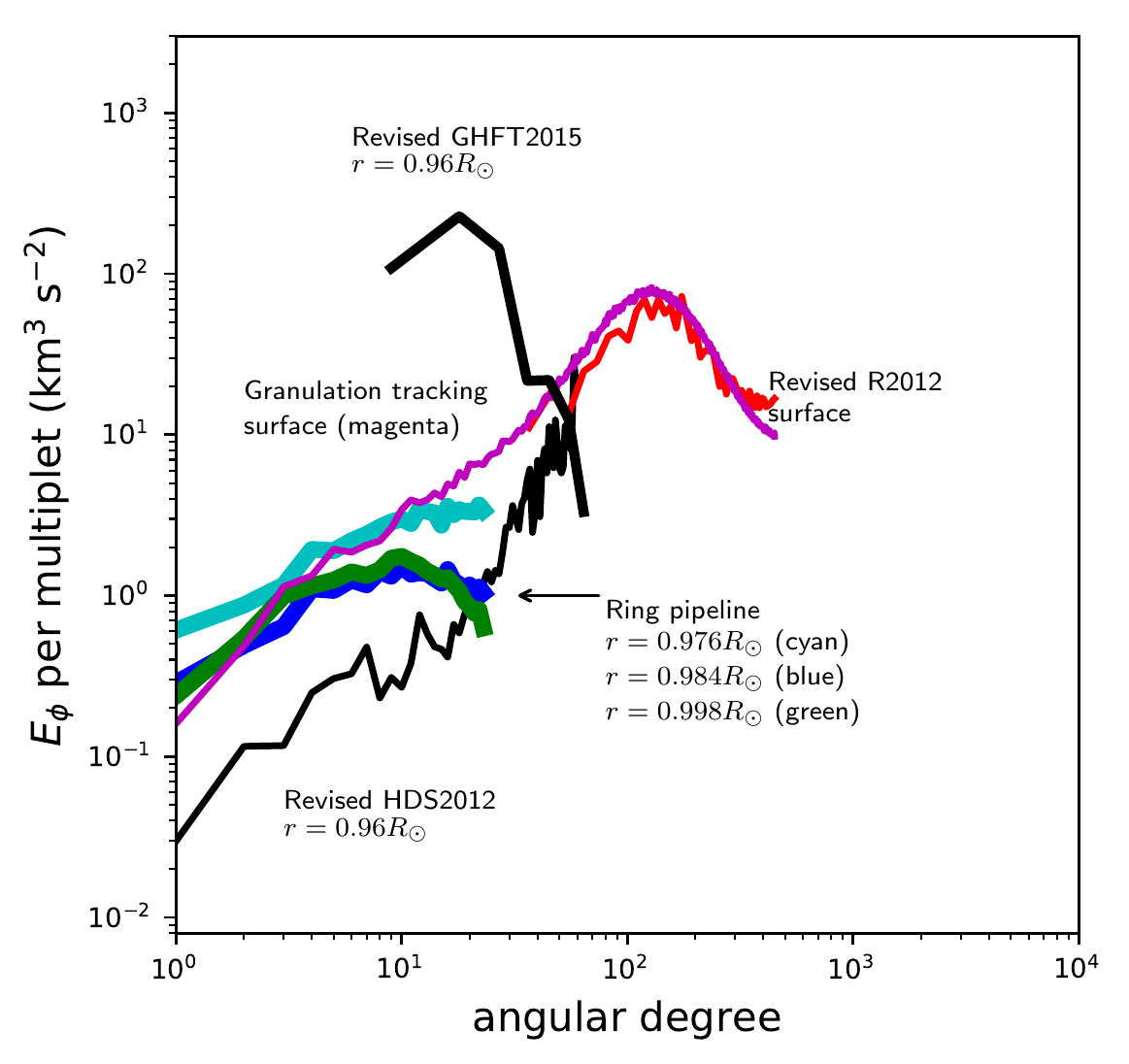}
\caption[Energy spectra of prograde flows, new estimates from LCT and RDA]{
New measurements of $E_\phi(\ell)$ from granulation tracking (magenta) and from the SDO/HMI ring-diagram pipeline ($1.1$~Mm, green; $11.1$~Mm, blue; $16.5$~Mm depth, cyan) along with the revised estimates from Sect.~\ref{sec.old_work} shown in Fig.~\ref{fig.E_of_k_full}.
}
\label{fig.E_of_k_new}
\end{figure}

Figure~\ref{fig.E_of_k_new} shows the $E_\phi$ of surface flows obtained from granulation tracking as described in Sect.~\ref{sec.new_surface_flows}. The formal error estimates (estimated from the scatter between Carrington rotations) are not shown as they are small compared to the differences between different methods.

At low $\ell$ the spectrum of surface flows is larger than the upper limit of \shravan for subsurface flows, but the shape of the curve is similar, although with a slightly lower slope. At the scale of supergranulation ($\ell \sim 120$), the spectrum of the surface flows measured here is similar to the spectrum from \roudier.

Figure~\ref{fig.E_of_k_new} also shows the average $E_\phi$ resulting from the SHT of the ring-pipeline Carrington maps (Sect.~\ref{sec.new_ring_flows}) for three different depths. The estimated errors (estimated from the scatter between Carrington rotations) are similar to the width of the line. The ring-diagram spectra $E_\phi$ extend up to $\ell = 23$, this is due to the spatial sampling of $7.5^\circ$. The tile size is $15^\circ$; the associated smoothing becomes more important with increasing $\ell$. This may be the reason for the flattening and downturn of the curves above $\ell \approx 10$.

\section{Conclusions and discussion}
\label{sec.conclusions}

\begin{figure}
\includegraphics[width=\linewidth]{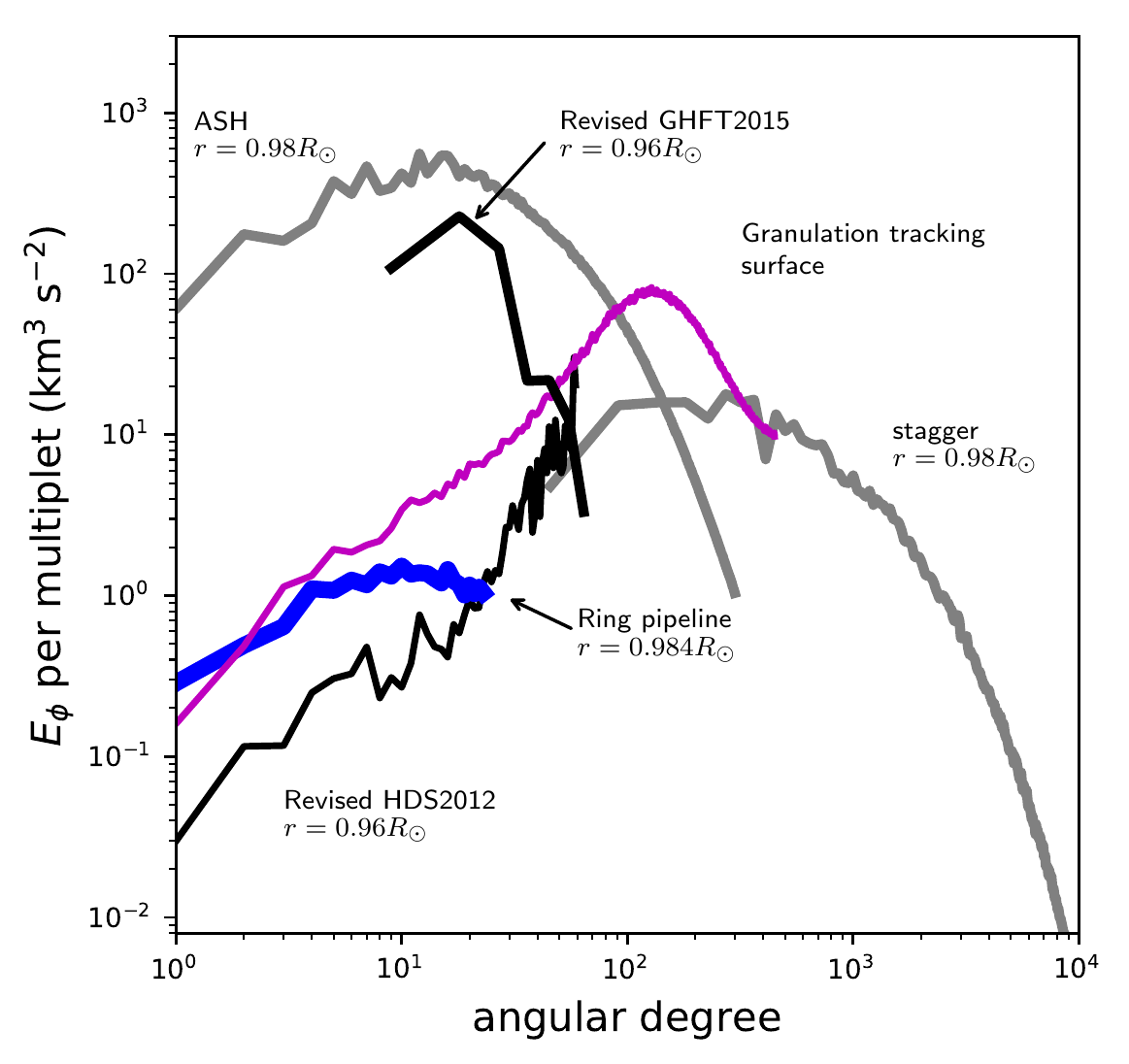}
\caption[Energy spectra of prograde flows, summary of the estimates]{
Summary of the estimates of $E_\phi$. As in Fig.~\ref{fig.E_of_k_new} the black curves show the revised estimates from \shravan and \greer. The figure also shows the new granulation-tracking measurements (magenta) and new measurements from the ring-diagram pipeline (blue). The curves corresponding to the ASH and stagger simulations as described in \gizon are shown in grey.
}
\label{fig.E_of_k_last}
\end{figure}

Figure~\ref{fig.E_of_k_last} summarizes our current understanding of the spectrum of surface and subsurface east-west velocities from observations and simulations. We removed the curve for \roudier for the sake of simplicity; this curve is compatible (within a factor of two) with the new surface measurements.

The two ring-diagram estimates of $E_\phi$ shown in Fig.~\ref{fig.E_of_k_last} are not compatible within the error estimates (see Fig.~5 from \greer; the error estimate on the rms $v_\phi$ at $\ell = 10$ is less than $10\%$). Preliminary work \citep{Nagashima2020} shows that the ring-fit parameter estimates from local power spectra by the ring-pipeline model {\tt rdfitf} \citep{Haber2000} and the multi-ridge fitting of \greer are not different enough to explain the factor of roughly one hundred difference between the respective estimates of $E_\phi$. We speculate the important difference between these two methods is the one-dimensional (depth only) tile-by-tile inversion employed by the pipeline and the three-dimensional (depth and horizontal directions) inversion used by \greer. A comparison of the two approaches in a setting where the correct answer is known is needed to resolve this difference.

The new ring-diagram estimates shown here are roughly a factor of ten above the upper limit from \shravan at the lowest few angular degrees. We speculate that this may be due to the treatment of noise: the new ring-diagram estimates do not include any attempt to remove the contribution of noise to the measured $E_\phi$; this is an important next step. In this sense, the ring-diagram measurements shown here are also upper limits.

The new ring-diagram estimates and the \shravan upper limits are both well below the spectrum of convection predicted by the simulations.

\paragraph{Acknowledgements}
This work used the SHTns library \citep{Schaeffer2013}, NumPy \citep{Oliphant2006} and matplotlib \citep{Hunter2007}. B.~Proxauf carried out the granulation tracking and ring-diagram analysis shown in Sect.~\ref{sec.new} and contributed to the interpretation of the results and to writing the manuscript. The SDO/HMI observations used here are courtesy of NASA/SDO and the HMI science team. We acknowledge (partial) support from ERC Synergy Grant WHOLE SUN \#810218.

\section{Appendix}

\subsection{SHT conventions}
\label{sec.sht}
\input{texfiles/main_convection_power/sht}
\subsection{Fourier conventions}
\input{texfiles/main_convection_power/fourier}

\subsection{Reproducing the \texorpdfstring{$S/N$}{S/N} fit from \shravan}
\label{sec.fitting}

\shravan used time-distance helioseismology to compute maps of east-west travel-time differences for consecutive time intervals of $0.1$~days. Let us denote these maps $\delta \tau_i(\theta,\phi)$. From these maps, travel-time maps corresponding to other time intervals were constructed by averaging these original maps. The variance $\sigma^2(T)$ of the resulting maps was then computed for $T = 0.1, 0.2, 0.3, 0.4, 0.5, 1$~days.

\begin{figure}
\includegraphics[width=\linewidth]{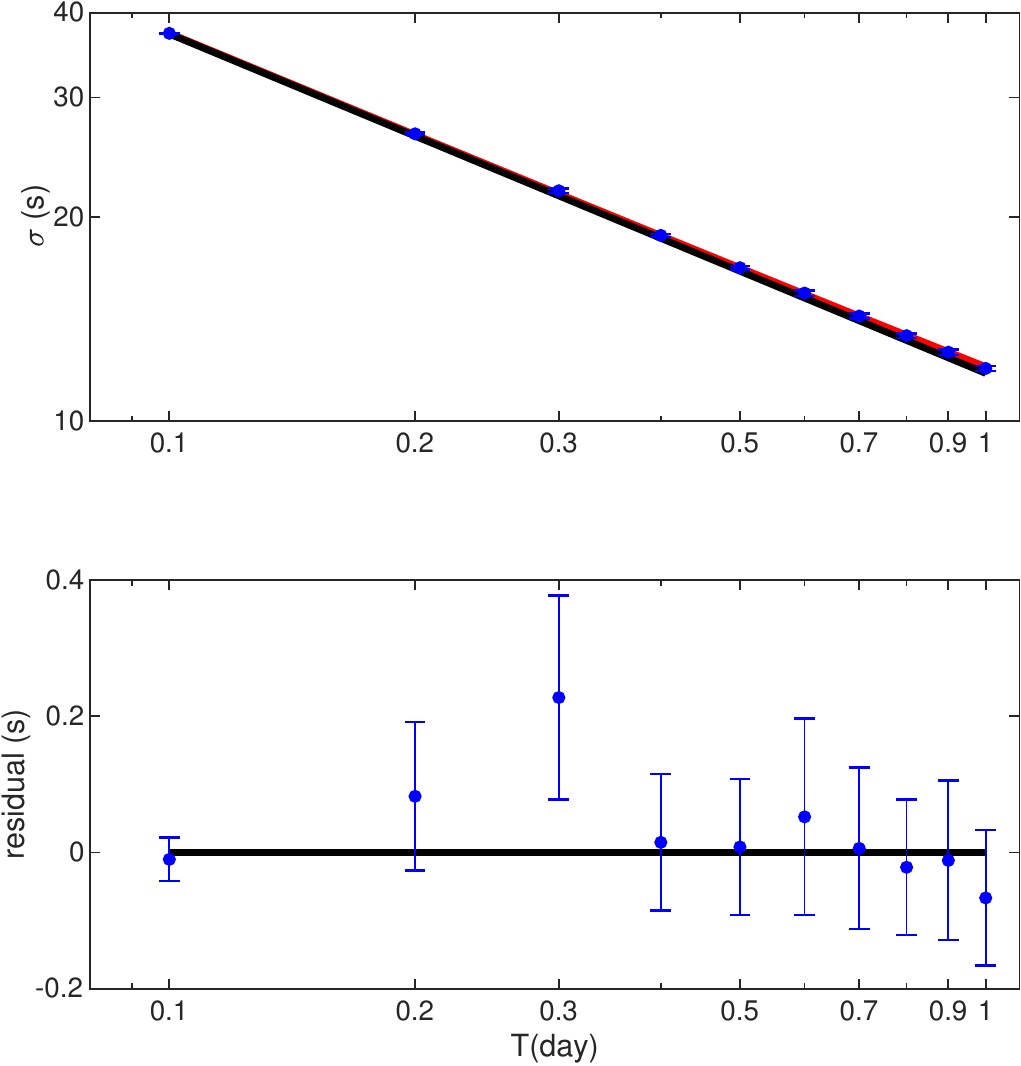}
\caption[Fit of the noise and signal contributions to the variance]{
Standard deviation of travel-time maps (top panel) and residuals from the fit $\sigma^2 = S^2 + N^2 / \left[ T/(1~{\rm day}) \right]$ (bottom panel) as functions of the averaging time $T$. In the fit, the term $S^2$ is the contribution to the variance from a time-independent flow and $N^2$ is the contribution from realization noise at $T = 1$~day. In the top panel the noise contribution is shown in black and the model including the contribution from the signal is shown in red.
}
\label{fig.variance_T}
\end{figure}

Figure~\ref{fig.variance_T} shows the variance in the travel-time maps as a function of the averaging time $T$. As expected for realization noise, the variance falls essentially as $1/T$. \shravan used the small deviation from a $1/T$ dependence to place an upper limit on the contribution of a time-independent signal to the variance. A least squares fit with the model $\sigma^2(T) = S^2 + N^2 / \left[ T/(1~\rm{day}) \right]$ yields $N = 11.77 \pm 0.02$~s and $S = 2.5 \pm 0.3$~s. This implies a noise-to-signal ratio of $N/S = 4.7 \pm 0.8$ at $T = 1$~day. This is in contrast with the value of $2$ used in \shravan.

\subsection{Calibration of \shravan travel times}

\begin{figure}
\includegraphics[width=\linewidth]{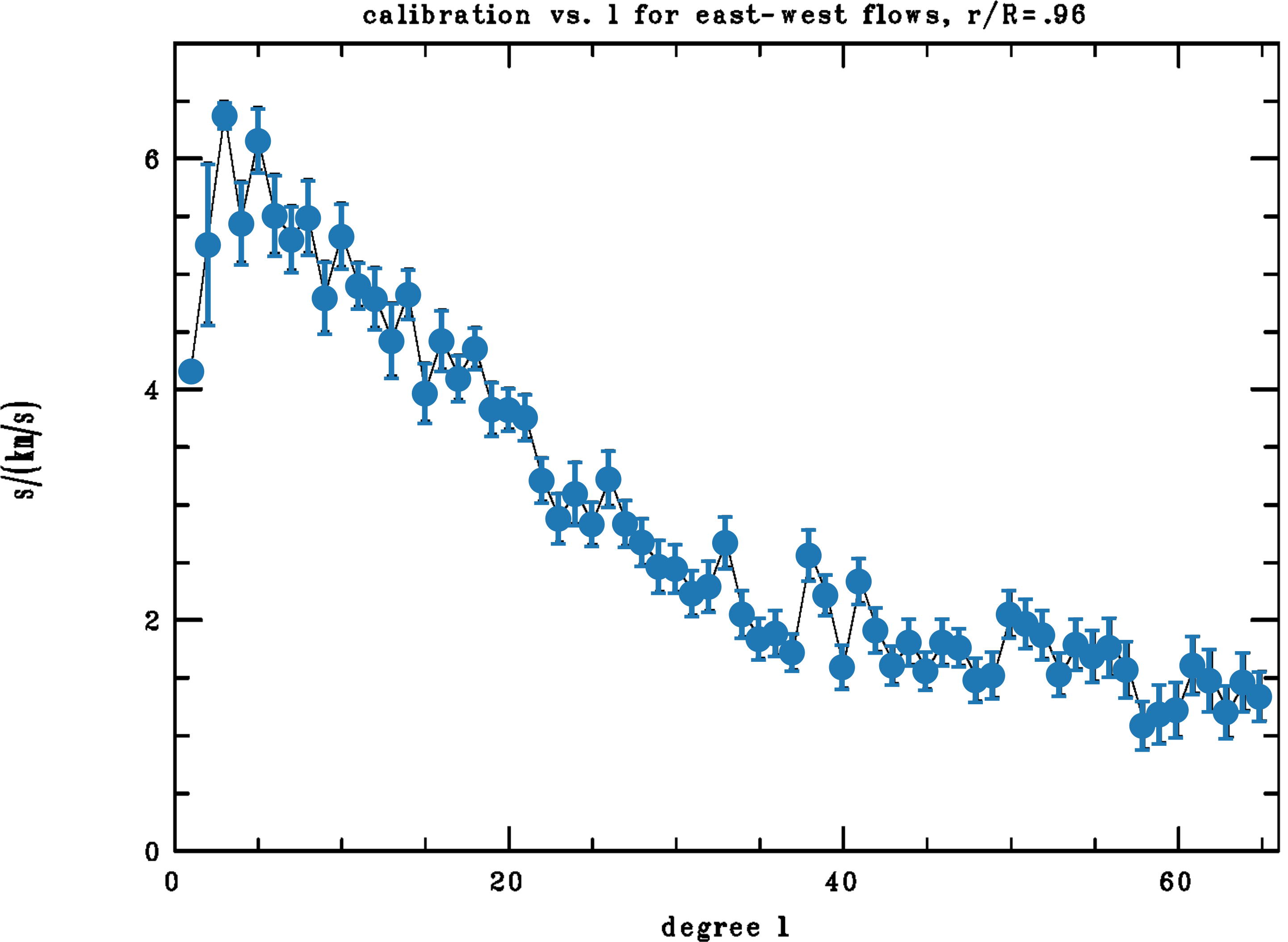}
\caption[Calibration from travel times to flows]{
Calibration curve for converting time-distance travel times to flows. The gray-scale image in the background shows Fig.~2 from the Supplementary Material of \shravan. The blue circles show the values used here.
}
\label{fig.calib}
\end{figure}

Figure~\ref{fig.calib} compares the original calibration curve with the calibration curve used in the current work. To obtain the values used here, we read values and error bars from the original figure from \shravan.

%% file: texfiles/main_convection_power/sht.tex
Here we use complex-valued spherical harmonic functions $Y_{\ell}^m$ that satisfy
\begin{equation}
\int_0^{2\pi}\!\!\int_0^{\pi} Y_{\ell}^{m*}(\theta,\phi) Y_{\ell'}^{m'}(\theta,\phi) \sin{\theta} \rmd\theta \rmd\phi = \delta_{\ell,\ell'} \delta_{m,m'} \; ,
\end{equation}
where $\theta$ is co-latitude and $\phi$ is longitude. The spherical harmonic transform of a function $f(\theta,\phi)$ is defined by:
\begin{equation}
f_{\ell,m} = \int_0^{2\pi}\!\!\int_0^{\pi} Y_{\ell}^{m*}(\theta,\phi) f(\theta,\phi) \sin{\theta} \rmd\theta \rmd\phi \; .
\end{equation}
In this paper, we will only consider functions $f$ that are real-valued. In this case, $f_{\ell,-m} = f^*_{\ell,m}$ and it is only necessary to compute the transform for $m \geq 0$. The inverse spherical harmonic transform is
\begin{equation}
\label{eq.synthesis}
f(\theta,\phi) = \sum_{\ell = 0}^{\infty} \sum_{m = -\ell}^{\ell} f_{\ell,m} Y_{\ell,m}(\theta,\phi) \; .
\end{equation}
The mean square value of a real-valued function $f$ over a sphere is related to the spherical harmonic coefficients $f_{\ell,m}$ by
\begin{equation}
\langle f^2 \rangle = \frac{1}{4\pi} \int_0^{2\pi} \!\! \int_0^{\pi} f^2(\theta,\phi) \sin{\theta} \rmd\theta \rmd\phi = \frac{1}{4\pi}\sum_{\ell = 0}^{\infty}\sum_{m = -\ell}^{\ell} |f_{\ell,m}|^2 \; .
\end{equation}
In practical applications, the sum over $\ell$ is truncated at some sufficiently large value of $\ell$. It is useful to define the power per multiplet:
\begin{equation}
\label{eq.weights}
P_\ell = \sum_{m = -\ell}^{\ell} |f_{\ell,m}|^2 = \sum_{m = 0}^{\ell} w_m |f_{\ell,m}|^2 \; ,
\end{equation}
where the weights $w_m = 2 - \delta_{m,0}$ include the contributions from $m < 0$ and avoid double counting the contribution $m = 0$. This definition implies that
\begin{equation}
\label{eq.sht_variance}
\langle f^2 \rangle = \frac{1}{4\pi}\sum_\ell P_\ell \; .
\end{equation}
The limits on the summation over $\ell$ are not written for the sake of simplicity. Summation over $\ell$ is taken over the full range.

\subsubsection{Connecting with \gizon}

To connect with the notation of \gizon
\begin{equation}
\label{eq_ephi_sht}
\sum_\ell E_\phi(\ell) = \frac{r}{2} \langle v_\phi^2\rangle = \frac{r}{8\pi} \sum_\ell P_\ell \; ,
\end{equation}
where $r$ is the radius at which the flow $v_\phi$ is measured. We can then identify $E_\phi(\ell)$ from \gizon with $r P_\ell / 8\pi$, where $P_\ell$ is the power per multiplet of the $v_\phi$ component of a flow field. 

\subsubsection{Connecting with \shravan}
\label{sec.sht_pnas}

We denote the spherical harmonic transform of \shravan with a tilde to distinguish this transform from the SHT employed elsewhere in this work. This transform is related to the previously defined transform by
\begin{equation}
\tilde{f}_{\ell,m} = \frac{1}{\sqrt{2\pi}} f_{\ell,m} \; .
\end{equation}
The power per mode is
\begin{equation}
\label{eq.shravan_power}
\tilde{P}_{\ell,m} = |\tilde{f}_{\ell,m}|^2 = \frac{1}{2\pi} |f_{\ell,m}|^2 \; .
\end{equation}

The average value of the square of $f$ over the sphere is
\begin{equation}
\langle f^2 \rangle = \frac{1}{2}\sum_\ell \sum_{m = 0}^{\ell} w_m \tilde{P}_{\ell,m} \; ,
\end{equation}
where the $w$ are defined above (Eq.~\ref{eq.weights}). For modes with $m > 0$, $w_m = 2$ and $\tilde{P}_{\ell,m}$ is the contribution of the mode $(\ell,m)$ to $\langle f^2 \rangle$. If we use $\delta f$ to denote $f$ after removing the longitude average of $f$ (this corresponds to removing the contribution of the $m = 0$ modes) then we have
\begin{equation}
\label{eq.pnas_rms}
\langle \left[ \delta f \right]^2 \rangle = \sum_\ell \sum_{m = 1}^{\ell} \tilde{P}_{\ell,m} \; .
\end{equation}

%% file: texfiles/main_convection_power/fourier.tex
In order to connect the rms flows in spherical geometry with the Cartesian-domain simulations, it is important to define the power in a way that allows direct comparisons. We use here the Fourier convention
\begin{equation}
f(\bk) = \sum_{\bf x} f(\bx) e^{-i\bk \cdot \bx} \; ,
\end{equation}
where $\bk$ is a two-dimensional horizontal wavevector. The position vector $\bx$ takes the values $(n,m) h_x$, where $h_x = h_y$ is the uniform grid spacing and $(n,m)$ is a pair of integers, each in the range $-N/2$ to $N/2 - 1$, where the even integer $N$ is the number of grid points. The wavevector, similarly, takes the values $(n,m) h_k$, where $h_k = 2\pi/(N h_x) = 2\pi/(N h_y)$ is the grid spacing in both $k_x$ and $k_y$. The inverse transform is
\begin{equation}
f(\bx) = \frac{1}{N^2}\sum_{\bf k} f(\bk) e^{i\bk \cdot \bx} \; .
\end{equation}
We distinguish the function $f(\bx)$ and its Fourier transform $f(\bk)$ only by the argument. For a real-valued function $f(\bx)$ the horizontal average of $f^2(\bx)$ is
\begin{equation}
\label{eq_mean_square_f}
\langle f^2 \rangle = \frac{1}{N^2} \sum_{\bx} f^2(\bx) = \frac{1}{N^4}\sum_{\bf k} |f(\bk)|^2 \; .
\end{equation}
The Cartesian equivalent of Eq.~\ref{eq.sht_variance} is
\begin{equation}
\langle f^2 \rangle = \sum_k P^{\rm fft}(k) \; ,
\end{equation}
where $P^{\rm fft}(k)$ is the sum of $|f(\bk)|^2/N^4$ for all grid points with $k - h_k/2 \le \|\bk\| < k + h_k/2$. The sum runs over non-negative integer values of $k/h_k$. We would like to define power per multiplet $P(\ell)$ such that
\begin{equation}
\langle f^2 \rangle = \sum_k P^{\rm fft}(k) = \sum_\ell P(\ell) \; .
\end{equation}
The contribution to $\langle f^2 \rangle$ per angular degree (i.e., for $k$ in an interval of length $1/r$) is $P^{\rm fft}(k)/(r h_k)$. This implies that
\begin{equation}
P(\ell) = \frac{1}{r h_k} P^{\rm fft}(\ell/r) \; ,
\end{equation}
where interpolation between values of $k$ is implied in the evaluation of $P^{\rm fft}(\ell/r)$.

To connect with the notation of \gizon
\begin{equation}
\sum_\ell E_\phi(\ell) = \frac{r}{2} \langle f^2 \rangle =\frac{1}{2 h_k} \sum_\ell P^{\rm fft}(\ell/r) \; ,
\end{equation}
where $r$ is the radius at which the flow field $f = v_\phi$ is measured. We can then identify $E_\phi(\ell)$ from \gizon with $1/(2 h_k) P^{\rm fft}(\ell/r)$, where, as described above, $P^{\rm fft}$ is the angle-integrated power spectrum of the east-west component of the flow field.

%% file: discussion.tex
\chapter{Discussion and outlook}
\label{chap_discussion}

\urlstyle{same}

Here I extend the analysis of the previous chapters. I first briefly adress the horizontal Rossby wave eigenfunctions as seen on the Sun. Additionally, I generalize the Rossby wave study in the radial vorticity (Chap.~\ref{chap_main_rossby}) to other observables: I study power spectra of the prograde and northward velocities $u_x$ and $u_y$, respectively, and of the horizontal divergence $\delta$. Regarding the energy spectrum of horizontal flows (Chap.~\ref{chap_main_convection}), I want to understand how these flows change in the presence of active regions. I thus study how the observed energy spectrum depends on solar activity. Finally, I outline possibilities for future research, both about solar Rossby waves and the energy spectrum of horizontal flows.

\section{Horizontal Rossby wave eigenfunctions on the Sun}

In order to obtain the horizontal eigenfunctions of solar Rossby waves, we can multiply the latitudinal eigenfunctions $C_m(\lambda)$ (Sect.~\ref{sect_latfunc}), shown in Figs.~\ref{fig_horiz_eigenfunction_all_mvals_real} and \ref{fig_horiz_eigenfunction_all_mvals_imag}, with the longitude dependence $e^{im\varphi}$ for each indiviual azimuthal order $m$. Here, however, I use eigenfunctions that were symmetrized in latitude.

\begin{figure}
\centering
\includegraphics[width=\textwidth]{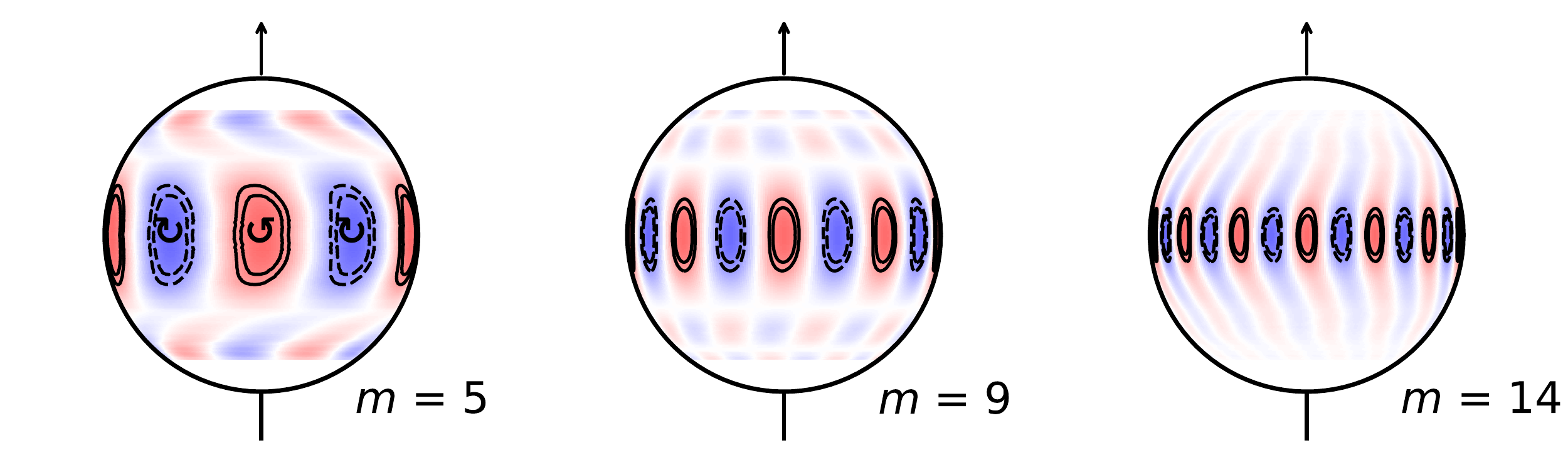}
\caption[Horizontal Rossby wave eigenfunctions on the Sun]{
Horizontal eigenfunctions of solar Rossby waves for the azimuthal orders $m = 5, 9$ and $14$ from LCT at the surface, as seen from the reference frame co-rotating at the equatorial rotation rate $\nu_{\text{eq}} = \Omega_{\text{eq}}/2\pi = \SI{453.1}{\nano\hertz}$. The solid and dashed lines indicate contours of the horizontal eigenfunctions (at $0.65$/$0.80$ and $-0.65$/$-0.80$, respectively). The counter-clockwise and clockwise arrows relate to the positive and negative radial vorticity of the horizontal eigenfunctions.}
\label{fig_horiz_eigenfunctions_rossby_waves_sphere}
\end{figure}

Fig.~\ref{fig_horiz_eigenfunctions_rossby_waves_sphere} shows the resulting horizontal eigenfunctions from LCT at the surface for $m = 5, 9$ and $14$. As can be seen, these eigenfunctions appear quite different from the sectoral spherical harmonics of early theoretical models (Fig.~\ref{fig_rossby_waves}). The horizontal eigenfunctions of solar Rossby waves are curved toward the retrograde direction for low $m$, they are nearly aligned in latitude and have a sign flip for intermediate $m$, and they are curved toward the prograde direction for high $m$. These different shapes are mainly due to the different sign of the imaginary part of the eigenfunctions for low and high $m$ (see Fig.~\ref{fig_horiz_eigenfunction_all_mvals_imag} and Sect.~\ref{sect_latfunc_results}). The presence of an imaginary part and thus curved horizontal eigenfunctions is possibly related to the interaction of viscous Rossby waves and latitudinal differential rotation, as described in \citet[][submitted]{Gizon2020}.

\section{Rossby waves as observed in the horizontal velocities}

I extend the HMI ring-diagram time series of $u_x$ and $u_y$ by roughly one and a half years to a total of more than nine years (May 19, 2010 to June 1, 2019). I process these flows as in Sect.~\ref{sect_data_methods_rda_overview} and compute the radial vorticity via Eq.~\ref{eq_definition_radial_vorticity}. The horizontal divergence is
\begin{equation}
\label{eq_horizontal_divergence}
\delta(t,r,\lambda,\varphi) = \frac{1}{r \cos\lambda} \frac{\partial u_x(t,r,\lambda,\varphi)}{\partial \varphi} + \frac{1}{r \cos\lambda} \frac{\partial (u_y(t,r,\lambda,\varphi) \cos\lambda)}{\partial\lambda}.
\end{equation}
I then obtain power spectra for the horizontal velocities $u_x$ and $u_y$ (after the subtraction of the longitude average), the horizontal divergence $\delta$, and the radial vorticity $\zeta$ (analogously to Chap.~\ref{chap_main_rossby}). The longer observation period implies a better frequency resolution.

\begin{sidewaysfigure}
\centering
\includegraphics[width=\textwidth]{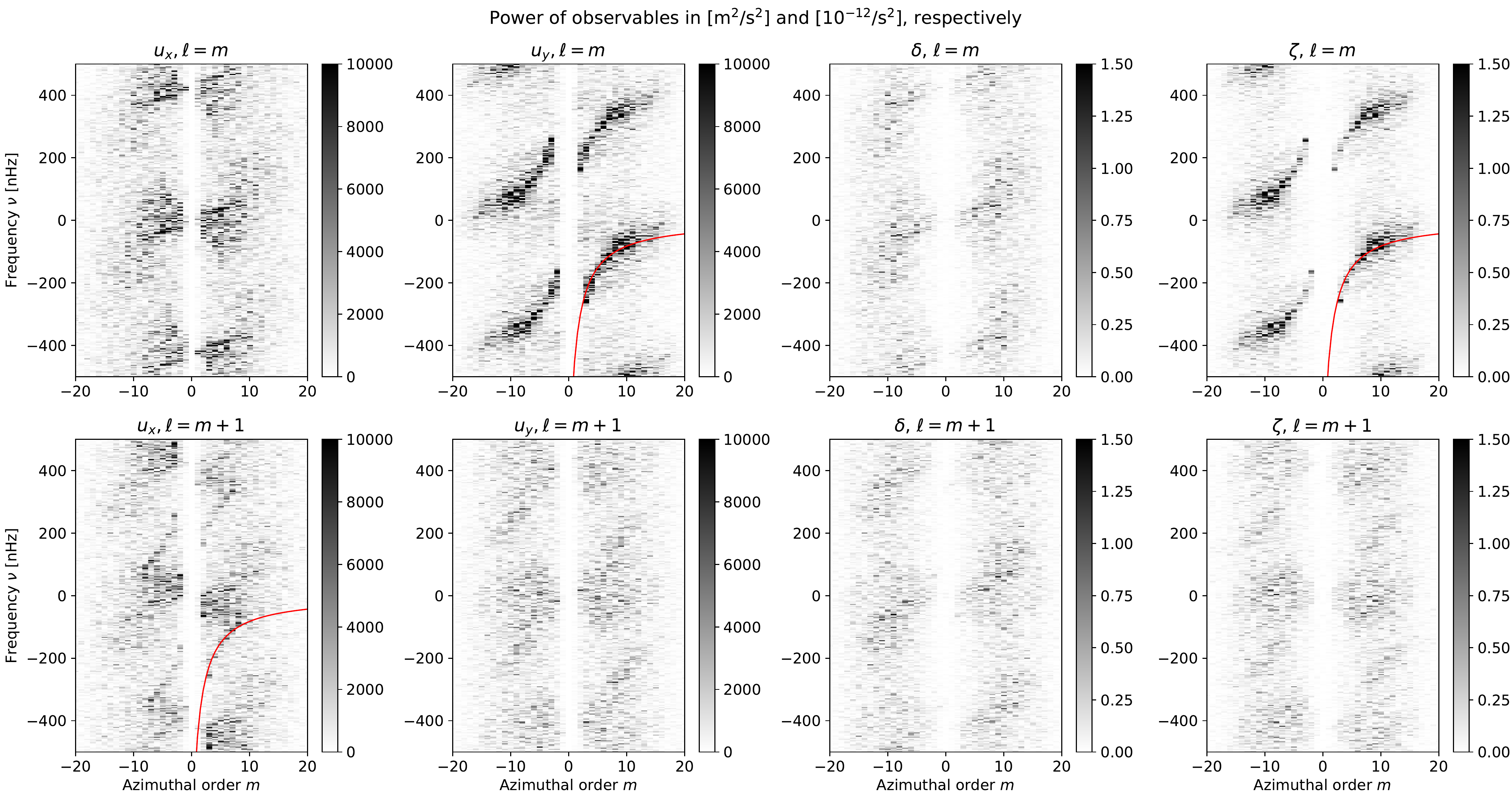}
\caption[Power spectra of horizontal velocities, horizontal divergence and radial vorticity]{
Power spectra of the prograde velocity $u_x$, northward velocity $u_y$, horizontal divergence $\delta$ and radial vorticity $\zeta$ (columns) versus frequency $\nu$ and azimuthal order $m$, for $\ell = m$ (top row) and $\ell = m + 1$ (bottom row), from RDA at a depth of $\SI{0.7}{\mega\metre}$. The red line shows the textbook Rossby wave dispersion relation (Eq.~\ref{eq_rossby_dispersion}) for the sectoral case ($\ell = m$).
}
\label{fig_power_spectra_sht}
\end{sidewaysfigure}

Figure~\ref{fig_power_spectra_sht} shows the resulting power spectra for $\ell = m$ and $\ell = m + 1$, for RDA data near the surface. The $\ell = m$ power spectrum for $\zeta$ is very similar to Fig.~\ref{fig_power_spectrum_2d}. The $\ell = m$ power spectrum for $u_y$ also contains Rossby waves (see \citet{Liang2019}). I do not observe $\ell = m = 2$ or $\ell = m = 1$ power near the frequencies predicted by the theoretical dispersion relation, neither in $\zeta$ nor in $u_y$, see \citet{Loeptien2018} and \citet{Liang2019}, respectively.

Rossby wave observations in $u_x$ have not yet been discussed in the literature. The $\ell = m + 1$ power spectrum for $u_x$, however, shows a ridge, which coincides in frequency with the $\ell = m$ Rossby wave power for $u_y$ and $\zeta$. The north-south anti-symmetry around the equator in $u_x$ is expected for sectoral Rossby waves, as is the north-south symmetry in $u_y$ and $\zeta$ (Sect.~\ref{sect_intro_rossby_waves}). These symmetries are visible in schematic illustrations of the flow field, see Fig.~\ref{fig_rossby_waves}, left panel.

I do not see north-south anti-symmetric Rossby waves in $u_y$ or $\zeta$ (or north-south symmetric Rossby waves in $u_x$). This is consistent with Chap.~\ref{chap_main_rossby} and the earlier observations by \citet{Loeptien2018} and \citet{Liang2019}. \citet{Loeptien2018} hypothesize that non-sectoral Rossby waves dissipate on much shorter timescales than sectoral ones, which may be why those modes are not observed. I also do not observe Rossby waves in $\delta$, similar to \citet{Loeptien2018}. This is consistent with the assumption of zero horizontal divergence used for the derivation of the theoretical Rossby wave dispersion relation (Sect.~\ref{sect_intro_rossby_waves}).

\section{Relation of energy spectra of horizontal flows to solar activity}

\subsection{Energy spectra of horizontal flows versus sunspot number}

Here I use LCT $u_x$ and $u_y$ velocities that have been averaged over $24$~hr in the Carrington reference frame, instead of the full time resolution of $30$~min (for computational reasons). I compute the energy spectra $E_x(\ell)$\footnote{In Chap.~\ref{chap_main_convection} the energy spectra of $u_x$ and $u_y$ flows are denoted as $E_\phi(\ell)$ and $E_\theta(\ell)$, respectively.} according to Eq.~\ref{eq_ephi_sht} for every time step and bin them in solar activity. For this I use the $13$-month smoothed monthly total sunspot number (SSN) \footnote{\url{http://sidc.be/silso/DATA/SN_ms_tot_V2.0.txt}} from the \textnormal{World Data Center - Sunspot Index and Long-term Solar Observations} \citep[WDC-SILSO,][]{Clette2014}, and split the energy spectra into three equal-sized bins (\textnormal{terciles}). I average $E_x(\ell)$ within the individual bins and also over all time steps at once.

\begin{figure}
\centering
\includegraphics[width=\textwidth]{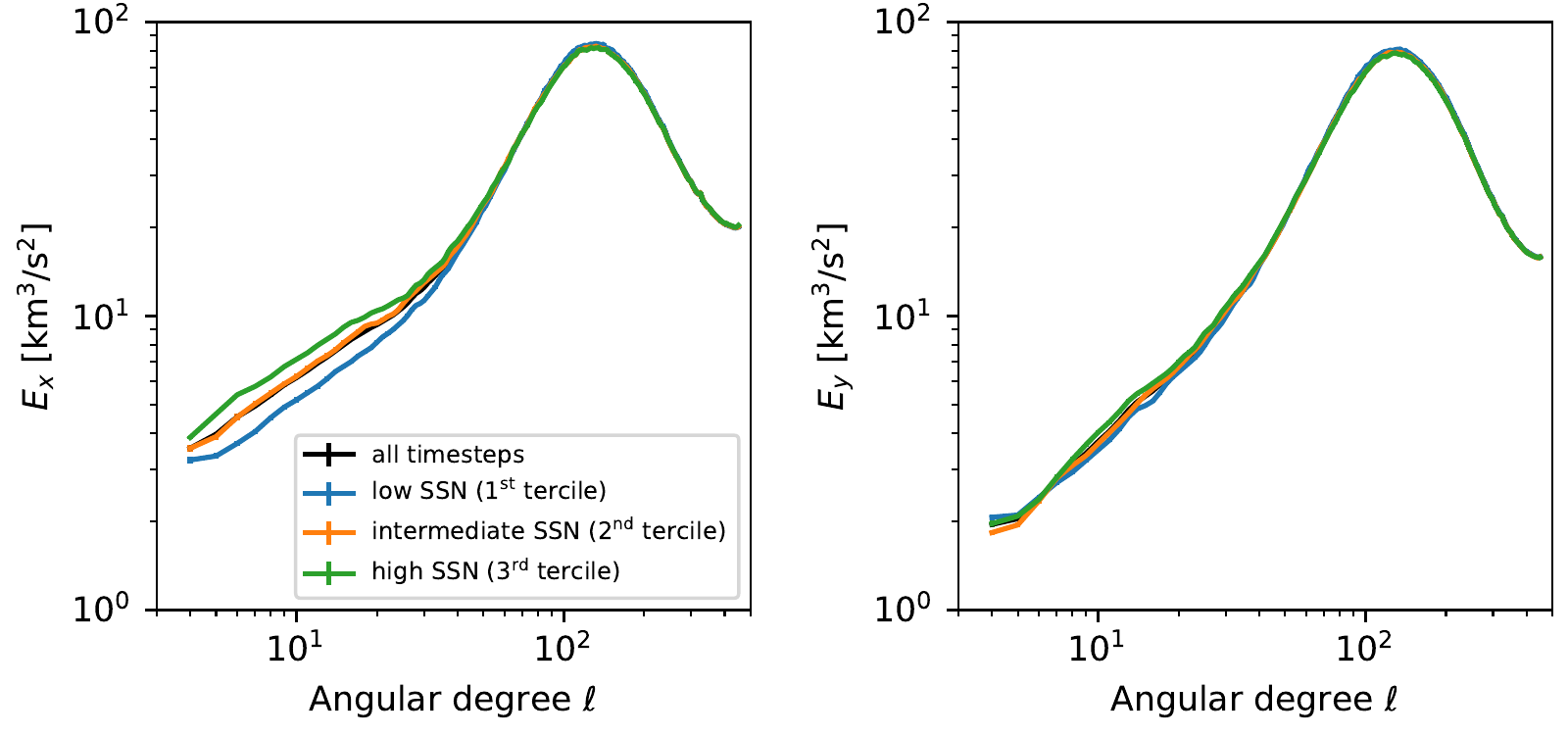}
\caption[Energy spectra of horizontal flows versus sunspot number]{
Left: Energy spectra $E_x(\ell)$ of $u_x$ velocities from LCT for different levels of solar activity. The black, blue, orange and green curves indicate the energy spectra averaged over all time steps, the first, second and third tercile of the sunspot number, respectively (see text). Right: Analogous results for energy spectra $E_y(\ell)$ of $u_y$ velocities from LCT. The error bars in both panels are comparable to the linewidths.
}
\label{fig_energy_spectra_ar_flows}
\end{figure}

The left panel of Fig.~\ref{fig_energy_spectra_ar_flows} shows that the energy spectrum for intermediate sunspot numbers is similar to that averaged over all time steps. At very large $\ell$, the energy spectra do not depend strongly on the sunspot number. However, at small to intermediate $\ell$ ($4 \lesssim \ell \lesssim 60$, i.e spatial extents of $2\pi/\ell \sim 6$-$\SI{90}{\degree}$), the energy increases with the sunspot number.

The energy ratio between high and low sunspot numbers is roughly $1.5$ at $\ell \sim 5$ and decreases to unity around $\ell \sim 60$. The excess power could potentially be due to flows related to active regions (Sect.~\ref{sect_intro_flows_active_regions}). The spatial scale of the largest excess, about $2\pi/\ell \sim \SI{72}{\degree}$, is much larger than flows up to $10$ to $\SI{15}{\degree}$ from the active region center \citep{Loeptien2017, Braun2019}. The size of groups of active regions might, however, affect the power spectrum of horizontal flows on spatial scales larger than this extent.

Around the supergranulation scale ($\ell \sim 120$), the power decreases with solar activity, with an energy ratio between high and low sunspot numbers of roughly $\SI{97}{\percent}$. Given that the power should be proportional to the area, this might be caused by a smaller area on the solar surface covered by supergranules for high solar activity (due to suppression of convection by strong magnetic fields within active regions). Assuming suppressed supergranulation inside the active regions (Fig.~\ref{fig_ar_masking_example}, within the red contours), the fractional disk area covered by supergranules is (on average) $\sim \SI{97}{\percent}$. This is in good agreement with the observed decrease of power.

\subsection{Effect of active regions on energy spectra of horizontal flows}

To check if flows around active regions indeed cause the observed excess power at high solar activity, I mask these flows. For this, I use HMI LOS magnetograms, tracked and averaged in time like the LCT velocity maps.

\begin{figure}
\centering
\includegraphics[width=\textwidth]{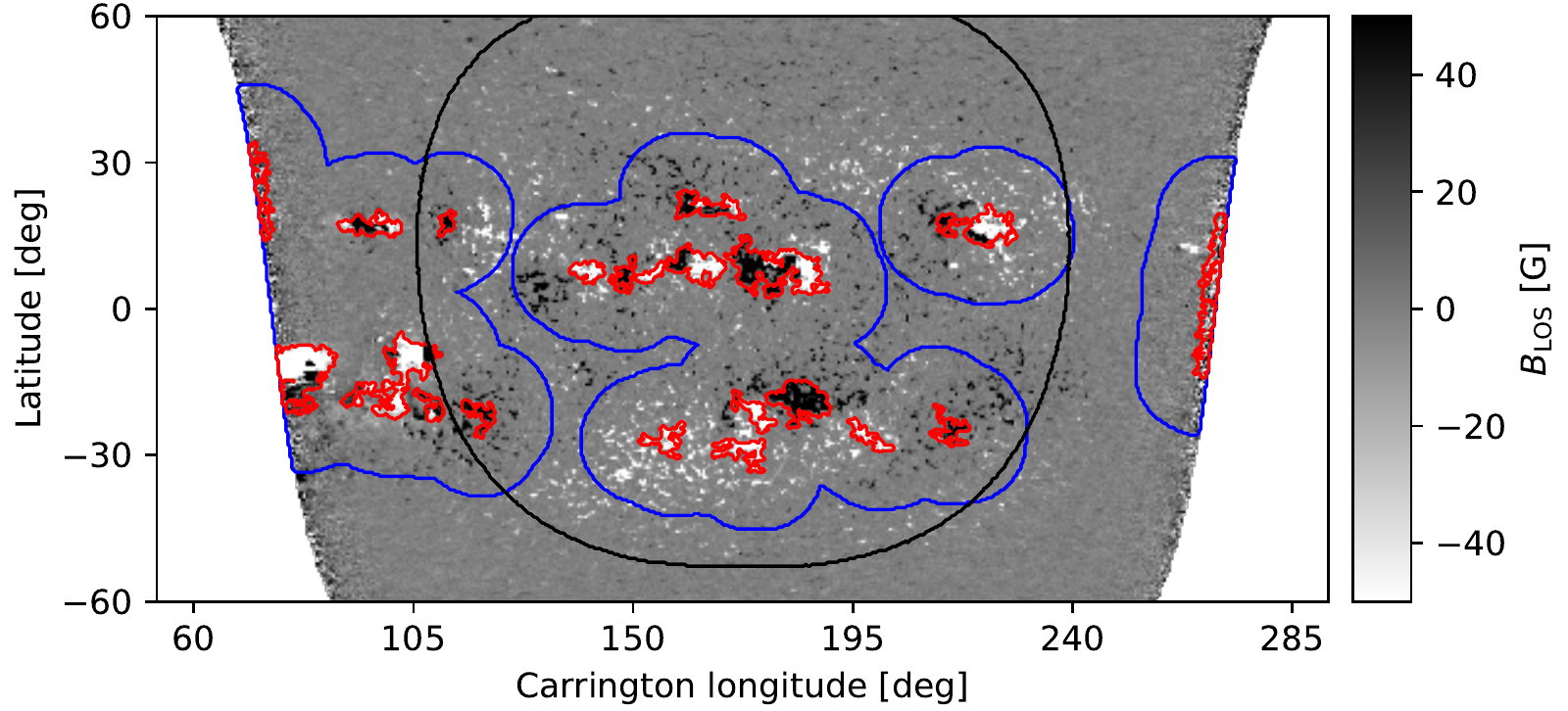}
\caption[Masking active regions and their related flows, example time step]{
Active region masking example. The line-of-sight magnetic field $B_{\textrm{LOS}}$ on September 24, 2012 is shown versus latitude and Carrington longitude. The contours indicate the border of the solar disk in the LCT velocity data for the same time step (black; it extends to $\SI{60}{\degree}$ from disk center) and the outer borders of masked active regions (red) and their surroundings (blue).
}
\label{fig_ar_masking_example}
\end{figure}

I apply a masking algorithm for active regions by \mbox{P.-L.}~Poulier. For every time step it creates pixel masks for three different regions (Fig.~\ref{fig_ar_masking_example}). First, there are the active regions themselves, defined via thresholds for the line-of-sight component of the magnetic field $B_{\textrm{LOS}}$ on the strength ($\geq 20$~G) and the flux ($\geq 10^{21}$~Mx). Second, there are rings around the active regions, which should contain the flows around the active regions. Third, there is the remaining solar disk, which should be the quiet Sun.

I distinguish between quiet-Sun maps (those without any active regions) and active region maps (with their pixel masks). Subsequently, I compute the energy spectra $E_x(\ell)$ according to Eq.~\ref{eq_ephi_sht} and average them, for both types of maps. The comparison of energy spectra for the active region maps to those for the quiet-Sun maps yields results that are qualitatively consistent with the previous sunspot number analysis: I observe a similar excess power at small to intermediate $\ell$ and a decrease of power at the supergranulation scale.

To study the origin of the excess power at intermediate spatial scales, I use quiet-Sun maps, where I replace pixels with those from active region maps, either within the interior or the surroundings of active regions (Fig.~\ref{fig_ar_masking_example}). I find that the increased power is mainly generated inside the active regions: Energy spectra for quiet-Sun maps with the interior of active regions replaced, are, between $\ell \sim 4$ and $20$, roughly $1.1$ to $1.3$ times larger than those for quiet-Sun maps with the surroundings of active regions replaced. Flows around active regions thus do not have a large effect on the energy spectrum of horizontal flows. Instead, flows inside active regions, such as the moat flow around sunspots, could be responsible for the excess power.

I also repeat the above analysis for the northward velocity $u_y$. The right panel of Fig.~\ref{fig_energy_spectra_ar_flows} shows the energy spectra $E_y(\ell)$ for the binning in sunspot number. The energy spectra of $u_y$ depend much less on solar activity than those of $u_x$. I note, however, that LCT velocity measurements at locations of high magnetic field strengths (beyond $500$~G) are unreliable \citep{Loeptien2017}.

\section{Outlook}

\subsection{Rossby waves}

Despite the new results presented in this thesis, both for solar Rossby waves and the energy spectrum of horizontal flows, many questions are still open. For example, the uncertainties on the depth dependence of Rossby waves from ring-diagram analysis are too large to constrain theories. Further research could be done with both theory (using models) and observations (using datasets that reach larger depths). For the latitude dependence of Rossby waves, improved modeling is needed to understand the observations. Theoretical research about Rossby waves is ongoing, but needs to take differential rotation, damping, and potentially the magnetic field into account. The dependence of the mode frequency on the differential rotation and the magnetic field is another topic for further Rossby wave studies. The excitation and damping of Rossby waves by convection should also be studied. 

\subsection{Convection}

The discrepancy between energy spectra of horizontal flows from simulations/multi-ridge-fitting ring-diagrams and time-distance helioseismology remains mostly unaffected by the new, consistent results that I obtained from LCT and RDA. Synthetic data could help to resolve the disagreement between the observational results, via a comparison of the expected energy spectra to those obtained with different observational methods. A better understanding of solar convection on large scales may help to study the origin of the convective conundrum. Apart from the prograde and northward velocities, energy spectra of the horizontal divergence and the radial vorticity could be studied. A similar analysis has been done by \citet{Hathaway2015b}. 

Regarding supergranulation, it is still unclear why it appears as a distinct scale in the energy spectrum of horizontal flows. Current models attribute the origin of the supergranulation scale to supergranulation being the largest scale of convection that is either driven by buoyancy \citep{Cossette2016} or that is not strongly influenced by the solar rotation \citep{Featherstone2016}.